\documentclass[aps,prb,superscriptaddress,twocolumn,preprintnumbers,amsmath,amssymb,floatfix]{revtex4-1}
\usepackage{color}
\usepackage{array}
\usepackage{amsmath}
\usepackage{amssymb}
\usepackage{amsbsy}
\usepackage{graphicx}
\usepackage{epstopdf}
\usepackage{hyperref}
\usepackage{float}
\usepackage{bibentry}
\usepackage{diagbox}
\usepackage{slashed}
\usepackage{framed}
\usepackage{leftidx}
\usepackage{braket}
\usepackage{subfig}

\usepackage{tikz}
\newcommand\BFGA[1][0.55]{
  \begin{tikzpicture}[scale=#1]
    \draw (-0.5,0.866)--(0.5,-0.866)--(-0.5,-0.866)--(0.5,0.866)--(-0.5,0.866);
  \node at (-0.5,0.866) [inner sep=0] {$\uparrow$};
      \node at (0.5,0.866)[inner sep=0] {$\downarrow$};
  \node at (-0.5,-0.866) [inner sep=0] {$\downarrow$};
  \node at (0.5,-0.866)[inner sep=0] {$\uparrow$};
  \end{tikzpicture}
  }
\newcommand\BFGB[1][0.55]{
  \begin{tikzpicture}[scale=#1]
    \draw (-0.5,0.866)--(0.5,-0.866)--(-0.5,-0.866)--(0.5,0.866)--(-0.5,0.866);
 \node at (-0.5,0.866) [inner sep=0] {$\downarrow$};
      \node at (0.5,0.866)[inner sep=0] {$\uparrow$};
  \node at (-0.5,-0.866) [inner sep=0] {$\uparrow$};
  \node at (0.5,-0.866)[inner sep=0] {$\downarrow$};
  \end{tikzpicture}
  }

\newcommand\bowtieA[1][0.55]{
  \begin{tikzpicture}[scale=#1]
    \draw (-0.5,0.866)--(0.5,-0.866)--(-0.5,-0.866)--(0.5,0.866)--(-0.5,0.866);
    \draw [opacity=0.3] (-1,-1.154)--(0,-0.577)--(0,0.577)--(-1,1.154);
    \draw [opacity=0.3] (0,0.577)--(1,1.154);
    \draw [opacity=0.3] (0,-0.577)--(1,-1.154);
  \node at (-0.5,0.866) [inner sep=0] {$\uparrow$};
      \node at (0.5,0.866)[inner sep=0] {$\downarrow$};
  \node at (-0.5,-0.866) [inner sep=0] {$\downarrow$};
  \node at (0.5,-0.866)[inner sep=0] {$\uparrow$};
    \node at (0,0.577) [inner sep=0] {$\scriptstyle \sigma_I^z$};
    \node at (0,-0.577) [inner sep=0] {$\scriptstyle \sigma_J^z$};
  \end{tikzpicture}
  }
\newcommand\bowtieB[1][0.55]{
  \begin{tikzpicture}[scale=#1]
    \draw (-0.5,0.866)--(0.5,-0.866)--(-0.5,-0.866)--(0.5,0.866)--(-0.5,0.866);
    \draw [opacity=0.3] (-1,-1.154)--(0,-0.577)--(0,0.577)--(-1,1.154);
    \draw [opacity=0.3] (0,0.577)--(1,1.154);
    \draw [opacity=0.3] (0,-0.577)--(1,-1.154);
 \node at (-0.5,0.866) [inner sep=0] {$\downarrow$};
      \node at (0.5,0.866)[inner sep=0] {$\uparrow$};
  \node at (-0.5,-0.866) [inner sep=0] {$\uparrow$};
  \node at (0.5,-0.866)[inner sep=0] {$\downarrow$};
    \node at (0,0.577) [inner sep=0] {$\scriptstyle -\sigma_I^z$};
    \node at (0,-0.577) [inner sep=0] {$\scriptstyle -\sigma_J^z$};
  \end{tikzpicture}
  }

\newcommand\pgA[1][0.2]{
  \begin{tikzpicture}[scale=#1]
    \draw (0,0)--(1,0)--(1.5,0.866)--(0.5,0.866)--(0,0);
    \draw (1,0)--(0.5,0.866);
  \end{tikzpicture}
  }
\newcommand\pgB[1][0.2]{
  \begin{tikzpicture}[scale=#1]
    \draw[rotate around={120:(0.75,0.433)},black] (0,0)--(1,0)--(1.5,0.866)--(0.5,0.866)--(0,0);
    \draw[rotate around={120:(0.75,0.433)},black] (1,0)--(0.5,0.866);
  \end{tikzpicture}
  }
\newcommand\pgC[1][0.2]{
  \begin{tikzpicture}[scale=#1]
    \draw[rotate around={-120:(0.75,0.433)}] (0,0)--(1,0)--(1.5,0.866)--(0.5,0.866)--(0,0);
    \draw[rotate around={-120:(0.75,0.433)}] (1,0)--(0.5,0.866);
  \end{tikzpicture}
  }
\newcommand\pA[1][0.3]{
  \begin{tikzpicture}[scale=#1]
    \draw [line width=0.1mm] (0,0)--(1,0);
    \draw [line width=0.6mm] (1,0)--(1.5,0.866);
    \draw [line width=0.1mm] (1.5,0.866)--(0.5,0.866);
    \draw [line width=0.6mm] (0.5,0.866)--(0,0);
    \draw [line width=0.1mm] (1,0)--(0.5,0.866);
  \end{tikzpicture}
  }
\newcommand\pB[1][0.3]{
  \begin{tikzpicture}[scale=#1]
    \draw [line width=0.6mm] (0,0)--(1,0);
    \draw [line width=0.1mm] (1,0)--(1.5,0.866);
    \draw [line width=0.6mm] (1.5,0.866)--(0.5,0.866);
    \draw [line width=0.1mm] (0.5,0.866)--(0,0);
    \draw [line width=0.1mm] (1,0)--(0.5,0.866);
  \end{tikzpicture}
  }

\newcommand\pC[1][0.3]{
  \begin{tikzpicture}[scale=#1]
    \draw [rotate around={120:(0.75,0.433)}] [line width=0.1mm] (0,0)--(1,0);
    \draw [rotate around={120:(0.75,0.433)}] [line width=0.6mm] (1,0)--(1.5,0.866);
    \draw [rotate around={120:(0.75,0.433)}] [line width=0.1mm] (1.5,0.866)--(0.5,0.866);
    \draw [rotate around={120:(0.75,0.433)}] [line width=0.6mm] (0.5,0.866)--(0,0);
    \draw [rotate around={120:(0.75,0.433)}] [line width=0.1mm] (1,0)--(0.5,0.866);
  \end{tikzpicture}
  }
\newcommand\pD[1][0.3]{
  \begin{tikzpicture}[scale=#1]
    \draw [rotate around={120:(0.75,0.433)}][line width=0.6mm] (0,0)--(1,0);
    \draw [rotate around={120:(0.75,0.433)}][line width=0.1mm] (1,0)--(1.5,0.866);
    \draw [rotate around={120:(0.75,0.433)}][line width=0.6mm] (1.5,0.866)--(0.5,0.866);
    \draw [rotate around={120:(0.75,0.433)}][line width=0.1mm] (0.5,0.866)--(0,0);
    \draw [rotate around={120:(0.75,0.433)}][line width=0.1mm] (1,0)--(0.5,0.866);
  \end{tikzpicture}
  }
\newcommand\pE[1][0.3]{
  \begin{tikzpicture}[scale=#1]
    \draw [rotate around={-120:(0.75,0.433)}] [line width=0.1mm] (0,0)--(1,0);
    \draw [rotate around={-120:(0.75,0.433)}] [line width=0.6mm] (1,0)--(1.5,0.866);
    \draw [rotate around={-120:(0.75,0.433)}] [line width=0.1mm] (1.5,0.866)--(0.5,0.866);
    \draw [rotate around={-120:(0.75,0.433)}] [line width=0.6mm] (0.5,0.866)--(0,0);
    \draw [rotate around={-120:(0.75,0.433)}] [line width=0.1mm] (1,0)--(0.5,0.866);
  \end{tikzpicture}
  }
\newcommand\pF[1][0.3]{
  \begin{tikzpicture}[scale=#1]
    \draw [rotate around={-120:(0.75,0.433)}][line width=0.6mm] (0,0)--(1,0);
    \draw [rotate around={-120:(0.75,0.433)}][line width=0.1mm] (1,0)--(1.5,0.866);
    \draw [rotate around={-120:(0.75,0.433)}][line width=0.6mm] (1.5,0.866)--(0.5,0.866);
    \draw [rotate around={-120:(0.75,0.433)}][line width=0.1mm] (0.5,0.866)--(0,0);
    \draw [rotate around={-120:(0.75,0.433)}][line width=0.1mm] (1,0)--(0.5,0.866);
  \end{tikzpicture}
  }
  
\newcommand\pgelA[1][0.8]{
  \begin{tikzpicture}[scale=#1]
    \draw (0,0)--(1,0)--(1.5,0.866)--(0.5,0.866)--(0,0);
    \draw [opacity=0.2] (1,0)--(0.5,0.866);
    \draw [opacity=0.2] (0,0.5774)--(0.5,0.2887)--(1,0.5774)--(1.5,0.2887);
    \draw [opacity=0.2] (0.5,-0.2887)--(0.5,0.2887);
    \draw [opacity=0.2] (1,0.5774)--(1,1.1547);
    \draw [rotate around={-120:(0.25,0.433)}] (0.25,0.433) ellipse (15pt and 2pt); 
    \draw [rotate around={-120:(1.25,0.433)}] (1.25,0.433) ellipse (15pt and 2pt); 
    \node at (0.5,0.2) [inner sep=0] {$\scriptstyle-\sigma_I^z$};
    \node at (0.9,0.65) [inner sep=0] {$\scriptstyle-\sigma_J^z$};
  \end{tikzpicture}
  }
\newcommand\pgelB[1][0.8]{
  \begin{tikzpicture}[scale=#1]
    \draw (0,0)--(1,0)--(1.5,0.866)--(0.5,0.866)--(0,0);
    \draw [opacity=0.2] (1,0)--(0.5,0.866);
    \draw [opacity=0.2] (0,0.5774)--(0.5,0.2887)--(1,0.5774)--(1.5,0.2887);
    \draw [opacity=0.2] (0.5,-0.2887)--(0.5,0.2887);
    \draw [opacity=0.2] (1,0.5774)--(1,1.1547);
    \draw (0.5,0) ellipse (15pt and 2pt); 
    \draw (1,0.866) ellipse (15pt and 2pt);
    \node at (0.5,0.2887) [inner sep=0] {$\scriptstyle\sigma_I^z$};
    \node at (1,0.5774) [inner sep=0] {$\scriptstyle\sigma_J^z$};
  \end{tikzpicture}
}

\newcommand\pgelC[1][0.8]{
  \begin{tikzpicture}[scale=#1]
    \draw (0,0)--(1,0)--(1.5,0.866)--(0.5,0.866)--(0,0);
    \draw [opacity=0.2] (1,0)--(0.5,0.866);
    \draw [opacity=0.2] (0,0.5774)--(0.5,0.2887)--(1,0.5774)--(1.5,0.2887);
    \draw [opacity=0.2] (0.5,-0.2887)--(0.5,0.2887);
    \draw [opacity=0.2] (1,0.5774)--(1,1.1547);
    \draw [rotate around={-120:(0.25,0.433)}] (0.25,0.433) ellipse (15pt and 2pt); 
    \draw [rotate around={-120:(1.25,0.433)}] (1.25,0.433) ellipse (15pt and 2pt); 
    \node at (0.5,0.2887) [inner sep=0] {$\scriptstyle\sigma_I^z$};
    \node at (1,0.5774) [inner sep=0] {$\scriptstyle\sigma_J^z$};
  \end{tikzpicture}
}
\newcommand\pgelD[1][0.8]{
  \begin{tikzpicture}[scale=#1]
    \draw (0,0)--(1,0)--(1.5,0.866)--(0.5,0.866)--(0,0);
    \draw [opacity=0.2] (1,0)--(0.5,0.866);
    \draw [opacity=0.2] (0,0.5774)--(0.5,0.2887)--(1,0.5774)--(1.5,0.2887);
    \draw [opacity=0.2] (0.5,-0.2887)--(0.5,0.2887);
    \draw [opacity=0.2] (1,0.5774)--(1,1.1547);
    \draw (0.5,0) ellipse (15pt and 2pt); 
    \draw (1,0.866) ellipse (15pt and 2pt);
    \node at (0.5,0.2887) [inner sep=0] {$\scriptstyle\sigma_I^z$};
    \node at (1,0.5774) [inner sep=0] {$\scriptstyle\sigma_J^z$};
  \end{tikzpicture}
}

\newcommand\pgelE[1][0.8]{
  \begin{tikzpicture}[scale=#1]
    \draw [rotate around={120:(0.75,0.433)}] (0,0)--(1,0)--(1.5,0.866)--(0.5,0.866)--(0,0);
    \draw [rotate around={120:(0.75,0.433)}] [opacity=0.2] (1,0)--(0.5,0.866);
    \draw [rotate around={120:(0.75,0.433)}] [opacity=0.2] (0,0.5774)--(0.5,0.2887)--(1,0.5774)--(1.5,0.2887);
    \draw [rotate around={120:(0.75,0.433)}] [opacity=0.2] (0.5,-0.2887)--(0.5,0.2887);
    \draw [rotate around={120:(0.75,0.433)}] [opacity=0.2] (1,0.5774)--(1,1.1547);
    \draw [rotate around={120:(0.75,0.433)}] [rotate around={-120:(0.25,0.433)}] (0.25,0.433) ellipse (15pt and 2pt); 
    \draw [rotate around={120:(0.75,0.433)}] [rotate around={-120:(1.25,0.433)}] (1.25,0.433) ellipse (15pt and 2pt); 
    \node  at (0.9,0.25) [inner sep=0] {$\scriptstyle-\sigma_I^z$};
    \node  at (0.5,0.6) [inner sep=0] {$\scriptstyle-\sigma_J^z$};
  \end{tikzpicture}
  }
\newcommand\pgelF[1][0.8]{
  \begin{tikzpicture}[scale=#1]
    \draw [rotate around={120:(0.75,0.433)}](0,0)--(1,0)--(1.5,0.866)--(0.5,0.866)--(0,0);
    \draw [rotate around={120:(0.75,0.433)}][opacity=0.2] (1,0)--(0.5,0.866);
    \draw [rotate around={120:(0.75,0.433)}][opacity=0.2] (0,0.5774)--(0.5,0.2887)--(1,0.5774)--(1.5,0.2887);
    \draw [rotate around={120:(0.75,0.433)}][opacity=0.2] (0.5,-0.2887)--(0.5,0.2887);
    \draw [rotate around={120:(0.75,0.433)}][opacity=0.2] (1,0.5774)--(1,1.1547);
    \draw [rotate around={120:(0.75,0.433)}](0.5,0) ellipse (15pt and 2pt); 
    \draw [rotate around={120:(0.75,0.433)}](1,0.866) ellipse (15pt and 2pt);
    \node at (0.9,0.25) [inner sep=0] {$\scriptstyle\sigma_I^z$};
    \node at (0.5,0.6) [inner sep=0] {$\scriptstyle\sigma_J^z$};
  \end{tikzpicture}
}

\newcommand\pgelG[1][0.8]{
  \begin{tikzpicture}[scale=#1]
    \draw [rotate around={-120:(0.75,0.433)}](0,0)--(1,0)--(1.5,0.866)--(0.5,0.866)--(0,0);
    \draw [rotate around={-120:(0.75,0.433)}][opacity=0.2] (1,0)--(0.5,0.866);
    \draw [rotate around={-120:(0.75,0.433)}][opacity=0.2] (0,0.5774)--(0.5,0.2887)--(1,0.5774)--(1.5,0.2887);
    \draw [rotate around={-120:(0.75,0.433)}][opacity=0.2] (0.5,-0.2887)--(0.5,0.2887);
    \draw [rotate around={-120:(0.75,0.433)}][opacity=0.2] (1,0.5774)--(1,1.1547);
    \draw [rotate around={-120:(0.75,0.433)}][rotate around={-120:(0.25,0.433)}] (0.25,0.433) ellipse (15pt and 2pt); 
    \draw [rotate around={-120:(0.75,0.433)}][rotate around={-120:(1.25,0.433)}] (1.25,0.433) ellipse (15pt and 2pt); 
    \node at (0.65,0.6) [inner sep=0] {$\scriptstyle-\sigma_I^z$};
    \node at (0.7,0.2) [inner sep=0] {$\scriptstyle-\sigma_J^z$};
  \end{tikzpicture}
  }
\newcommand\pgelH[1][0.8]{
  \begin{tikzpicture}[scale=#1]
    \draw [rotate around={-120:(0.75,0.433)}](0,0)--(1,0)--(1.5,0.866)--(0.5,0.866)--(0,0);
    \draw [rotate around={-120:(0.75,0.433)}][opacity=0.2] (1,0)--(0.5,0.866);
    \draw [rotate around={-120:(0.75,0.433)}][opacity=0.2] (0,0.5774)--(0.5,0.2887)--(1,0.5774)--(1.5,0.2887);
    \draw [rotate around={-120:(0.75,0.433)}][opacity=0.2] (0.5,-0.2887)--(0.5,0.2887);
    \draw [rotate around={-120:(0.75,0.433)}][opacity=0.2] (1,0.5774)--(1,1.1547);
    \draw [rotate around={-120:(0.75,0.433)}](0.5,0) ellipse (15pt and 2pt); 
    \draw [rotate around={-120:(0.75,0.433)}](1,0.866) ellipse (15pt and 2pt);
    \node at (0.8,0.7) [inner sep=0] {$\scriptstyle\sigma_I^z$};
    \node at (0.8,0.2) [inner sep=0] {$\scriptstyle\sigma_J^z$};
  \end{tikzpicture}
}

\newcommand\elA[1][0.4]{
  \begin{tikzpicture}[scale=#1]
     \draw (0,0)--(1,0);
     \draw (0.5,0) ellipse (20pt and 4pt);  
  \end{tikzpicture}
}
\newcommand\elB[1][0.4]{
  \begin{tikzpicture}[scale=#1]
     \draw[rotate around={120:(0.5,0)}] (0,0)--(1,0);
     \draw[rotate around={120:(0.5,0)}] (0.5,0) ellipse (20pt and 4pt);  
  \end{tikzpicture}
}
\newcommand\elC[1][0.4]{
  \begin{tikzpicture}[scale=#1]
     \draw[rotate around={-120:(0.5,0)}] (0,0)--(1,0);
     \draw[rotate around={-120:(0.5,0)}] (0.5,0) ellipse (20pt and 4pt);  
  \end{tikzpicture}
}

\newcommand\elAp[1][0.4]{
  \begin{tikzpicture}[scale=#1]
     \draw (0,0)--(1,0);
     \draw [densely dashed] (0.5,0) ellipse (20pt and 4pt);  
  \end{tikzpicture}
}
\newcommand\elBp[1][0.4]{
  \begin{tikzpicture}[scale=#1]
     \draw[rotate around={120:(0.5,0)}] (0,0)--(1,0);
     \draw[densely dashed] [rotate around={120:(0.5,0)}] (0.5,0) ellipse (20pt and 4pt);  
  \end{tikzpicture}
}
\newcommand\elCp[1][0.4]{
  \begin{tikzpicture}[scale=#1]
     \draw[rotate around={-120:(0.5,0)}] (0,0)--(1,0);
     \draw[densely dashed] [rotate around={-120:(0.5,0)}] (0.5,0) ellipse (20pt and 4pt);  
  \end{tikzpicture}
}

\newcommand\elApp[1][0.4]{
  \begin{tikzpicture}[scale=#1]
     \draw (0,0)--(1,0);
     \draw [densely dotted] (0.5,0) ellipse (20pt and 4pt);  
  \end{tikzpicture}
}
\newcommand\elBpp[1][0.4]{
  \begin{tikzpicture}[scale=#1]
     \draw[rotate around={120:(0.5,0)}] (0,0)--(1,0);
     \draw[densely dotted] [rotate around={120:(0.5,0)}] (0.5,0) ellipse (20pt and 4pt);  
  \end{tikzpicture}
}
\newcommand\elCpp[1][0.4]{
  \begin{tikzpicture}[scale=#1]
     \draw[rotate around={-120:(0.5,0)}] (0,0)--(1,0);
     \draw[densely dotted] [rotate around={-120:(0.5,0)}] (0.5,0) ellipse (20pt and 4pt);  
  \end{tikzpicture}
}

\newcommand\ellA[1][0.2]{
  \begin{tikzpicture}[scale=#1]
     \draw (0,0)node[inner sep=0]{$\scriptstyle\pmb\uparrow$}--(2,0)node[inner sep=0]{$\scriptstyle\pmb\downarrow$};
  \end{tikzpicture}
}
\newcommand\ellAm[1][0.2]{
  \begin{tikzpicture}[scale=#1]
     \draw (0,0)node[inner sep=0]{$\scriptstyle\pmb\downarrow$}--(2,0)node[inner sep=0]{$\scriptstyle\pmb\uparrow$};
  \end{tikzpicture}
}
\newcommand\ellB[1][0.2]{
  \begin{tikzpicture}[scale=#1]
     \draw[rotate around={120:(1,0)}] (0,0)node[inner sep=0]{$\scriptstyle\pmb\uparrow$}--(2,0)node[inner sep=0]{$\scriptstyle\pmb\downarrow$};
  \end{tikzpicture}
}
\newcommand\ellBm[1][0.2]{
  \begin{tikzpicture}[scale=#1]
     \draw[rotate around={120:(1,0)}] (0,0)node[inner sep=0]{$\scriptstyle\pmb\downarrow$}--(2,0)node[inner sep=0]{$\scriptstyle\pmb\uparrow$};
  \end{tikzpicture}
}
\newcommand\ellC[1][0.2]{
  \begin{tikzpicture}[scale=#1]
     \draw[rotate around={-120:(1,0)}] (0,0)node[inner sep=0]{$\scriptstyle\pmb\uparrow$}--(2,0)node[inner sep=0]{$\scriptstyle\pmb\downarrow$};
  \end{tikzpicture}
}
\newcommand\ellCm[1][0.2]{
  \begin{tikzpicture}[scale=#1]
     \draw[rotate around={-120:(1,0)}] (0,0)node[inner sep=0]{$\scriptstyle\pmb\downarrow$}--(2,0)node[inner sep=0]{$\scriptstyle\pmb\uparrow$};
  \end{tikzpicture}
}

\newcommand\ellAp[1][0.2]{
  \begin{tikzpicture}[scale=#1]
     \draw (0,0)node[inner sep=0]{$\scriptstyle\pmb\uparrow$}--(2,0)node[inner sep=0]{$\scriptstyle\pmb\uparrow$};
  \end{tikzpicture}
}
\newcommand\ellApm[1][0.2]{
  \begin{tikzpicture}[scale=#1]
     \draw (0,0)node[inner sep=0]{$\scriptstyle\pmb\downarrow$}--(2,0)node[inner sep=0]{$\scriptstyle\pmb\downarrow$};
  \end{tikzpicture}
}
\newcommand\ellBp[1][0.2]{
  \begin{tikzpicture}[scale=#1]
     \draw[rotate around={120:(1,0)}] (0,0)node[inner sep=0]{$\scriptstyle\pmb\uparrow$}--(2,0)node[inner sep=0]{$\scriptstyle\pmb\uparrow$};
  \end{tikzpicture}
}
\newcommand\ellBpm[1][0.2]{
  \begin{tikzpicture}[scale=#1]
     \draw[rotate around={120:(1,0)}] (0,0)node[inner sep=0]{$\scriptstyle\pmb\downarrow$}--(2,0)node[inner sep=0]{$\scriptstyle\pmb\downarrow$};
  \end{tikzpicture}
}
\newcommand\ellCp[1][0.2]{
  \begin{tikzpicture}[scale=#1]
     \draw[rotate around={-120:(1,0)}] (0,0)node[inner sep=0]{$\scriptstyle\pmb\uparrow$}--(2,0)node[inner sep=0]{$\scriptstyle\pmb\uparrow$};
  \end{tikzpicture}
}
\newcommand\ellCpm[1][0.2]{
  \begin{tikzpicture}[scale=#1]
     \draw[rotate around={-120:(1,0)}] (0,0)node[inner sep=0]{$\scriptstyle\pmb\downarrow$}--(2,0)node[inner sep=0]{$\scriptstyle\pmb\downarrow$};
  \end{tikzpicture}
}

\newcommand\intersectinglines[1][0.06]{
  \begin{tikzpicture}[scale=#1]
    \draw (0,0)--(6,0);
    \draw (3,-2)--(3,2);
  \end{tikzpicture}
  }
\newcommand\intersectinglinesA[1][0.07]{
  \begin{tikzpicture}[scale=#1]
    \draw (0,0)node[anchor=east]{\scriptsize{i}}--(6,0)node[anchor=west]{\scriptsize{j}};
    \draw (3,-2)node[anchor=north]{\scriptsize{I}}--(3,2)node[anchor=south]{\scriptsize{J}};
  \end{tikzpicture}
  }
\newcommand\intersectinglinesAp[1][0.07]{
  \begin{tikzpicture}[scale=#1]
    \draw [densely dashed] (0,0)node[anchor=east]{\scriptsize{i}}--(6,0)node[anchor=west]{\scriptsize{j}};
    \draw (3,-2)node[anchor=north]{\scriptsize{I}}--(3,2)node[anchor=south]{\scriptsize{J}};
  \end{tikzpicture}
  }
\newcommand\intersectinglinesApp[1][0.07]{
  \begin{tikzpicture}[scale=#1]
    \draw [densely dotted] (0,0)node[anchor=east]{\scriptsize{i}}--(6,0)node[anchor=west]{\scriptsize{j}};
    \draw (3,-2)node[anchor=north]{\scriptsize{I}}--(3,2)node[anchor=south]{\scriptsize{J}};
  \end{tikzpicture}
  }
  \usepackage{wasysym}
\newcommand\bindingterm[1][0.07]{
  \begin{tikzpicture}[scale=#1]
    \draw (0,0)node[circle, fill, inner sep=1pt, label=left:\scriptsize{I}]{}--(3,0)node[circle, fill, inner sep=0.7pt,gray, label=above:\scriptsize{i}]{}--(6,0)node[circle, fill,inner sep=1pt, label=right:\scriptsize{J}]{};
  \end{tikzpicture}
  }

\begin{document}
\title{Dyonic Lieb-Shultz-Mattis Theorem  and Symmetry Protected Topological Phases in Decorated Dimer Models}
\author{Xu Yang}
\affiliation{Department of Physics,
Boston College, Chestnut Hill, MA 02467, USA}

\author{Shenghan Jiang}
\affiliation{Department of Physics,
Boston College, Chestnut Hill, MA 02467, USA}

\author{Ashvin Vishwanath}
\affiliation{Department of Physics, Harvard University, Cambridge, MA 02138, USA}

\author{Ying Ran}
\affiliation{Department of Physics,
Boston College, Chestnut Hill, MA 02467, USA}

\begin{abstract}{We consider 2+1D lattice models of interacting bosons or spins, with both magnetic flux and fractional spin in the unit cell.  We propose and prove a modified Lieb-Shultz Mattis (LSM) theorem in this setting, which applies even when the spin in the enlarged magnetic unit cell is integral. There are two nontrivial outcomes for gapped ground states that preserve all symmetries. In the first case, one necessarily obtains a symmetry protected topological (SPT) phase with protected edge states. This allows us to readily construct models of SPT states by decorating dimer models of Mott insulators to yield SPT phases, which should be useful in their physical  realization. In the second case, exotic bulk excitations, i.e. topological order, is necessarily present. While both scenarios  require fractional spin in the lattice unit cell, the second requires that the symmetries protecting the fractional spin is related to that involved in the magnetic translations. Our discussion encompasses the general notion of fractional spin (projective symmetry representations) and magnetic flux (magnetic translations tied to a symmetry generator). The resulting SPTs display a dyonic character in that they associate charge with symmetry flux, allowing the flux in the unit cell to screen the projective representation on the sites. We provide an explicit formula that encapsulates this physics, which identifies a specific set of allowed SPT phases. }
\end{abstract}
\maketitle
\tableofcontents
\section{Introduction}\label{sec:introduction}
The Lieb Shultz Mattis (LSM) theorem \cite{Lieb:1961p407}, appropriately generalized to higher dimensions\cite{Oshikawa:2000p1535,Paramekanti2004luttinger, hastings2004lieb, Hastings:2005p824}, requires that a gapped spin system with fractional spin (eg. S=1/2) per unit cell possess excitations with fractional statistics (anyon) and fractional quantum numbers (topological order), if all symmetries (including lattice translations) are preserved. This has served as a powerful principle to diagnose exotic phases such as the fractional quantum Hall effect, and quantum spin liquids. Furthermore, in some cases the nature of the resulting topological order can be further constrained by the microscopic data \cite{Zaletel:2014p,cheng2015translational}.

In recent years there has been an explosion of activity on  symmetry protected topological (SPT) phases, which feature protected boundary modes although the bulk is short range entangled (SRE) and in contrast to the situation above, is free of anyon excitations. These include phases like topological insulators, which can be captured by free fermion models\cite{Qi:2011p1057,Hasan:2010p3045}, as well as intrinsically interacting phases \cite{Affleck:1987p799,Chen:2012p1604,chen2013symmetry} A natural question to ask is - are there setting where the microscopic data alone would enforce an SPT phase, in a fashion analogous to the LSM theorem? If so, for a particular set of microscopic data,  can we further characterize precisely which kinds of SPT orders are mandated?

These questions are answered in the present work. We show that SPT order {\em must} arise when the following conditions are met. The first ingredient is  magnetic translation symmetry, that is an enlargement of the unit cell due to the non-commutativity of the primitive translation operations. Second, we require that the primitive unit cell (ignoring the noncommutativity) does not admit a trivial insulating phase. This is arranged by requiring a projective representation at each lattice site. Finally, we need some compatibility conditions between these two ingredients that allow, among other conditions, that the enlarged unit cell to be effectively at integer filling, what admits a short range entangled ground state.  The latter is then shown to be an SPT. Furthermore  for 2+1D bosonic systems we explicitly calculate the allowed SPTs compatible with the microscopic specifications. In addition we construct exactly soluble lattice models of this phenomenon to demonstrate the validity of our conclusions. This general principle should aid in the search for SPTs in realistic settings and exposes anew aspect of the interplay between symmetry and topology. 

To give some simple plausibility arguments as to how microscopic details can enforce SPT order, consider  free fermions in a magnetic field,  when the filling fraction (ratio of particle density to magnetic flux density) is an integer. Then, an integer number of  Landau levels will be filled, leading to a Chern insulator - which is a SRE topological phase with gapless edge states. Even in the presence of a lattice , one can establish a similar connection between the Hall conductance $\sigma_{xy}$, the flux $n_\phi$ and electron filling in the unit cell  $n_e$  \cite{dana1985quantised, LuRanOshikawa} which has been extended to the case of time reversal symmetric topological insulators \cite{wu2017symmetry}.  

To state our result more precisely, we consider a two dimensional lattice where the unit translations obey: $T_xT_yT_x^{-1}T_y^{-1} = g$, where $g$ is an element of the symmetry group $G$. This generalizes the notion of a magnetic translation, particles acquire a phase factor depending on their $g$ charge. We assume $g$ is in the center of the symmetry group $G$ (i.e. commutes with all other elements), but otherwise consider a general $G$, which can either be discrete or continuous, Abelian or nonAbelian, and can include time reversal implemented by an antiunitary representatation. Furthermore, in each unit cell a projective representation of the symmetry group labeled by  `$\alpha$' is present. We derive a formula which provides a necessary and sufficient condition on these inputs to allow for a SRE phase, and determine constraints on the resulting SPT. Physically, this formula demands that a symmetry flux $g$ inserted into this system will precisely generate a projective representation that can screen `$\alpha$' \cite{wu2017symmetry}.  

Let us give two physical pictures to view this filling and flux enforced SPTs. First we describe a vortex condensation based picture, for a system of lattice bosons with a conserved U(1) charge, with flux  $n_\phi$  and filling $n_b$  per lattice unit cell. Although our paper focuses on having projective representations per site (rather than fractional filling) this example will be useful to build intuition.  It is well known that a conventional insulator can be thought of as a condensate of vortices. However, for fractional filling $n_b$,  the vortices see a fractional flux per unit cell \cite{lee1989anyon}, and their condensate will break lattice symmetries. Similarly, the bosons themselves cannot condense without breaking lattice symmetries due to the fractional flux $n_\phi$. {\em However} the bound state of a vortex and $p$  bosons  may be able to propagate freely if: $n_b \pm p n_\phi \in {\mathcal Z}$ is an integer. The resulting object is a boson for $p$ even which can then condense giving rise to  a SRE and symmetric insulator. These are nothing but the Bosonic Integer quantum Hall insulators at $\nu=n_b/n_\phi = p$ \cite{chen2013symmetry,Lu:2013classification, senthil2013integer}. Note, here the condensing particle carries unit vorticity and hence the resulting insulator is free of topological order \cite{balents1999dual} and also preserves the U(1) symmetry since the condensing charge is attached to vorticity. A generalization of this result to include arbitrary symmetry groups is the main result of this paper. An interesting exception occurs for $p=1$, which is realized for example when one has bosons at half filling (or a projective representation of $U(1)\rtimes Z_2$), and a $\pi$ flux in each unit cell. The doubled unit cell is at integer filling. At first sight it appears we can obtain an insulator by condensing the vortex-charge composite which sees no net flux in the unit cell. However, this composite is a fermion and cannot be condensed. This is also seen by a flux threading argument \cite{LuRanOshikawa} that constrains such SRE phases to have $\sigma_{xy} = {\rm odd\ integer}$, which is impossible for a SRE topological phase of bosons \cite{Lu:2013classification,senthil2013integer}.  Interestingly, this result continues to hold if the $U(1)$ is broken to a discrete symmetry as shown below.  

A second perspective is to begin in a topologically ordered phase with fractionalized excitations and consider confining all exotic excitations by an appropriate anyon condensate. For example, for bosons at half filling, one could obtain toric code (Z$_2$) topological order where the $e$ particle carries half charge \cite{Wenbook}.  The $m$ particle however sees the fractional charge density as background flux and cannot condense while preserving spatial symmetries. This is the situation in the absence of magnetic translations, where the LSM theorem enforces topological order for gapped symmetric states. However, once we allow for magnetic translations with $g$ charge, a way out to an SRE phase may become available. The $m$ particle, bound to a $g$ charge that sees the magnetic flux, forms  a composite object that may condense uniformly and confine the topological order. At the same time, this leads to an SPT phase since the condensing anyon carries nontrivial symmetry charge \cite{jiang2017anyon,duivenvoorden2017entanglement}. Indeed this picture will allows us to construct models of such LSM enforced SPT phases as we describe below.

Before discussing construction of models, it may be helpful to give a few examples. Consider a system of degenerate doublets (``S=1/2") on sites of a square lattice. This site degeneracy may arise from spin rotation invariance ($SO(3)$), or  even just as Kramers degeneracy protected by time reversal $Z_2^T$ symmetry. Now consider an additional $Z_2$ symmetry which is invoked in defining the magnetic translations, i.e. we have a fully frustrated Ising model on the same lattice. According to our results, in both these situations SRE ground states are possible but must be SPT phases. While the SPT phase is unique for the second case of Kramers doublets of $Z_T\times Z_2$, in the former case of $SO(3)\times Z_2$ there is more than one SPT phase possible. Interestingly, if we consider a minor modification of the $Z_T\times Z_2$ model, such that the doublets on each site are non Kramers pairs, protected by the combination of the two symmetries, then {\em no} SRE ground state exists (and hence no SPT exists) that respects all symmetries. These examples are discussed in detail in Section \ref{sec:examples} which also introduces models that realize them. 

In constructing models, the first step is to begin in the deconfined phase of a discrete lattice gauge theory (or of a dimer model). Then, one way to obtain a confined phase is by decorating the electric field lines with domain walls of a global symmetry. This identification implies that we have condensed the composite of magnetic flux and symmetry charge. The resulting confined phase is potentially an SPT if the electric charges are associated with the appropriate symmetry fractionalization \cite{jiang2017anyon,duivenvoorden2017entanglement}. However, to obtain an LSM enforced SPTs the situation is different since they involve fractional spin on the sites. In a dimer model this corresponds to having an odd number of dimers associated with a unit cell, in which case we cannot decorate them with regular domain walls (which should be closed loops). {\em However} if the global symmetry is also associated with flux in the unit cell (for example a fully frustrated Ising model), the two kinds of frustration cancel each other out, and one can still achieve this decoration of electric field line. This is discussed explicitly  in Section \ref{sec:BFG}, for a specific model and the resulting state is shown to be the desired SPT. The model there is one of hardcore bosons on the Kagome lattice tuned to half filling by particle hole symmetry, previously introduced by Balents Fisher and Girvin \cite{balents2002fractionalization}. While their focus was on a Z$_2$ spin liquid phase, we decorate their model with an additional Z$_2$ symmetry realized by a fully-frustrated Ising model. The combination is shown to realize an LSM enforced SPT phase with gapless edge states, but a short range entangled bulk.

Finally in Section \ref{sec:MainResults} we discuss the problem for general symmetry groups, and derive the necessary and sufficient conditions for SRE phases to emerge and identify the class of SPTs that must be realized. Proofs can be found in the appendices.
 
\section{A simple model realizing SPT phase}\label{sec:BFG}
Our discussion starts from a concrete microscopic model realizing an SPT phase. The beauty of this model is its simplicity, which only includes two-spin and three spin interactions. It turns out that the crucial features of this model can be systematically generalized which form the main results of the current study.

The model constructed below (see Eq.(\ref{eq:decoratedBFG_full})) is based on the Balents-Fisher-Girvin(BFG) model\cite{balents2002fractionalization}. The original BFG model\cite{balents2002fractionalization} is a model with spin-1/2 residing on Kagome lattice. It is the low energy effective Hamiltonian if we take the $J_z\gg J_{\perp}$ limit of the following $XXZ$ Hamiltonian
\begin{equation}
H^{XXZ}=J_{\perp}\sum\limits_{\small\hexagon}[(\sum\limits_{i\in\hexagon}S^x_i)^2+(\sum\limits_{i\in\hexagon}S^y_i)^2-3]+J_z\sum\limits_{\hexagon}(\sum\limits_{i\in\hexagon}S^z_i)^2,
\end{equation}
which has a spin-liquid ground state for $J_z\gg J_{\perp}$ with deconfined spinons as confirmed by various numerical methods \cite{sheng2005numerical,Isakov2006spin}.

Let's then take a look at the low energy effective Hamiltonian. The limit $J_z\gg J_{\perp}$ ensures that $S^z_{\hexagon}=0$ for every hexagon and the resulting Hamiltonian in this low energy manifold takes the following ring-exchange form
\begin{equation}\label{eq:originalBFG}
H^{BFG}=-J_{\text{ring}}\sum\limits_{\bowtie}(\big|\raisebox{-4mm}\BFGB\big\rangle \big\langle\raisebox{-4mm}\BFGA\big|+h.c.),
\end{equation}
with $J_{ring}=J_{\perp}^2/J_z$. 

\begin{figure}
\includegraphics[width=0.48\textwidth]{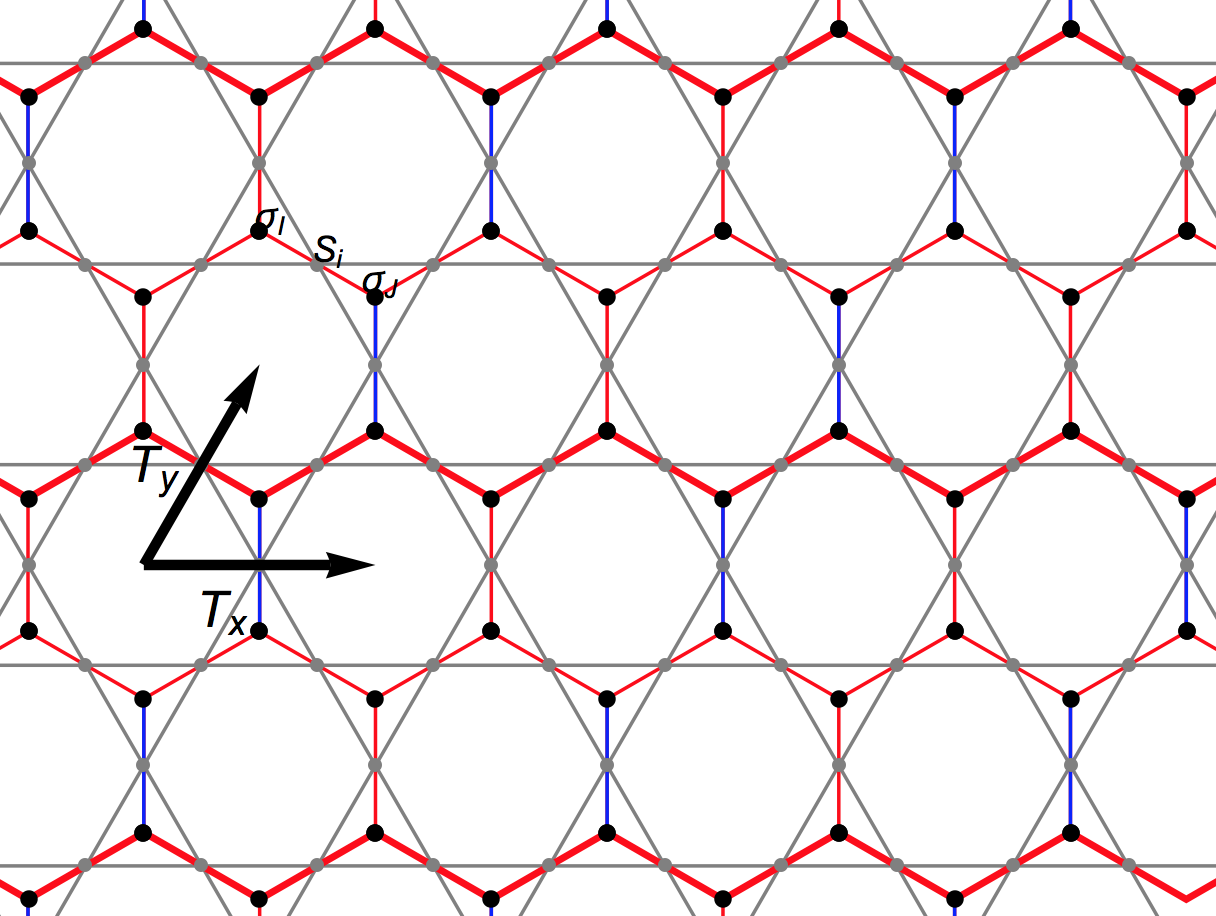};
\caption{(color online) Degrees of freedom in the decorated BFG model. The Ising d.o.f. $\sigma_I$ live on the honeycomb lattice and the spin d.o.f. $S_i$ lives on the Kagome lattice. The Ising coupling signs $s_{IJ}=+1$ on red bonds, and $s_{IJ}=-1$ on blue bonds. The thick red bonds represent the ``$y$-odd zigzag chains'' used in Eq.~\eqref{eq:BFG_magn_trans}}
\label{fig:decoratedBFG}
\end{figure}

Let's then decorate the $XXZ$ model by putting a layer of Ising spins $\sigma$ inside every triangle of the Kagome lattice, which comprises a honeycomb lattice. The Ising spins are in a transverse field, \textit{i.e.},
\begin{equation}
H^{Ising}=h\sum\limits_{I}\sigma_I^x,
\end{equation}
with Ising spin $\sigma_I$ living on the honeycomb lattices labeled by $I$.

We then couple these two layers through a binding term
\begin{equation}
H^{binding}=-\sum\limits_{\bindingterm}\lambda S_i^z\cdot(s_{IJ}\sigma^z_I\sigma^z_J),
\end{equation}
where the summation is over all the bonds $IJ$ on honeycomb lattice with $S_i$ at the bond center. The sign $s_{IJ}=\pm 1$ are frustrated in the sense that $\prod\limits_{I,J\in\hexagon}s_{IJ}=-1$. We have specifically chosen a choice of $s_{IJ}$ in Fig.~\ref{fig:decoratedBFG}. The binding term binds spin-up with Ising happy bond ($s_{IJ}\sigma_I^z\sigma_J^z=+1$) and spin-down with Ising un-happy bond ($s_{IJ}\sigma_I^z\sigma_J^z=-1$).

The full Hamiltonian we are considering is then given by (see Fig.~\ref{fig:decoratedBFG})
\begin{equation}\label{eq:decoratedBFG_full}
H=H^{XXZ}+H^{binding}+H^{Ising}.
\end{equation}

One can divide $H$ into two parts
\begin{equation}
\begin{split}
&H_0=J_z\sum\limits_{\hexagon}(S^z_{\small\hexagon})^2-\sum\limits_{\bindingterm}\lambda S_i^z(s_{IJ}\sigma^z_I\sigma^z_J).\\
&H_1=J_{\perp}\sum\limits_{\hexagon}[(S^x_{\hexagon})^2+(S^y_{\hexagon})^2-3]+\sum\limits_{I}h\sigma_I^x.
\end{split}
\end{equation}

Considering the the limit where $J_z,\lambda \gg J_{\perp},h$, we can first deal with $H_0$ and then treat $H_1$ as a perturbation. All the terms in $H_0$ commutes with each other and hence all the eigen-states and eigen-energies are known for $H_0$. In fact, there is a two-to-one mapping from the ground state sector to the low energy sector of the BFG model (i.e. $\{S_i^z\}$ configurations satisfying 3 $S^z=+1/2$  per hexagon). We will consider periodic boundary conditions, and the Hilbert space of the original BFG model has four topological sectors labeled by parities of the $\prod_{k\in \mathcal{C}}2S_k^z$ around the non-contractable loops $\mathcal{C}$ (which is just the non-contractable vison flux line\cite{fradkin2013field}). This mapping only map onto one specific topological sector since $\prod_{k\in \mathcal{C}}2S_k^z$ is identified with  $\prod_{IJ\in \mathcal{C}}s_{IJ}$ due to $H^{binding}$. The preimage of any low energy $\{S_i^z\}$ configuration inside this topological sector are two states $\ket{\{S_i^z,+\}}$ and $\ket{\{S_i^z,-\}}$ (related to each other by a global Ising flip).


It turns out that the effective Hamiltonian in the parameter regime where $J_z\gg \lambda\gg J_{\perp},h$ and $\frac{h^2}{\lambda^2}\gg \frac{J_{\perp}}{J_z}$ has the following form (see Appendix.~\ref{app:BFG} for detailed calculation)
\begin{equation}\label{eq:decoratedBFG}
H^{deco.BFG}=-\frac{10J_{\perp}^2 h^2}{9J_z\lambda^2}\sum\limits_{\bowtie}(\big|\raisebox{-4mm}\bowtieB\big\rangle \big\langle\raisebox{-4mm}\bowtieA\big|+h.c.).
\end{equation}

Note that the kinetic term in this effective Hamiltonian is the ring exchange term of four spins at the ends of each bowtie as in the original BFG model combined with the flipping term of the two Ising d.o.f. within this bowtie, such that the constraint $S_i^z(s_{IJ}\sigma^z_I\sigma^z_J)=1$ is still satisfied everywhere.

\begin{figure}
\includegraphics[width=0.48\textwidth]{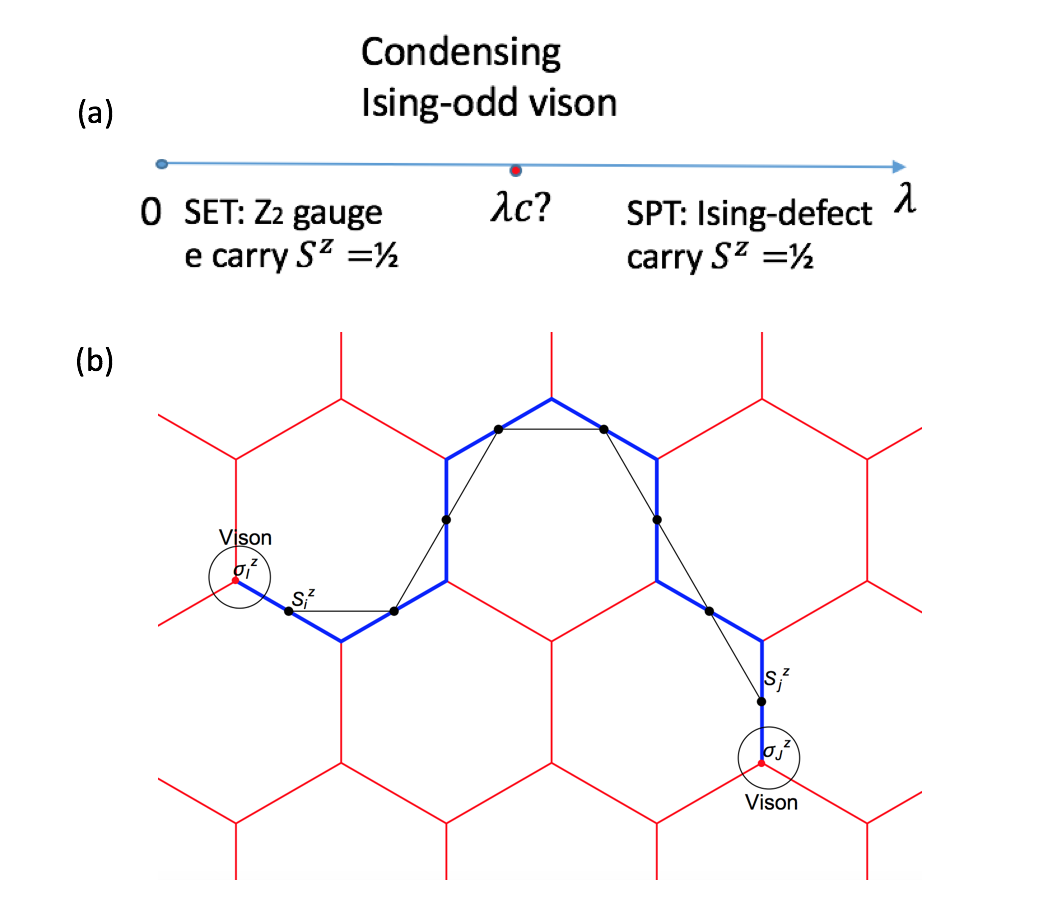}
\caption{(color online) (a) Schematic phase diagram of the decorated BFG model by tuning $\lambda$. We have already fixed $J_z\gg J_{\perp},h$. In the limit $\lambda\rightarrow 0$ the Ising layer is decoupled and the ground state is just that of the original BFG model with $Z_2$ topological order. This is an SET state with spinon carrying $S^z=1/2$. When $\lambda$ is tuned to be within the parameter regime where $J_z\gg\lambda\gg J_{\perp},h$ and $\frac{h^2}{\lambda^2}\gg \frac{J_{\perp}}{J_z}$, we have an SPT state with Ising defect carrying $S^z=1/2$ as discussed in the main text. There is a possible direct phase transition triggered by the condensation of Ising-odd visons at some intermediate $\lambda_c$. (b) A schematic view of vison condensation. The honeycomb lattice where Ising d.o.f. lives is shown and the spin d.o.f. lives in the bond center. Two visons are created at the ends $I,J$ of the string operator $\sigma_I^z\sigma^z_J\prod\limits^j_{k=i}\!\!\!\!\!\!\!\!\!\!\longrightarrow 2S^z_k.$ with $S_k^z$ runs over all the black dot shown in the graph. Alternatively we can view the string operator as the product of bond variable $S_i^z\sigma_I^z\sigma^z_K$ along the thick blue bonds. Due to the constraint $S_i^z(s_{IK}\sigma_I^z\sigma_K^z)=1$, the string operator will yield a factor (product of $s_{IK}$'s along the thick blue bonds) when acting on the ground state wave-function, which means the visons are condensed and the topological order is killed. Note that the condensed visons in the present case are dressed by local $\sigma^z$ operator and hence carry the quantum number of $Z_2$ Ising symmetry, which result in an SPT state.}
\label{fig:vison_condensation}
\end{figure}
We shall then prove that the ground state of the decorated BFG model is a symmetric short-range entangled SPT state. In fact, using the mapping $\mathcal{P}$ between the Hilbert space of the decorated BFG model and the original BFG model 
\begin{equation}
    \mathcal{P}: (\ket{\{S_i^z,+\}}+\ket{\{S_i^z,-\}})/\sqrt{2}\rightarrow\ket{\{S_i^z\}}.
\end{equation}
Such a mapping is clearly an isometry. Next we notice that
\begin{equation}
    \mathcal{P}H^{deco.BFG}\mathcal{P}^{-1}=H^{BFG},
\end{equation}
with the identification $J_{ring}=\frac{10J_{\perp}^2h^2}{9J_z\lambda^2}$, which can be proven by directly comparing the matrix elements on the two sides. 

Therefore the spectrum of $H^{deco.BFG}$ within the Ising-even sector is exactly the same as that of $H^{BFG}$ inside a specific topological sector, which is known to be gapped. And the ground state $\ket{\psi}$ of $H^{BFG}$, should also be mapped to the ground state $\ket{\psi^{deco.}}$ of $H^{deco.BFG}$. However there is still one possibility that there exists a state in the Ising-odd sector with exactly the same energy as $\ket{\psi^{deco.}}$, which features the Ising symmetry breaking. This possibility is ruled out because $\ket{\psi^{deco.}}$ has no long-range order in $\sigma^z$ as will be discussed in the context of vison condensation.  

The ground state of the original BFG model has $Z_2$ topological order which supports vison and spinon excitations. In the original BFG model, the vison excitations live in the honeycomb lattice and are created at the ends of the string of $S_k^z$ operators\cite{balents2002fractionalization}, \textit{i.e.}, 
\begin{equation}
v_Iv_J=\prod^{j}_{k=i}\!\!\!\!\!\!\!\!\!\!\longrightarrow2S^z_k,
\end{equation}
where two visons are created at $I$ and $J$ (see Fig.~\ref{fig:vison_condensation}).

We can see that the visons are condensed in $\ket{\psi^{deco.}}$. The vison operator $v^{deco.}$ at site $i$ should now be dressed by the local Ising operators $\sigma^z_I$ with $I$ around $i$ to obtain
\begin{equation}
v^{deco.}_Iv^{deco.}_J=\sigma_I^z\sigma^z_J\prod^j_{k=i}\!\!\!\!\!\!\!\!\!\!\longrightarrow 2S^z_k.
\end{equation}

With the constraint that $S_i^z(s_{IJ}\sigma^z_I\sigma^z_J)\equiv 1$ and the fact that intermediate $\sigma^z$ squared to 1, we know that $v^{deco.}_Iv^{deco.}_J$ must yield a constant (depending only on the product of $s_{IJ}$'s along the vison string) when acting on $\ket{\psi^{deco.}}$, see Fig.~\ref{fig:vison_condensation} for an illustration.

Now it is clear that the correlator $\sigma_I^z\sigma_J^z$ is short-range because we have
\begin{equation}
\begin{split}
&|\bra{\psi^{deco.}} \sigma^z_I\sigma_J^z\ket{\psi^{deco.}}|=|\bra{\psi^{deco.}} \prod^j_{k=i}\!\!\!\!\!\!\!\!\!\!\longrightarrow 2S^z_k\ket{\psi^{deco.}}|,\\
&=|\bra{\psi}\prod^j_{k=i}\!\!\!\!\!\!\!\!\!\!\longrightarrow 2S^z_k\ket{\psi}|,
\end{split}
\end{equation}
where the last correlator exhibits exponential decay since visons are deconfined in the original BFG model (the last equality holds because $\mathcal{P}$ commutes with the string operator).

The above discussions feature the physical picture of the condensation of visons carrying Ising quantum number, which kills the $Z_2$ topological order. We will soon show that the resulting phase is an SPT phase, which is exactly a realization of the anyon condensation mechanism to obtain SPT phases proposed in Ref.~\onlinecite{jiang2017anyon}. In fact, if we start from the decoupling limit with $\lambda=0$ and gradually increase $\lambda$ with all other couplings fixed, we should be able to see two phases: an SET phase with spinon carrying $S^z=1/2$ when $\lambda$ small and an SPT phase resulting from the condensation of Ising-odd visons when $\lambda$ is in the parameter regime $J_z\gg\lambda\gg J_{\perp},h$ and $\frac{h^2}{\lambda^2}\gg \frac{J_{\perp}}{J_z}$. This two phases might be related by a continuous phase transition at some intermediate $\lambda_c$. See Fig.~\ref{fig:vison_condensation} for an illustration. 

\begin{figure}
\includegraphics[width=0.48\textwidth]{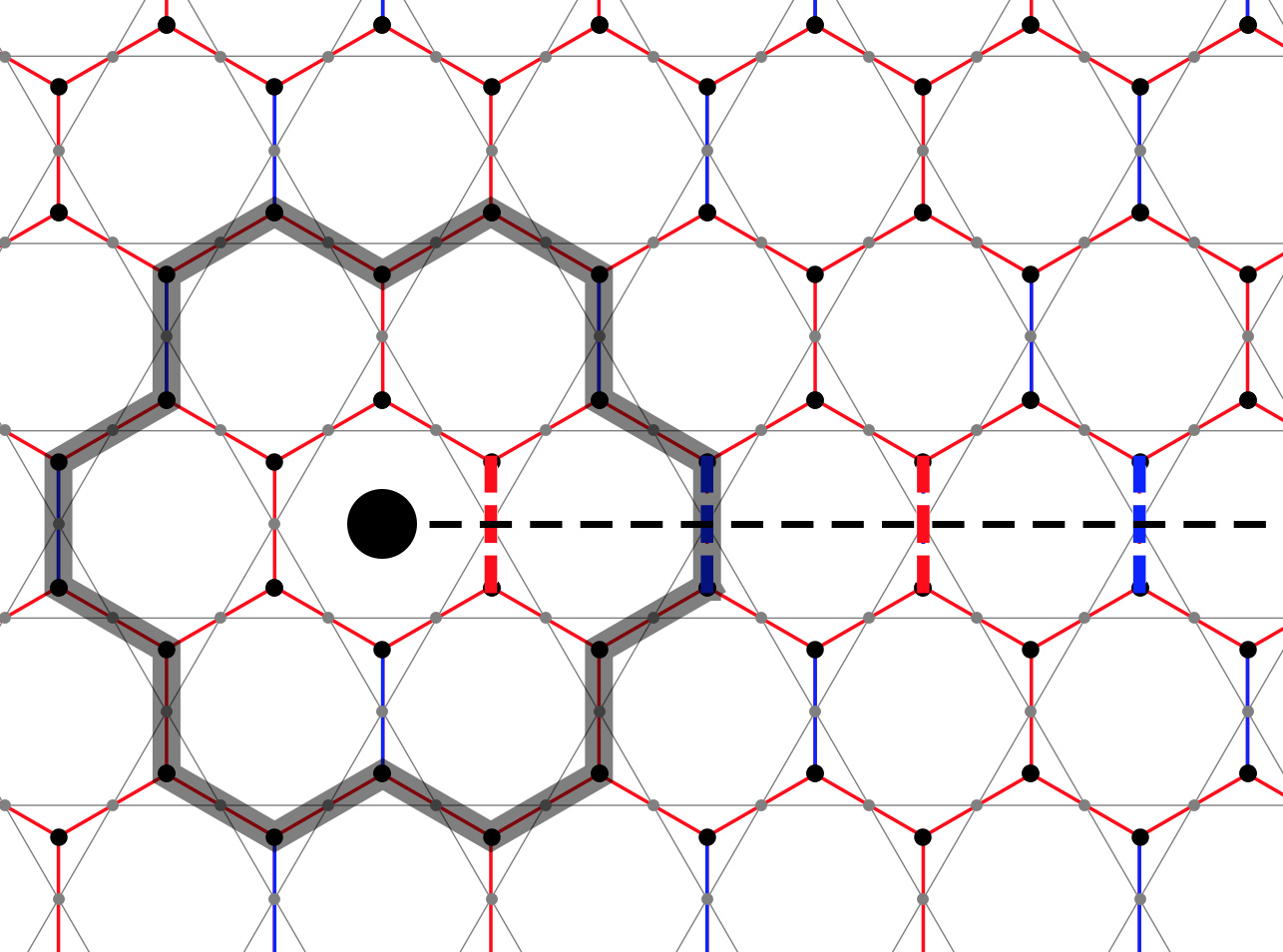};
\caption{(color online) An pair of Ising defects (only one is shown) is created at the end points of the branch cut (dashed black line) after modifying the original Hamiltonian $H$ in Eq.(\ref{eq:decoratedBFG_full}) into $H'$. The sign $s_{IJ}$ is flipped in $H'$ along the branch cut comparing with the original model. (red bond: $s_{IJ}=+1$, blue bond: $s_{IJ}=-1$) For any loop $\mathcal{C}$ enclosing the Ising defect as the gray loop shown here, the product $\prod s_{IJ}$ around the loop flips sign comparing with the original model.  In order that $H'^{binding}$ does not cost extra energy, the spin should be flipped wherever $s_{IJ}$ changes its sign. As a result, the total $S^z$ around the loop $\mathcal{C}$ is changed by an odd integer. The result is that Ising defect is topologically bound with a half-integer spin. See the discussion in the main text.}
\label{fig:Ising_defect_BFG}
\end{figure}

One way to see that the ground state of the decorated BFG model Eq.(\ref{eq:decoratedBFG}) is an SPT phase is to consider the Ising defects, which turn out to be topologically bound with $S^z=\pm 1/2$ --- a projective representation of the symmetry group (see discussion below). In order to introduce Ising defects we need to take a branch cut and modify the terms straddling the branch cut such that only one side is conjugated by the Ising symmetry $\sigma^x$. The net effect is that for the bonds $IJ$ crossed by the branch cut, the sign of $s_{IJ}$ is flipped. See Fig.~\ref{fig:Ising_defect_BFG} for an illustration. To compute the $S^z$ quantum number carried by the Ising defect it is convenient to introduce the equivalent hard-core boson description: $n_i\equiv S^z_i+1/2$. Now let's take a loop $\mathcal{C}$ enclosing one of the Ising defect and measure the total charge within the area $\mathcal{D}$ bounded by $\mathcal{C}$. This is done by the following U(1) transformation 
\begin{equation}\label{eq:bosonnumber}
\prod_{\hexagon\in \mathcal{D}}(\prod_{i\in\hexagon}e^{i\frac{\theta}{2}n_i})=\prod_{i\in \mathcal{C}}e^{i\frac{\theta}{2}n_i}\cdot \prod_{i\in \mathcal{D}/\mathcal{C}}e^{i\theta n_i},
\end{equation}
from which we know that the fractional charge part is determined by boson numbers on the boundary only.

In order that the binding term after modification $H'^{binding}$ does not cost energy, the boson number $n_i$ on the bond $IJ$ across the branch cut should be changed by 1. Since $\mathcal{C}$ only crosses Ising defect line odd times, the total boson numbers around $\mathcal{C}$ should be changed by $1(mod\ 2)$. From Eq.~\eqref{eq:bosonnumber}, this amounts to the change of charge within $\mathcal{D}$ by $\frac{1}{2}(mod\ 1)$. This fact doesn't depend on the position of the branch cut or the position of $\mathcal{C}$ we have chosen, indicating that the extra $\frac{1}{2}$ charge is bounded with the Ising defect. In the original spin language it means Ising defect carries spin $S^z=\pm 1/2$.

\textbf{Summary:} Before proceeding let's pay close attention to the symmetry property of the Hamiltonian in Eq.~\eqref{eq:decoratedBFG_full}. 
This model has the following $Z_{2g}\times(U(1)\rtimes Z_{2h}) $ onsite symmetry  
\begin{enumerate}
\item Spin-rotation symmetry $U(1)$: $\prod\limits_ie^{i\theta (S_i^z+\frac{1}{2})},$ with $\theta\in [0,2\pi)$.
\item Spin-flip and Ising-flip symmetry $Z_{2h}=\{I,h\}$: $\prod\limits_i S_i^x\prod\limits_{I\in A}\sigma^x_I$, which flips all the spins on Kagome lattice and Ising d.o.f on $A$ sublattice of the honeycomb lattice. This symmetry operation leaves the Hamiltonian in Eq.~\eqref{eq:decoratedBFG_full} invariant.
\item Ising symmetry $Z_{2g}=\{I,g\}$: $\prod\limits_I\sigma^x_I$, which flips all the Ising d.o.f. on the honeycomb lattice.
\end{enumerate}

The onsite symmetry group has the direct product structure $G=Z_{2g}\times (U(1)\rtimes Z_{2h})$. In addition, every unit cell (three kagome sites and two honeycomb sites) carries a nontrivial projective representation of $U(1)\rtimes Z_{2h}$ (i.e. a half-integer spin). Naively one would suspect that the LSM theorem would rule out a symmetric short-range-entangled state in such a system. However, due to the frustrated nature of the gluing term, we actually have magnetic translation symmetry $T_x,T_y$ in the system instead of usual translation $T_{x}^{orig.},T_y^{orig.}$, which can be written as
\begin{equation}\label{eq:BFG_magn_trans}
\begin{split}
&T_x=(\prod\limits_{I\in\text{y-odd zigzag chains}}\sigma_I^x)T_x^{orig.},\\
& T_y=T_y^{orig.},
\end{split}
\end{equation}
which satisfy the magnetic translation algebra
\begin{equation}
T_xT_yT_x^{-1}T_y^{-1}=g.
\end{equation}
Therefore it is still possible for us to have an SPT state. And we also know that the $g$-defect carries $S^z=\pm 1/2$, which has the same projective representation as that carried by the unit-cell. Below we will find that this is not merely a coincidence and there is a deep connection between the patterns of short-range entanglement and the projective representation carried by a unit-cell related by the magnetic translation algebra.

\section{Main Results}
\label{sec:MainResults}
Our main results are captured in two theorems.  Theorem-I is easier to state but less general. Theorem-II is more general but is more mathematically involved to state.

Consider a two-dimensional \emph{bosonic} quantum system respecting an onsite symmetry group $G$ (which could contain time-reversal), and $g$ is a unitary symmetry element in the center of $G$ (i.e., $g$ commutes with any element in $G$). The system respects a ``magnetic'' translation symmetry group generated by $T_x,T_y$ satisfying the algebra: 
\begin{align}
 T_xT_yT_x^{-1}T_y^{-1}=g,\label{eq:mag_transl}
\end{align}
where $T_x,T_y$ are assumed to be the usual translation operations combined with certain site-dependent onsite unitary transformations. We further assume that the physical degrees of freedom (d.o.f.) in each real space unit cell (\emph{not} the enlarged magnetic unit cell) form a nontrivial projective representation $\alpha$ of $G$, specified by a 2-cocycle: $\forall a,b\in G, \alpha(a,b)\in U(1)$ and $\alpha\in  H^2(G,U(1))$. Precisely, the unitary or antiunitary transformation $U_a,U_b$ of $a,b\in G$ satisfy:
\begin{align}
 U_a \leftidx{^a}U_b=\alpha(a,b) U_{ab},\label{eq:alpha}
\end{align}
where the left-superscript $a$ in $\leftidx{^a}U_b$ denotes the group action of $a$ on $U_b$: if $a$ is unitary (antiunitary), then $\leftidx{^a}U_b=U_b$ ($\leftidx{^a}U_b=U_b^{*}$, i.e. complex conjugation of $U_b$).

We ask the following question: is it possible for such a system to have a short-range entangled(SRE) gapped ground state without breaking symmetries? 

Here we use the definition of SRE states following Ref.~\onlinecite{chen2013symmetry}; i.e., those are gapped quantum phases that can be deformed into the trivial product state via local unitary transformations. Note that if the system respects usual translational symmetries, this would be impossible: constrained by a generalized  Lieb Shultz Mattis theorem\cite{Lieb:1961p407,hastings2004lieb,Oshikawa:2000p1535}, the nontrivial projective representation per unit cell indicates that a gapped liquid ground state necessarily features topological order. Here because the system respects a magnetic translation symmetry, it is possible that a SRE liquid ground state exists. We give sufficient and necessary conditions for such a liquid phase to exist, and show that this SRE liquid phase must be an SPT phase.

In the presence of an onsite symmetry group $G$, focusing on bosonic systems specified by Eq.(\ref{eq:mag_transl},\ref{eq:alpha}), we have the following theorems:

\begin{framed}
\textbf{Theorem-I}: Here we further assume that $G=G_1\times Z_N$ where $Z_N$ is the finite abelian subgroup generated by $g$. The quantum system above can have a SRE liquid ground state if and only if the two conditions below are both satisfied. Such a liquid phase is necessarily a nontrivial SPT phase because the $g$-symmetry defect must carry the projective representation $\alpha$.
\begin{enumerate}
 \item $\alpha^N\simeq \mathbf{1} \in H^2(G,U(1))$, i.e. $N$ of these projective representations fuse into a regular representation.
 \item The group function (which maps elements of the symmetry group to phases, while preserving the group relations) $\gamma^{\alpha}_g(a)\equiv\frac{\alpha(g,a)}{\alpha(a,g)}$, $\forall a\in G$ is a trivial 1-cocycle (or equivalently, a trivial one-dimensional representation): \\
 $\gamma^{\alpha}_g\simeq \mathbf{1} \in H^1(G,U(1))$.
\end{enumerate}
\end{framed}
What is the physical meaning of these two conditions?  The first condition ensures that the enlarged magnetic unit cell does not have projective representations. Interestingly, this is not sufficient to ensure an SRE phase. Additionally, condition 2 must be satisfied, which essentially states that the symmetry involved in magnetic translations, $g$, can be chosen to commute with all other projective group actions in a proper gauge.
 
\begin{framed}
\textbf{Theorem-II}: Here we do not make extra assumptions on $G$. The quantum system above can have a SRE liquid ground state if and only if there exists a 3-cocyle $\omega_0$: $\forall a,b,c\in G,\omega_0(a,b,c)\in U(1)$ and $\omega_0\in H^3(G,U(1))$, such that the group function $\delta^{\omega_0}_g(a,b)\equiv \frac{\omega_0(g,a,b)\omega_0(a,b,g)}{\omega_0(a,g,b)}$ is 2-cycle equivalent to $\alpha^{-1}$: $\delta^{\omega_0}_g\simeq \alpha^{-1}\in H^2(G,U(1))$. Such a liquid phase is necessarily a nontrivial SPT phase because the $g$-symmetry defect must carry the projective representation $\alpha$. The possible nontrivial SPT phases form a coset from the classification point of view (see Remark below).
\end{framed}

\textbf{Remark}: it is straightforward to show that $\forall \alpha \in H^2(G,U(1))$, $\gamma^{\alpha}_g\in H^1(G,U(1))$. And similarly $\forall \omega \in H^3(G,U(1))$, $\delta^{\omega}_g\in H^2(G,U(1))$. The mappings $\gamma_g: H^2(G,U(1))\rightarrow H^1(G,U(1))$ and $\delta_g: H^3(G,U(1))\rightarrow H^2(G,U(1))$ reducing a $n$-cocycle to a $(n-1)$-cocycle are the so-called slant-products in mathematical context. $\gamma_g$ and $\delta_g$ preserve the multiplication relation in the cohomology group. In particular, there is a subgroup $\mathcal{A}_g\in H^{3}(G,U(1))$ such that $\forall\omega\in\mathcal{A}_g, \delta^{\omega}_g\simeq \mathbf{1}\in H^{2}(G,U(1))$, (i.e., $\mathcal{A}_g$ the kernal of the mapping $\delta_g$).

When the condition in Theorem-II is satisified, $\omega_0$ must be a nontrivial element in $H^3(G,U(1))$ because $\alpha$ is nontrivial by assumption. And the realizable SPT phases form a coset from the classification point of view. More precisely, the 3-cocyle characterizing the SRE liquid phase must be one of the element in the following coset: $\omega_0\cdot\mathcal{A}_g$.

\textbf{Outline of the proof}: The proof of these theorems is a combination of a pumping argument of entanglement spectra and derivations/constructions based on a recently developed symmetric tensor-network formulation\cite{jiang2017anyon}, which we outline here. Basically, if a SRE liquid phase exists, by the pumping argument of entanglement spectra one knows that \emph{the $g$-symmetry-defect in this phase must carry the projective representation $\alpha$, and consequently this phase must be an SPT phase}. This physical observation can be further justified by calculations based on symmetric tensor-networks, leading to the following mathematical result: if the 3-cocycle characterizing the SRE liquid phase as $\omega\in H^3(G,U(1))$, then magnetic translation symmetry dictates $\delta^{\omega}_g\simeq\alpha^{-1}$, which is exactly the same mathematical condition for the $g$-symmetry-defect carrying the projective representation $\alpha$. In addition, based on the symmetric tensor-network formulation, for any $\omega$ satisfying $\delta^{\omega}_g\simeq\alpha^{-1}$, a SRE liquid phase characterized by $\omega$ respecting the magnetic translation symmetry can be constructed. These prove that the conditions in Theorem-II are \emph{necessary and sufficient} for the SRE liquid phase to exist. In addition, when $\omega_0$ exists, because a 3-cocycle $\omega\in H^3(G,U(1))$ satisfies $\delta^{\omega}_g\simeq\alpha^{-1}$ if and only if $\omega\in \omega_0\cdot\mathcal{A}_g$, the coset structure in the Remark is also established.

Theorem-I is just a special case of Theorem-II. Namely when $G=G_1\times Z_N$, one can show that if and only if the two conditions in Theorem-I is satisified does the condition in Theorem-II is satisfied. The condition-(1) in Theorem-I is well anticipated. If condition-(1) is not satisfied, then physical degrees of freedom form a nontrivial projective representation of $G$ even in the enlarged magnetic unit cell ($N$ times larger than original unit cell), and the generalized  Lieb Shultz Mattis theorem\cite{Lieb:1961p407,hastings2004lieb,Oshikawa:2000p1535} already forbids a SRE liquid phase to exist. The condition-(2) is less obvious and more interesting, which puts additional constraints for the existence of a SRE liquid phase (see example-(4) below).

Before going into the details of the proof, let us consider a few simple examples to see the applications of the Theorems and the Remark.

\subsection{Examples}\label{sec:examples} In these examples, the element $g$ in the magnetic translation algebra Eq.(\ref{eq:mag_transl}) generates a $Z_2^{Ising}\equiv \{I, g\}$ Ising symmetry group. For instance, a fully frustrated Ising model on the square lattice would satisfy this magnetic translation symmetry. The symmetry-enforced SPT phases in example-(1,2,3) will be demonstrated via a class of decorated quantum dimer models, which are exactly solvable at the Rokhsar-Kivelson points\cite{rokhsar1988superconductivity}.

\textbf{(1) $\mathbf{G=SO(3)\times Z_2^{Ising}}$, and a spin-1/2 per unit cell:} Namely, the projective representation $\alpha$ per unit cell is nontrivial because only the $SO(3)$ part is projectively represented, and the Ising and the spin-rotation still commute: $\alpha(g, a)=\alpha(a,g),\; \forall a\in G$. Clearly the two conditions in Theorem-I are both satisified. First, two spin-1/2's fuse into a regular $SO(3)$ representation, and $\gamma_g^{\alpha}(a)=1,\forall a\in G$.

\emph{According to Theorem-I, at least one SRE liquid phase can exist and must be an SPT phase in which the $g$-symmetry-defect carries a half-integer spin.} To understand how many SPT phases are possibly realized, one can follow the Remark. \footnote{SPT phases protected by $G=SO(3)\times Z_2^{Ising}$ form a group $H^3(SO(3)\times Z_2^{Ising},U(1))$. The Kunneth formula gives: $H^3(SO(3)\times Z_2^{Ising},U(1))=H^3(SO(3),U(1))\times H^3(Z_2,U(1))\times H^2(SO(3),Z_2)=Z\times Z_2\times Z_2$. Following the Remark, it is straightforward to show that only the $Z_2$ index in $H^2(SO(3),Z_2)$ is enforced to be nontrivial. (Namely $\omega_0$ in Theorem-(2) can be chosen to be the nontrivial element in $H^2(SO(3),Z_2)$,and the kernal $\mathcal{A}_g=H^3(SO(3),U(1))\times H^2(SO(3),Z_2)=Z\times Z_2$.)} The result is that among all possible SPT phases classified by $H^3(SO(3)\times Z_2^{Ising},U(1))=Z\times Z_2^2$, only one of the $Z_2$ indices is enforced to be nontrivial. And there are many distinct SPT phases that can be realized, which form a coset $\omega_0\cdot \mathcal{A}_g$, where $\mathcal{A}_g=Z\times Z_2$. \emph{In particular, after gauging the $Z_2^{Ising}$ symmetry, one may obtain either the toric-code or double-semion topological order, depending on which SPT phase is realized.}

\textbf{(2) $\mathbf{G=Z_2^T\times Z_2^{Ising}}$, and a Kramer doublet per unit cell:} Here $Z_2^T=\{I,\mathcal{T}\}$ is the time-reversal symmetrg group. Denoting the Ising and time-reversal transformations on the physical d.o.f. in one unit cell as $U_g$, and $U_{\mathcal{T}}$ (antiunitary), the projective representation $\alpha$ per unit cell satisfies:
\begin{align}
 U_g^2&=1,& U_{\mathcal{T}}U_{\mathcal{T}}^*&=-1, &U_{\mathcal{T}} U^*_g&=U_g U_{\mathcal{T}}.\label{eq:Kramer}
\end{align}
For instance, this algebra is satisfied if $U_g=\sigma_x$ and $U_{\mathcal{T}}=i \tau_y$ for a four-dimensional local Hilbert space (upon which $\sigma$ and $\tau$ Pauli matrices act). One can check that the two conditions in Theorem-(1) are both satisfied, and thus \emph{at least an SRE liquid phase can exist and must be an SPT phase in which the $g$-symmetry-defect carries the projective representation $\alpha$ (a Kramer-doublet) above.} 

Naively this example is very similar to the example-(1). However there is an important difference. In this example, \emph{only one SPT phase can be realized} --- following the Remark, this is because the kernel subgroup $\mathcal{A}_g$ is the trivial $Z_1$ group. \footnote{Following the Kunneth formula: $H^3(Z_2^T\times Z_2^{Ising},U(1))=H^3(Z_2^T,U(1))\times H^3(Z_2,U(1))\times H^2(Z_2^T,Z_2)=Z_1\times Z_2\times Z_2$. In this example, the $Z_2$ index in $H^2(Z_2^T,Z_2)$ is enforced to be nontrivial, and the $Z_2$ index in $H^3(Z_2,U(1))$ is enforced to be trivial. This is because here $\omega_0$ in Theorem-2 is the nontrivial element in $H^2(Z_2^T,Z_2)$, and the kernel subgroup $\mathcal{A}_g=Z_1$.} \emph{After gauging the $Z_2^{Ising}$ symmetry, one must obtain a toric code topological order.} This realizable SPT phase is topologically identical to the one obtained by decorating Ising domain walls with the $Z_2^T$ Haldane chains\cite{chen2014symmetry}.

\textbf{(3) $\mathbf{G=Z_2^T\times Z_2^{Ising}}$, and a non-Kramer doublet per unit cell:} Here the projective representation $\alpha$ per unit cell satisfies:
\begin{align}
 U_g^2&=1,& U_{\mathcal{T}}U_{\mathcal{T}}^*&=1, &U_{\mathcal{T}} U^*_g&=-U_g U_{\mathcal{T}}.\label{eq:non_Kramer}
\end{align}
For instance, $U_g=\sigma_x$ and $U_{\mathcal{T}}=\sigma_z$ on a two-dimensional local Hilbert space would satisfy this algebra. One can check that the two conditions in Theorem-(1) are both satisfied, and thus \emph{at least an SRE liquid phase can exist and must be an SPT phase in which the $g$-symmetry-defect carries the projective representation $\alpha$ (a Kramer-doublet) above.} 

Similar to example-(2), \emph{there is only one realizable SPT phase}. This SPT phase turns out to be the non-trivial Levin-Gu SPT phase protected by the $Z_2^{Ising}$ alone. \footnote{Now in this example, $\omega_0$ in Theorem-2 is the nontrivial element in $H^3(Z_2,U(1))$, and the kernel subgroup $\mathcal{A}_g=Z_1$.} Namely, here \emph{after gauging the $Z_2^{Ising}$ symmetry, one must obtain a double-semion topological order. }

\textbf{(4) $\mathbf{G=Z'_2\times Z_2^{Ising}}$, and a projective representation per unit cell:} In this example, $Z'_2=\{I,h\}$ is another unitary Ising symmetry group. The projective representation $\alpha$ satisfies:
\begin{align}
 U_g^2&=1, & U_h^2&=1,&U_gU_h&=-U_hU_g.
\end{align}
For instance, $U_g=\sigma_x$, $U_h=\sigma_z$ realize this algebra. Two of such projective representations fuse into a regular representation of $G$, so the condition-(1) in Theorem-(1) is satisified. But one can show that the condition-(2) is \emph{not} satisfied:
\begin{align}
 \gamma_g^{\alpha}(h)=-1,
\end{align}
i.e., $\gamma_g^{\alpha}$ is a nontrivial 1-cocycle. Therefore according to Theorem-(1), \emph{a SRE liquid phase is not possible.} Without breaking symmetry, this suggests that topological order is inevitable for gapped systems. This is a somewhat surprising result. If one views the system using the enlarged magnetic unit cell, there is no reason why a SRE liquid is not allowed.

\section{Decorated Quantum Dimer Models for SPT phases}
Closely related to the decorated-BFG model in Sec.\ref{sec:BFG}, in this section we describe a class of exactly solvable models realizing symmetry-enforced SPT phases. These models are constructed by decorating quantum dimer models(QDM) with relevant physical degrees of freedom, whose ground states can be exactly solved at the corresponding Rokhsar-Kivelson point\cite{rokhsar1988superconductivity}. Although this class of models can be generalized to other lattices, here we will focus on the decoration of the QDM on the triangular lattice\cite{moessner2001resonating,qi2015double}. In particular, we will construct models realizing the symmetry-enforced SPT phases in example-(1,2,3) in Sec.\ref{sec:examples}

\subsection{$G=SO(3)\times Z_2^{Ising}$, a spin-1/2 per unit cell}\label{sec:SO3_Ising_models}
Continuing with discussions in example-(1) in Sec.\ref{sec:examples}, in the presence of onsite global symmetry $G=SO(3)\times Z_2^{Ising}$, we consider quantum systems with one spin-$1/2$ per unit cell in two spatial dimensions respecting the Ising magnetic translation symmetry Eq.(\ref{eq:mag_transl}). Note that we will reserve symbols $T_x,T_y$ for the magnetic translations, and use $T_x^{orig.},T_y^{orig.}$ to represent the original translations. 

We will construct two exactly solvable models (model-A and model-B) respecting the symmetry described above featuring SRE liquid ground states. Although the Ising defects in both models carry half-integer spins, the two models are in distinct SPT phases. The simplest way to understand their difference is that, after gauging the Ising symmetry, model-A has toric-code topological order while model-B has double-semion topological order.

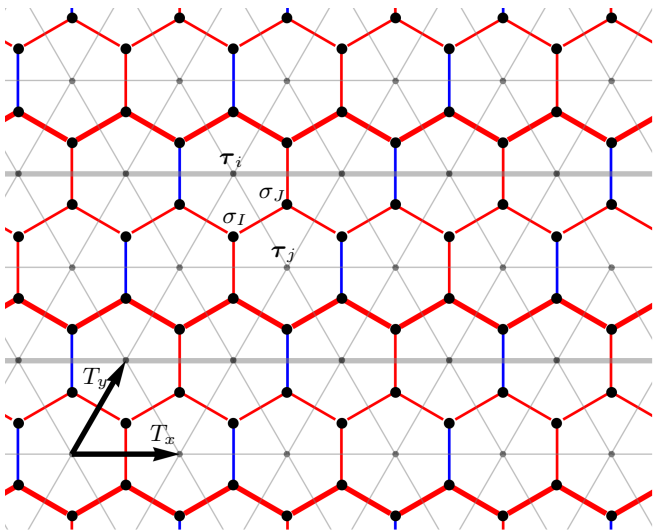
\begin{figure}
\begin{tikzpicture}
\node {\includegraphics[width=0.48\textwidth]{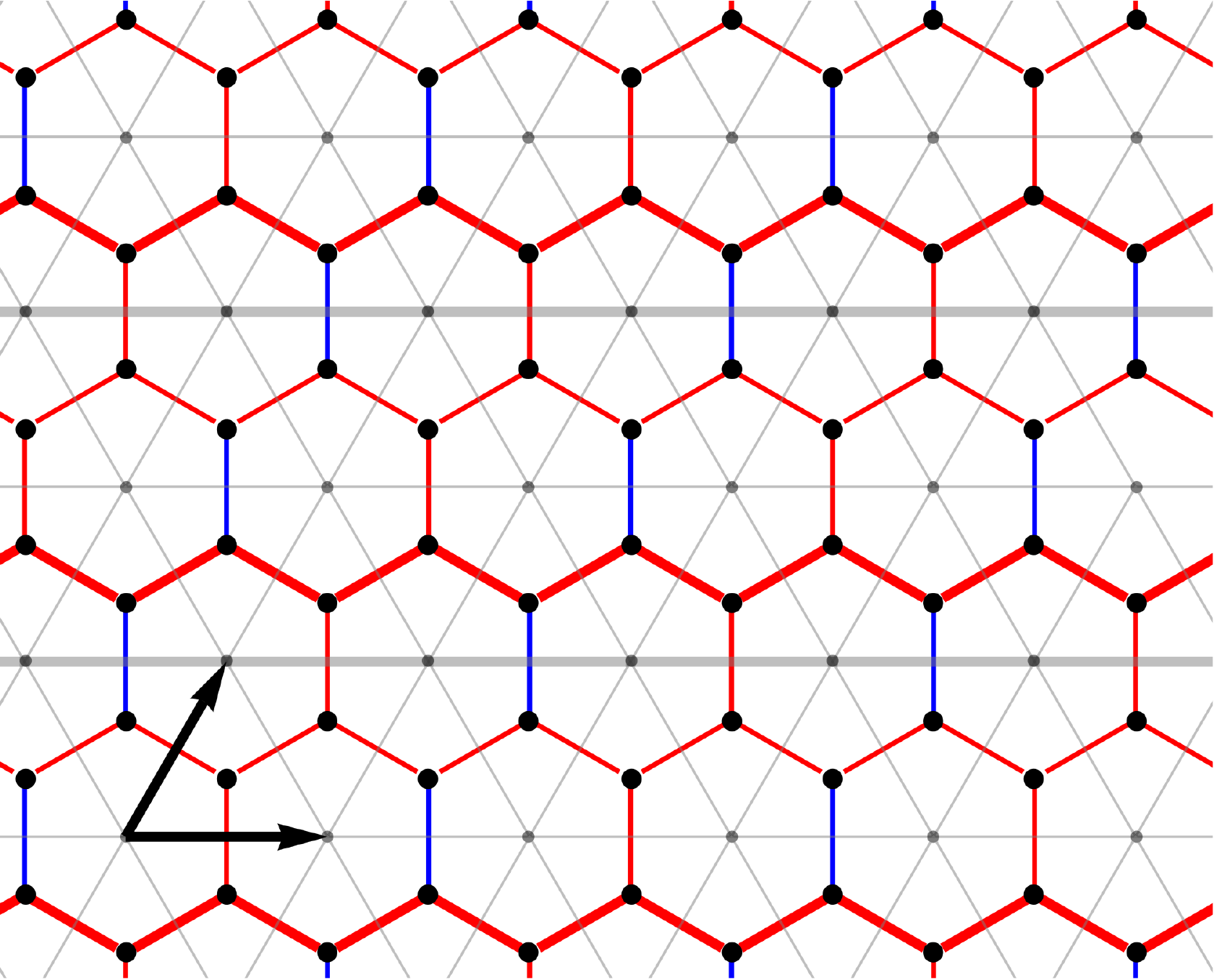}};
\node at (-2.2,-2.2) [inner sep=0] {$T_x$};
\node at (-3.1,-1.45) [inner sep=0] {$T_y$};
\node at (-1.25,0.67) [inner sep=0] {$\sigma_I$};
\node at (-0.75,1.) [inner sep=0] {$\sigma_J$};
\node at (-0.6,0.2) [inner sep=0] {$\boldsymbol{\tau}_j$};
\node at (-1.3,1.45) [inner sep=0] {$\boldsymbol{\tau}_i$};
\end{tikzpicture}
\caption{(color online) The Ising d.o.f. $\sigma$ live on the honeycomb lattice and the spin d.o.f. $\boldsymbol{\tau}$ lives on the triangular lattice. The Ising coupling signs $s_{IJ}=+1$ on red bonds, and $s_{IJ}=-1$ on blue vertical bonds. The thick red bonds represent the ``$y$-odd zigzag chains'' used in Eq.(\ref{eq:Ising_mag_transl},\ref{eq:modified_Ising_mag_transl} ). The thick gray horizontal bonds on the triangular lattice represent the ``$y$-odd rows'' used in Eq.(\ref{eq:modified_Ising_mag_transl}). }
\label{fig:model-A}
\end{figure}

We start with constructing model-A. This model contains two sets of degrees of freedom (d.o.f.): the Ising-d.o.f. $\sigma_I$ which live on a honeycomb lattice and the spin-$1/2$-d.o.f. $\mathbf{S}_i=\boldsymbol{\tau}_i/2$ which live on the triangular lattice formed by centers of the hexagons, as shown in Fig.\ref{fig:model-A}. The Hamiltonian of model-A contains three terms:
\begin{align}
 H=H^{Ising}+H^{binding}+H^{A}.\label{eq:model-A}
\end{align}
where $H^{Ising}$ is simply a frustrated nearest-neighbor Ising model:
\begin{align}
 H^{Ising}=-K\sum_{\langle I J\rangle} s_{IJ}\sigma_I^{z}\sigma_J^{z}.\label{eq:H_Ising}
\end{align}
Here $\sigma_I^z,\sigma_J^z$ are the Ising spins living on the honeycomb sites labled by $I,J$, the coupling constant $K>0$, and $s_{IJ}=\pm 1$ defined as in Fig.\ref{fig:model-A}. $H^{binding}$ is an interaction between the Ising-d.o.f. and spin-$1/2$-d.o.f., which commutes with $H^{Ising}$:
\begin{align}
H^{binding}=-\lambda \sum_{\intersectinglinesA} \frac{1}{2}(1-s_{IJ}\sigma_I^z\sigma_J^z)\cdot \hat P_{\mathbf{S}_i+\mathbf{S}_j=0},\label{eq:H_binding}
\end{align}
where $\lambda>0$, $i,j$ labels the sites on the triangular lattice, and the summation of ``$\intersectinglines$'' is over all intersection points between the triangular lattice and the honeycomb lattice as shown in Fig.\ref{fig:model-A}. $\hat P_{\mathbf{S}_i+\mathbf{S}_j=0}\equiv \frac{1}{4}- \mathbf{S}_i\cdot \mathbf{S}_j$ is the operator projecting the two spin-1/2's on site-$i$ and site-$j$ into a spin singlet. 

$H^{A}$ is more complicated and will be given in Eq.(\ref{eq:HA}). It is straightforward to checked that $H$ respects the Ising symmetry $U_g\equiv\prod_I \sigma_I^x$, the spin-rotation symmetry generated by $\sum_i \mathbf{S}_i$. $H$ also respects magnetic translation operations: 
\begin{align}
T_x&=(\prod_{I\in \mbox{\scriptsize{y-odd zigzag chains}}}\sigma_I^x )\cdot T_x^{orig.},&T_y&=T_y^{orig.},\label{eq:Ising_mag_transl}
\end{align} 
(see Fig.\ref{fig:model-A}), and the magnetic translation algebra Eq.(\ref{eq:mag_transl}) is satisfied. We will show that the ground state of $H$ is in a gapped liquid phase without topological order. According to our general results, this ground state must be in an SPT phase, which we will show momentarily.

The physical consequence of $H^{Ising}$ and $H^{binding}$ is to provide a highly degenerate low energy manifold, which will be lifted by $H^{A}$. To understand the low energy manifold, let us firstly consider $H^{Ising}$. Because every hexagonal plaquette frustrated, there will be at least one bond in each plaquette that is energetically unhappy. Namely the ground state manifold of $H^{Ising}$ is formed by all possible Ising configurations satisfying the ``one-unhappy-Ising-bond-per-plaquette'' condition. 

$H^{binding}$ further constrains the spin-1/2 d.o.f. in the low energy manifold. It has effect only on the Ising unhappy bonds ($s_{IJ}\sigma_I^z\sigma_J^z=-1$), and energetically binds the two spin-1/2's near the unhappy Ising-bond into a spin singlet. The degenerate ground state manifold of $H^{Ising}+H^{binding}$ is now clear: it is formed by all such quantum states satisfying the ``one-unhappy-Ising-bond-per-plaquette'' condition and the two neighboring spin-1/2's normal to every unhappy Ising-bond form a spin singlet. 

It is well-known that the Ising configurations satisfying the ``one-unhappy-Ising-bond-per-plaquette'' condition are intimately related to the Hilbert space of the QDM\cite{moessner2001ising,moessner2001short,moessner2000two}. Pictorically, any such an Ising configuration can be mapped to a dimer covering (with one dimer per site) on the triangular lattice by assigning a dimer crossing the unhappy Ising bond. The effect of $H^{binding}$ is simply to energetically binds the two spin-1/2's in each dimer into a spin-singlet. Namely if the Ising configuration is given, the state of spin-1/2 d.o.f. is also fixed. 

Similar to the model in Sec.\ref{sec:BFG}, there is a two-to-one mapping from the ground state manifold of $H^{Ising}+H^{binding}$ to the QDM Hilbert space, since two low energy states differ by a global Ising transformation map into the same state in the QDM. Second, these states only map to one specific topological sector of the QDM: The parity of the number of dimers crossing a loop is simply given by the sign of the product $\prod_{\langle IJ\rangle\in loop} s_{IJ}$.

In fact, these relations between the Ising d.o.f. and the dimer d.o.f. can be viewed as the well-known duality mapping between quantum Ising models and $Z_2$ gauge theories\cite{kogut1979introduction}. The Ising paramagnet phase is dual to the deconfined $Z_2$ gauge phase in the QDM, while the Ising ordered phase is dual to the confined $Z_2$ gauge theory. More precisely, the ``unhappy-Ising-bond'' is dual to the electric flux line (i.e., the dimer) in the gauge theory, and the ``one-unhappy-Ising-bond-per-plaquette'' condition in the low energy manifold on the Ising side is dual to the ``one-$Z_2$-gauge-charge-per-site'' condition in the QDM. Here, the only new ingredient apart from this well-known duality is that there are also spin-1/2 d.o.f. on the Ising side. But due to $H^{binding}$, these spin-1/2's form a pattern of spin singlets fixed by the Ising d.o.f., and consequently the duality mapping is not modified after a convention of the spin-singlet signs is given (see below). 

After introducing $H^{A}$, we will see that the degeneracy in the low energy manifold will be lifted and the unique ground state on a torus sample is formed. The usual QDM Hamiltonian on the triangular lattice is:
\begin{align}
 H^{TC}_{QDM}&=-t\sum_{\mbox{\scriptsize{plaquettes}}}\big(|\pA\rangle\langle\pB|+h.c.\big)\notag\\
 &+v\sum_{\mbox{\scriptsize{plaquettes}}}\big(|\pA\rangle\langle\pA|+|\pB\rangle\langle\pB|\big),\label{eq:TC_QDM}
\end{align}
where the summation is over all plaquettes (rhombi): ``$\pgA$'', ``$\pgB$'', ``$\pgC$''. The ground states of this model are exactly known at the RK-point given by $t=v>0$, and the superscript $TC$ is highlighting that the topological order is toric-code\cite{Kitaev:2003p2,moessner2001resonating} like in the deconfined phase (i.e., the usual $Z_2$ gauge theory). At this point $H^{TC}_{RK}$ can be rewritten as a summation of projectors\cite{moessner2001resonating}:
\begin{align}
 H^{TC}_{RK}&=t\sum_{\mbox{\scriptsize{plaquettes}}}\big(|\pA\rangle-|\pB\rangle\big)\big(\langle\pA|-\langle\pB|\big).
\end{align}
Clearly the equal weight superposition of all dimer coverings within any fixed topological sector $|\Phi^{TC}_{RK}\rangle=\sum_c |c\rangle$ ($c$ labels possible dimer coverings) is one ground state of $H^{TC}_{RK}$ since it is annihilated by all projectors.

Based on the duality mapping, $H^{TC}_{QDM}$ is mapped to $H^A$. Any dimer covering $|c\rangle$ will be mapped to two Ising configurations $|c_1\rangle$ and $|c_2\rangle$ distinct from each other by a global Ising transformation. One can further choose a translationally symmetric sign convention for the spin-singlets on the nearest neighbor bonds along the three orientations:
\begin{align}
 |\elA\rangle&\equiv \frac{1}{\sqrt{2}}\big(|\ellA\rangle -|\ellAm\rangle\big),\notag\\
 |\raisebox{-2mm}\elB\rangle&\equiv \frac{1}{\sqrt{2}}\big(\big|\raisebox{-2mm}\ellB\big\rangle -\big|\raisebox{-2mm}\ellBm\big\rangle\big),\notag\\
 |\raisebox{-2mm}\elC\rangle&\equiv \frac{1}{\sqrt{2}}\big(\big|\raisebox{-2mm}\ellC\big\rangle -\big|\raisebox{-2mm}\ellCm\big\rangle\big)\rangle.\label{eq:singlet_dimer}
\end{align}
With this sign convention, given a $c$, $|c_i\rangle$ ($i=1$ or $2$) fully determines a state in the ground state manifold of $H^{Ising}+H^{binding}$ by replacing the dimer configuration by the corresponding spin-singlet configuration. In addition, $\{|c_1\rangle,|c_2\rangle\}$ for all $c$ form a complete orthornormal basis in this manifold.

$H^A$ is defined as:
\begin{align}
 H^{A}=&-t\sum^{\sigma_I^z,\sigma_J^z=\pm1}_{\mbox{\scriptsize{plaquettes}}}\left(\big|\raisebox{-4mm}\pgelA\big\rangle \big\langle\raisebox{-4mm}\pgelB\big|+h.c.\right)\notag\\
 &+v\sum^{\sigma_I^z,\sigma_J^z=\pm1}_{\mbox{\scriptsize{plaquettes}}} \left(\big|\raisebox{-4mm}\pgelC\big\rangle \big\langle\raisebox{-4mm}\pgelC\big|\right.\notag\\
  &\;\;\;\;\;\;\;\;+\left.\big|\raisebox{-4mm}\pgelD\big\rangle \big\langle\raisebox{-4mm}\pgelD\big|\right)\label{eq:HA}
\end{align}
Note that the $t$-term also flips the two Ising spins inside the plaquette. At the RK point $t=v>0$, $H^A$ can again be written as a summation of projectors:
\begin{align}
 H^{A}_{RK}=t\sum^{\sigma_I^z,\sigma_J^z=\pm1}_{\mbox{\scriptsize{plaquettes}}}&\left(\big|\raisebox{-4mm}\pgelA\big\rangle -\big|\raisebox{-4mm}\pgelB\big\rangle \right)\notag\\
 &\cdot \left(\big\langle\raisebox{-4mm}\pgelA\big|-\big\langle\raisebox{-4mm}\pgelB\big|\right)\label{eq:A_RK}
\end{align}
To study the ground state of the total Hamiltonian $H$, it is suffice to focus on the degenerate ground state manifold of $H^{Ising}+H^{binding}$, and clearly $H^{A}$ acts within this manifold. In addition, it is straightforward to show that $|\Phi^A_{RK}\rangle=\sum_{c}(|c_1\rangle+|c_2\rangle)$, i.e, the equal weight superposition of all states in this manifold, is a ground state of $H^A_{RK}$ because it is annihilated by every projector in Eq.(\ref{eq:A_RK}). $|\Phi^A_{RK}\rangle$ is clearly a fully symmetric liquid wavefunction.

It is known that for the QDM Eq.(\ref{eq:TC_QDM}), the RK point of $H$ is exactly at a first-order phase transition boundary between a deconfined gapped liquid phase ($v<t$) and a staggered valence bond solid phase ($v>t$)\cite{moessner2001resonating}. Based on the duality mapping, the model-A is in a fully symmetric gapped liquid phase for $v_c<v<t$ with a unique ground state on torus. In the limit of $K,\lambda\gg v,t$, $v_c$ is given by the same critical value $v_c\approx 0.7 t$ as in the original QDM\cite{moessner2001resonating}. More precisely, in the global Ising-even sector of the Hilbert space of model-A, this mapping to the Hilbert space of the QDM is given by $1/\sqrt{2}(|c_1\rangle+|c_2\rangle)\rightarrow |c\rangle$, and clearly $H^A$ is mapped to $H^{TC}_{QDM}$. Namely, the full energy spectrum of model-A in the Ising-even sector has a one-to-one correspondence with the full energy spectrum of $H^{TC}_{QDM}$. In addition, in the Ising paramagnetic phase, ground state in the global Ising-odd sector of model-A has a finite excitation energy which is the same as the energy cost of a $Z_2$ gauge flux in the QDM.

Next we show that the liquid phase $v_c<v<t$ in model-A is an SPT phase because the Ising defects carry half-integer spins. Similar to the discussion in Sec.\ref{sec:BFG} (see Fig.\ref{fig:Ising_defect_BFG}), after a pair of Ising defects are spatially separated the original Hamiltonian $H$ is modified into $H'$. Comparing with $H$, the $s_{IJ}$ flips sign in $H'$ whenever the bond $I-J$ crosses the branch cut. Namely, for any loop on the honeycomb lattice enclosing a single Ising defect, the product $\prod s_{IJ}$ along the loop changes sign. In order not to cost $H'^{binding}$ energy, the parity of the number of dimers crossing this loop also flips. Consequently, an Ising defect is topologically bound with a monomer (an unpaired site on the triangular lattice). This monomer clearly carries a half-integer spin in model-A, which can be detected by the local spin susceptibility at low temperatures.

Next, we demonstrate a different symmetry-enforced SPT phase using the model-B defined as follows:
\begin{align}
 H=H^{Ising}+H^{binding}+H^B.\label{eq:model-B}
\end{align}
Comparing with the model-A in Eq.(\ref{eq:model-A}), only the last term is modified:
\begin{widetext}
\begin{align}
 H^B=&v\sum^{\sigma_I^z,\sigma_J^z=\pm1}_{\mbox{\scriptsize{plaquettes}}} \left(\big|\raisebox{-4mm}\pgelC\big\rangle \big\langle\raisebox{-4mm}\pgelC\big|+\big|\raisebox{-4mm}\pgelD\big\rangle \big\langle\raisebox{-4mm}\pgelD\big|\right)+\sum^{\sigma_I^z,\sigma_J^z=\pm1}_{\hspace{-1mm}\pgA}\left(-it\big|\raisebox{-4mm}\pgelA\big\rangle \big\langle\raisebox{-4mm}\pgelB\big|+h.c.\right)\notag\\
 &+\sum^{\sigma_I^z,\sigma_J^z=\pm1}_{\hspace{-1mm}\pgB}\left(-it\big|\raisebox{-4mm}\pgelE\big\rangle \big\langle\raisebox{-4mm}\pgelF\big|+h.c.\right)
 +\sum^{\sigma_I^z,\sigma_J^z=\pm1}_{\hspace{-1mm}\pgC}\left(-it\big|\raisebox{-6mm}\pgelG\big\rangle \big\langle\raisebox{-6mm}\pgelH\big|+h.c.\right)
 \label{eq:HB}
\end{align}
\end{widetext}
One can straightforwardly check that the model-B defined in Eq.(\ref{eq:model-B},\ref{eq:HB}) also respects the Ising symmetry, the $SO(3)$ spin-rotation symmetry, and the magnetic translation symmetry Eq.(\ref{eq:mag_transl}). Below we show that in a finite parameter regime $v'_c<v<t$, the model-B is in a gapped liquid phase, and this phase is another SPT phase.

In the ground state manifold of $H^{Ising}+H^{binding}$, the duality transformation maps $H^B$ into the following QDM Hamiltonian:
\begin{align}
 &H^{DS}_{QDM}=v\sum_{\mbox{\scriptsize{plaquettes}}}\big(|\pA\rangle\langle\pA|+|\pB\rangle\langle\pB|\big)\notag\\
 &+\sum_{\pgA,\pgB,\pgC}-it\big(|\pA\rangle\langle\pB|+|\pC\rangle\langle\pD|+|\raisebox{-2mm}\pE\rangle\langle\raisebox{-2mm}\pF|\big)+h.c.\label{eq:DS_QDM}
\end{align}
This QDM was firstly introduced and studied in Ref.~\onlinecite{qi2015double}, where the exactly solvable RK point $t=v$ has been shown to be adjacent a gapped liquid phase for $v\lesssim t$. Interestingly, this phase was demonstrated to have a double-semion topological order (the superscript $DS$ here is to highlight this fact). By the duality mapping, we know that in a finite parameter regime $v'_c<v<t$, the model-B is in a gapped liquid phase.

Similar to previous disussion on the model-A, it is straightforward to show that the Ising defect in the gapped liquid phase of model-B also carries half-integer spin, so it is also an SPT phase. To see the difference from the SPT phase realized in model-A, let us consider the Ising symmetry only. It is known that Ising symmetry itself can protect two paramagetic phases: the trivial phase and the SPT phase. Levin and Gu pointed out\cite{levin2012braiding} that the duality mapping maps the usual Ising paramagnet to the toric-code topological order, while the nontrivial Ising SPT phase maps to the double semion topological order. Consequently, the SPT phases realized in model-A and model-B are different because the Ising symmetry alone already distinguishes them.

\subsection{$G=Z_2^T\times Z_2^{Ising}$}\label{sec:Z2T_Ising_models}
Here we demonstrate symmetry-enforced SPT phases outlined in example-(2) and (3) in Sec.\ref{sec:examples}. Unlike the $G=SO(3)\times Z_2^{Ising}$ case, here we show that symmetry conditions fully determine the SPT phase. The models below have the same Hilbert space as in the $G=SO(3)\times Z_2^{Ising}$ case (i.e., $\sigma_I$ on the honeycomb lattice and $\boldsymbol{\tau}_i$ on the triangular lattice), but with different symmetries defined.

\textbf{A Kramer doublet per unit cell}: A simple generalization is for example-(2) (i.e., a Kramer-doublet per unit cell) where we can recycle the model-A. Namely, defining the Ising symmetry $U_g=\prod_I\sigma_I^x$ as before and the antiunitary time-reversal symmetry as $U_{\mathcal{T}}=e^{i\pi\mathbf{S}^y}=i\boldsymbol{\tau}^y$, clearly model-A respect all the required symmetries. In addition, we have one Kramer doublet $\boldsymbol{\tau}$ per unit cell. According to our discussion in example-(2), this SRE liquid phase realized in $v_c<v<t$ must be an SPT phase in which the Ising defect carries a Kramer doublet, which is obviously realized in model-A. In addition, after gauging $Z_2^{Ising}$ symmetry, one necessarily obtains the toric-code topological order, which is confirmed in model-A. 

On the other hand, model-B, gauging which gives double-semion topological order, explicitly breaks the time-reversal symmetry defined above.

\begin{figure}
 \includegraphics[width=0.48\textwidth]{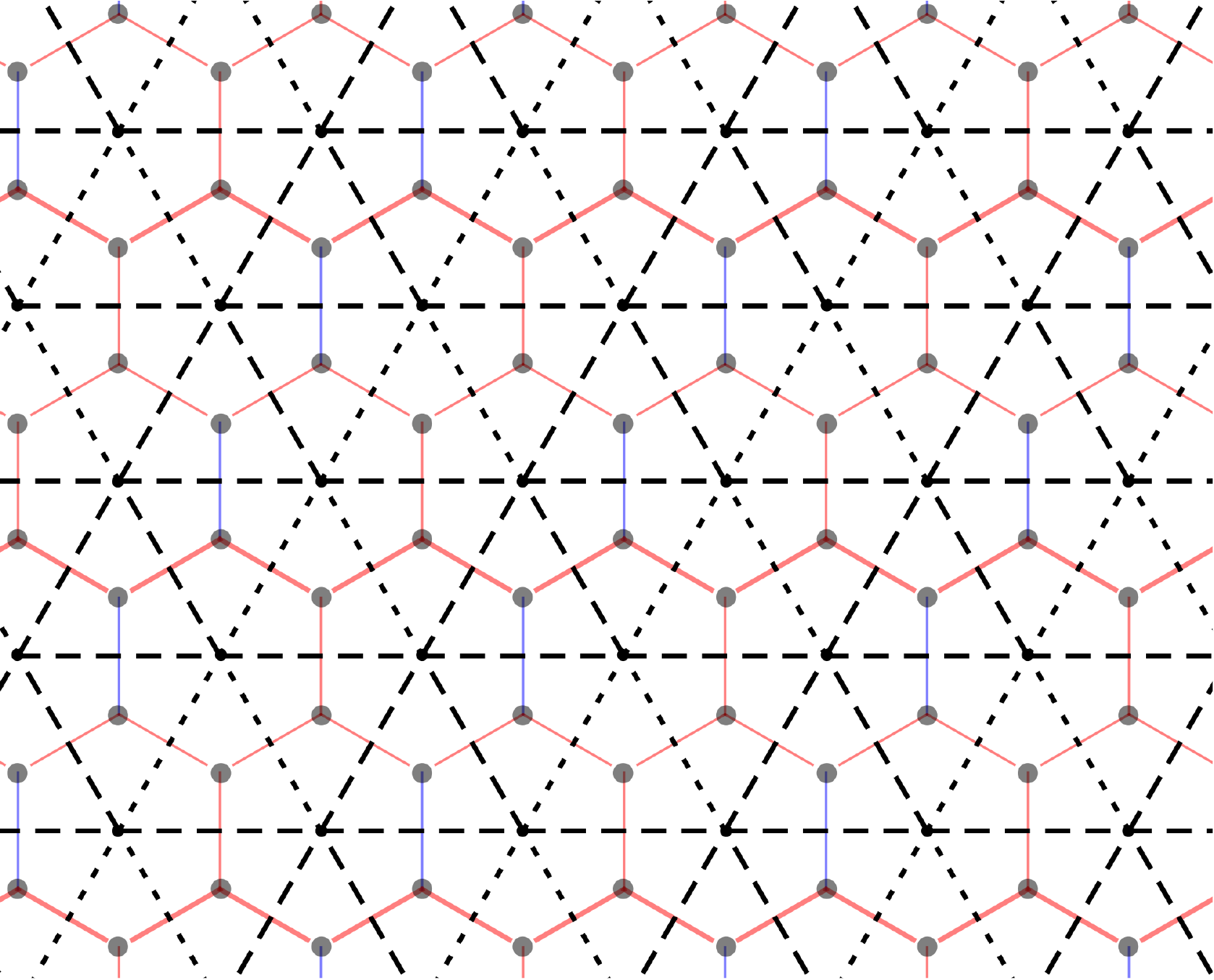}
 \caption{To construct model in Eq.(\ref{eq:tilde_model_B}), the dimer states living on the nearest neighbor bonds on the triangular lattice have a spatial dependent pattern: the dimer states living on the dashed bonds are defined in Eq.(\ref{eq:dashed_singlet_dimer}), and those living on the dotted bonds are defined in Eq.(\ref{eq:dotted_singlet_dimer}).}
 \label{fig:tilde_symmetric_dimers}
\end{figure}

\textbf{A non-Kramer doublet per unit cell}: Now let us move on to example-(3) in Sec.\ref{sec:examples}. In order to construct a model realizing the non-Kramer doublet projective representation defined in Eq.(\ref{eq:non_Kramer}), let us define the following symmetry operations:
\begin{align}
 U_g&=\boldsymbol{\tau}^z\prod_I\sigma_I^x ,&U_\mathcal{T}&=\boldsymbol{\tau}^x.\label{eq:U_non_Kramer}
\end{align}
Consequently we have one non-Kramer doublet $\boldsymbol{\tau}$ per unit cell. The model Hamiltonian in this example will be given by
\begin{align}
 H=H^{Ising}+\tilde H^{binding}+\tilde H^B,\label{eq:tilde_model_B}
\end{align}
where $H^{Ising}$ is given in Eq.(\ref{eq:H_Ising}), $\tilde H^{binding}$ is in Eq.(\ref{eq:tilde_H_binding}). $\tilde H^B$ has the same form as $H^B$ in Eq.(\ref{eq:HB}), but with a modified interpretations of the dimers: replacing the $|\elA\rangle$'s defined in Eq.(\ref{eq:singlet_dimer}) by $|\elAp\rangle$'s and $|\elApp\rangle$'s defined in Eq.(\ref{eq:dashed_singlet_dimer},\ref{eq:dotted_singlet_dimer}) depending the dimer positions (see discussions below and Fig.\ref{fig:tilde_symmetric_dimers}). Eventually we will show that in a finite regime $v \lesssim t$ this model features a SRE gapped liquid ground state which is the Levin-Gu Ising SPT phase, consistent with discussions in Sec.\ref{sec:examples}.

We will construct a model similar to $H=H^{Ising}+H^{binding}+...$, but respecting a magnetic translation operations different from Eq.(\ref{eq:Ising_mag_transl}) since $U_g$ is now different. In particular, we define magnetic translations:
\begin{align}
 T_x&=\big(\prod_{i\in \mbox{\scriptsize{y-odd row}}}\boldsymbol{\tau}_i^z\big)\big(\prod_{I\in \mbox{\scriptsize{y-odd zigzag chains}} }\sigma_I^x\big)T_x^{orig.},\notag\\
 T_y&=T_y^{orig.},\label{eq:modified_Ising_mag_transl}
\end{align}
(see Fig.\ref{fig:model-A}), which satisifies Eq.(\ref{eq:mag_transl}) with $g=\prod_i\boldsymbol{\tau}_i^z\prod_I\sigma_I^x$ following Eq.(\ref{eq:U_non_Kramer}). Although $H^{Ising}$ still respects all the symmetries, $H^{binding}$ does not respect $g$ and $T_x$. This is because the usual spin singlets $|\elA\rangle$ in the projector $\hat P_{\mathbf{S}_i+\mathbf{S}_j=0}=|\elA\rangle\langle \elA|$ does not respect $g$ and $T_x$. We therefore need to modify these dimer states and the projectors.

We define the following dimer states formed by the $\boldsymbol{\tau}$ spins:
\begin{align}
 |\elAp\rangle&\equiv \frac{1}{\sqrt{2}}\big(|\ellAp\rangle +|\ellApm\rangle\big),\notag\\
 |\raisebox{-2mm}\elBp\rangle&\equiv \frac{1}{\sqrt{2}}\big(\big|\raisebox{-2mm}\ellBp\big\rangle +\big|\raisebox{-2mm}\ellBpm\big\rangle\big),\notag\\
 |\raisebox{-2mm}\elCp\rangle&\equiv \frac{1}{\sqrt{2}}\big(\big|\raisebox{-2mm}\ellCp\big\rangle +\big|\raisebox{-2mm}\ellCpm\big\rangle\big)\rangle.
 \label{eq:dashed_singlet_dimer}
\end{align}
Clearly these dimer states are both Ising and time-reversal even according to Eq.(\ref{eq:U_non_Kramer}).

Under the magnetic translation $T_x$ in Eq.(\ref{eq:modified_Ising_mag_transl}), dimer states with $|\raisebox{-2mm}\elBp\rangle$ and $|\raisebox{-2mm}\elCp\rangle$ connecting even and odd rows on the triangular lattice will transform into the following states:
\begin{align}
 |\raisebox{-2mm}\elBpp\rangle&\equiv \frac{1}{\sqrt{2}}\big(\big|\raisebox{-2mm}\ellBp\big\rangle -\big|\raisebox{-2mm}\ellBpm\big\rangle\big),\notag\\
 |\raisebox{-2mm}\elCpp\rangle&\equiv \frac{1}{\sqrt{2}}\big(\big|\raisebox{-2mm}\ellCp\big\rangle -\big|\raisebox{-2mm}\ellCpm\big\rangle\big)\rangle.
 \label{eq:dotted_singlet_dimer}
\end{align}
Note that these dotted dimer states are Ising even, but \emph{time-reversal odd} according to Eq.(\ref{eq:U_non_Kramer}). The magnetic translational symmetric assignment of the dimer states is given in Fig.(\ref{fig:tilde_symmetric_dimers}): the states $|\elAp\rangle$, $|\raisebox{-2mm}\elBp\rangle$, $|\raisebox{-2mm}\elCp\rangle$ are assigned on the dashed bonds, while the states $|\raisebox{-2mm}\elBpp\rangle$, $|\raisebox{-2mm}\elCpp\rangle$ are assigned on the dotted bonds.

By replacing the spin singlets in $\hat P_{\mathbf{S}_i+\mathbf{S}_j=0}=|\elA\rangle\langle \elA|$ by the corresponding $|\elAp\rangle\langle \elAp|$ and $|\elApp\rangle\langle \elApp|$ in the spatial dependent fashion in Fig.\ref{fig:tilde_symmetric_dimers}, we modify $H^{binding}$ naturally as
\begin{align}
 \tilde H^{binding}&=-\lambda \sum_{\intersectinglinesAp} \frac{1}{2}(1-s_{IJ}\sigma_I^z\sigma_J^z)\cdot \big(|\elAp\rangle\langle\elAp|\big)\notag\\
 &-\lambda \sum_{\intersectinglinesApp} \frac{1}{2}(1-s_{IJ}\sigma_I^z\sigma_J^z)\cdot \big(|\elApp\rangle\langle\elApp|\big),\label{eq:tilde_H_binding}
\end{align}

Now $H^{Ising}+\tilde H^{binding}$ respects all the required symmetries, but one still need terms like $H^A$ or $H^B$ to lift the ground state degeneracy to reach a SRE gapped liquid phase. In order to preserve the Ising and magnetic translation symmetry, it is natural to replace the dimer states $|\elA\rangle$ by the corresponding $|\elAp\rangle$ and $|\elApp\rangle$. Let us denote the resulting modified Hamiltonians as  $\tilde H^A$ or $\tilde H^B$. Both modified models $H^{Ising}+\tilde H^{binding}+\tilde H^A$ and $H^{Ising}+\tilde H^{binding}+\tilde H^B$ are solvable, by duality mapping to the $H^{TC}_{QDM}$ and $H^{DS}_{QDM}$ respectively. These two models both give SRE gapped ground states in the regime $v\lesssim t$ respecting the magnetic translation symmetry and the Ising symmetry, gauging which give toric-code and double semion topological order respectively. 

Finally, let us consider the time-reversal symmetry in Eq.(\ref{eq:non_Kramer}). Importantly, $|\elAp\rangle$'s are time-reversal even while $|\elApp\rangle$ are time-reversal odd. As shown in the pattern Fig.\ref{fig:tilde_symmetric_dimers}, any dimer resonant term like $|\pA\rangle\langle\pB|$ will involve an odd number of $|\elApp\rangle$ states. Consequently, \emph{only $\tilde H^B$ is time-reversal symmetric, while $\tilde H^A$ explicitly breaks the time-reversal.} In fact according to Theorem-I, it is impossible to have a SRE liquid respecting all the required symmetries and gauging the Ising symmetry gives a toric code topological order.

\section{Proof of Theorems}
Here we present a combination of physical argument and mathematical derivations based on symmetric tensor network formulation\cite{perez2010characterizing,Zhao:2010p174411,Singh:2010p50301,Singh:2011p115125, Singh:2012p195114,Bauer:2011p125106,Weichselbaum:2012p2972,jiang2015symmetric,jiang2017anyon}. We will focus on Theorem-II, and in Appendix\ref{app:theorem-i} we show that Theorem-I can be viewed as its special case. 

We need to show the condition in Theorem-II is necessary and sufficient for a SRE liquid phase to exist, which must be an SPT phase. To show this condition is necessary, we consider such a SRE liquid phase and the pumping of the entanglement spectra during an adiabatic process in Sec.\ref{sec:entanglement_pumping}, leading to an observation that a $g$-symmetry defect must carry a projective representation $\alpha$. In a SRE liquid phase characterized by 3-cocycle $\omega$, we use the symmetric tensor-network formulation in Sec.\ref{sec:SPT_constraints} to establish that the projective representation carried by a $g$-symmetry defect is given by $(\delta_g^{\omega})^{-1}$. Since the projective representation carried by a symmetry defect is physical and independent of formulation, together with the pumping argument, the necessary condition in Theorem-II is established \emph{independent of formulation}. 

As a complementary calculation, we also explicitly compute the projective representation carried by a $g$-symmetry defect in a SRE liquid phase representable within the symmetry tensor-network formulation in Appendix.\ref{app:mag_transl_proj_rep}, which turns out to be $\alpha$, consistent with the previous discussion\cite{zaletel2014detecting}.

To further show that the condition is also sufficient, we will show that for any 3-cocycle $\omega_0$ satisfying $(\delta_g^{\omega_0})^{-1}\simeq \alpha$, generic symmetric tensor network wavefunctions representing a SRE liquid phase characterized by $\omega_0$ can be constructed in Sec.\ref{sec:SPT_constructions}.

\subsection{Entanglement Pumping argument}\label{sec:entanglement_pumping}
Here let us assume a SRE liquid phase exist. It is straightforward to show that after a local unitary transformation (a site dependent $U_g$ action) without changing the physical action of $G$, one can always choose a gauge in which $T_y=T_y^{orig.}$ and $T_x=(\prod_{y-odd}U_g)\cdot T_x^{orig.}$. Then we consider putting this SRE liquid on a infinite cylinder $C$ along the $x$-direction, with $L_y$ number of unit cells across the $y$-direction loop. We choose $L_y=l_y\cdot N+d_y$ where $l_y, d_y$ are integers, $g^N=I$ and $0 \leq d_y<N$. Note that if $d_y\neq 0$ this choice of $L_y$ will explicitly break the $T_x$ symmetry. 

This infinite long cylinder can be viewed as a one-dimensional system, and for a large enough $L_y$ the onsite symmetry $G$ will be respected. We will study the entanglement spectrum of this one dimensional system at a particular cut $x_0+1/2$. A SRE liquid respecting $G$ symmetry dictates that entanglement states at this cut carry a particular projective representation $\xi$ of $G$\cite{Pollmann:2010p64439,Chen:2011p35107}. 

\begin{figure}
\includegraphics[width=0.48\textwidth]{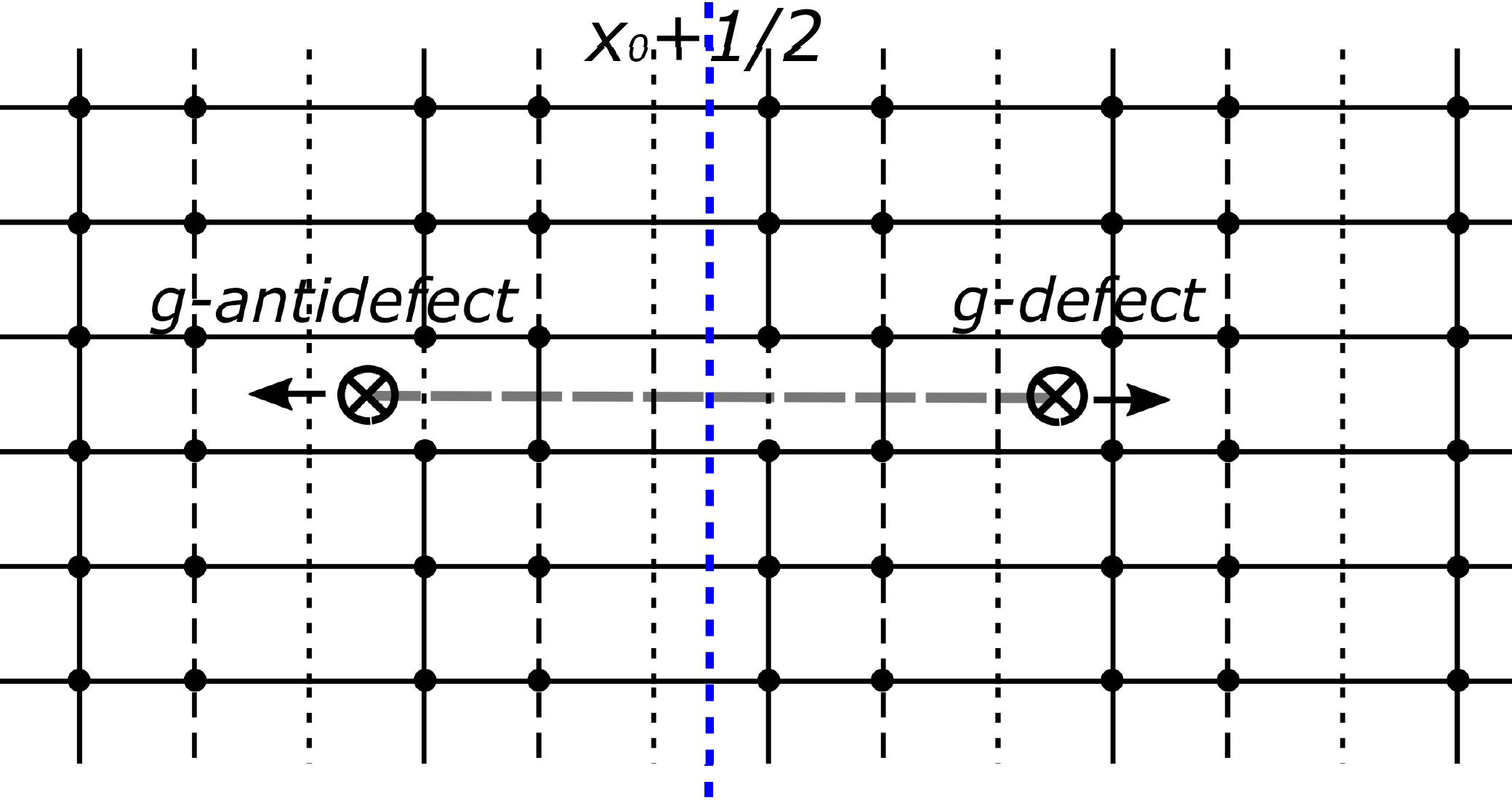}
\caption{Illustration of adiabatically separating a pair of g-defect/antidefect along the $x$-direction with $g^3=I$. For simplicity, one may imagine Hamiltonian to host nearest neighbor (NN) terms. Along the $x$-direction, due to the magnetic translation symmetry Eq.(\ref{eq:mag_transl}), the NN interactions on the vertical bonds have a three-unit-cell periodicity (solid,dashed and dotted bonds). While the $g$-defect crosses the entanglement cut at $x_0+1/2$, the Hamiltonian along the branch cut (dashed gray line) is effectively translated along $x$-direction by one unit cell. After separating such pairs of defects for every row, the final Hamiltonian is related to the original Hamiltonian by $T_x^{orig.}$.}
\label{fig:entanglement_pumping}
\end{figure}

Next, we adiabatically create a $g$-symmetry defect/anti-defect pair at a given $y$-coordinate and separate them to infinity along the $x$-direction. After repeating this procedure for every $y$-coordinate, we totally move $L_y$ number of $g$-symmetry defects acrosses the entanglement cut $x_0+1/2$. As shown in Fig.\ref{fig:entanglement_pumping}, the final Hamiltonian is related with the original Hamiltonian by the \emph{original} translation operation $T^{orig.}_x$. Therefore, the final entanglement states at the cut $x_0+1/2$ is equivalent to the initial entanglement states at a different cut: $x_0-1/2$.

Because the initial entanglement states at $x_0-1/2$ differs from the initial entanglement states at $x_0+1/2$ by a column of unit cells along $y$-direction, we conclude that the final entanglement eigenstates at $x_0+1/2$ must carry the $\alpha^{d_y}\cdot\xi$ projective representation. This pumping of the entanglement projective representation can only be explained by the projective representation $\tilde \alpha$ carried by a $g$-symmetry defect, and $(\tilde \alpha)^{L_y}\simeq \alpha^{d_y}$. But because $N$ $g$-defect fuse into a trivial object, we must have $(\tilde \alpha)^N\simeq 1$. Consequently $\tilde\alpha=\alpha$.

\subsection{Symmetry-enforced constraints on SPT cocycles}\label{sec:SPT_constraints}
Here we consider an SPT wavefunction represented using the symmetric tensor-network formulation\cite{jiang2017anyon}. The advantage of this formulation is that it allows us to introduce symmetry defects conveniently. 

The local symmetry transformation of an onsite symmetry $a$ on a $g$-defect is given by the application of of $U_a$ inside a disk $D$ covering the $g$-defect, together with an application of unitary operations on the virtual degrees of freedom on the boundary of $D$. This boundary operation should be defined in such a way that after these two operations, no excitation is created near the boundary of $D$. In Appendix \ref{app:defect_proj_rep} we explictly constructed such boundary operations. With these boundary operations, we explicitly show that the projective representaion carried by the $g$-defect is given by $(\delta^{\omega}_g)^{-1}$ in an SPT phase characterized by the 3-cocycle $\omega$, which is given without proof in Ref.~\onlinecite{zaletel2014detecting}.

\subsection{Generic constructions of Symmetry-enforced SPT wavefunctions}\label{sec:SPT_constructions}
Our strategy here is to start from an SPT state characterized by a 3-cocycle $\omega$ with a regular representation of $G$ per unit cell and respecting the usual translation symmetry $T_x^{orig.} , T_y^{orig.}$. Such a state can be generically represented using the symmetric tensor-network formulation\cite{jiang2017anyon}. In particular, the symmetric tensor-network needs to satisfy a collection of algebraic equations (constraints). Then we show that after properly modifying these algebraic constraints, the new tensor-network will respect the magnetic translation symmetry $T_x,T_y$, at the same time the physical d.o.f. must carry a projective representation $(\delta^{\omega}_g)^{-1}$ per unit cell (otherwise the wavefunction vanishes). The tensor-network states satisfying these modified constraints are generic constructions of the SPT states in Theorem-II.

The details of the construction can be found in Appendix \ref{app:generic_construction}.

\section{Discussion}
Generalized Hastings-Oshikawa-Lieb-Schultz-Mattis theorems put strong constraints on possible symmetric quantum states of matter. In particular, in the presence of translation symmetry and a projective representation of the onsite symmetry group per unit cell, it is impossible to have a gapped short-range entangled (SRE) symmetric ground state. In this paper we discuss that in the presence of magnetic translation symmetry, gapped SRE symmetric ground states could exist, which are enforced to be symmetry protected topological (SPT) phases. Focusing on bosonic systems in two spatial dimensions, we provide the generic necessary and sufficient condition for such symmetry-enforced SPT phases to occur in Theorem-I and II. When the condition is satisified, we sharply characterize the coset structure of the realizable SPT phases in the Remark. 

The condition-(2) in Theorem-I is particular non-obvious. It states that if symmetries protecting the fractional spin (projective representation) per unit cell and those generating the magnetic translations fail to commute with one another, then SRE liquid state is impossible even if the fractional spins fuse into an integer spin (regular representation) in the magnetic unit cell. 

In addition, we design a class of decorated quantum dimer models realizing some of these symmetry-enforced SPT phases, which are exactly solvable at the corresponding Rokhsar-Kivelson points. A particularly simple model realizing a symmetry-enforced SPT phase is given in Sec.\ref{sec:BFG} by coupling a Balents-Fisher-Girvin spin liquid with a layer of pure-transverse-field Ising spins via three-spin interactions. This model also demonstrates the route to obtain SPT phases via condensing anyons in SET phases\cite{jiang2017anyon,duivenvoorden2017entanglement}. 

It is interesting to consider the situation of fermions with magnetic translation symmetries, in which case (generalized) Hastings-Oshikawa-Lieb-Schultz-Mattis theorem apply for fractional filled systems with regular translation symmetries. In fact, earlier works\cite{dana1985quantised,LuRanOshikawa} establish the magnetic translation symmetry protected integer Hall conductivity, and a recent work by Wu et.al. studied the magnetic translation enforced quantum spin Hall insulators in fractionally filled fermionic systems\cite{wu2017symmetry}. Connecting with these works, the present work focuses on bosonic systems with projective representation per unit cell, but obtains systematic results.

YR would like to thank Yuan-Ming Lu and Masaki Oshikawa for inspiring related collaborations. XY, SJ and YR are supported by NSF under Grant No. DMR-1151440. AV is funded by a Simons Investigator award.

\appendix
\section{Perturbation study of the decorated Balents-Fisher-Girvin model}\label{app:BFG}
\begin{widetext}
The Hamiltonian for the decorated BFG model can be split into two parts
\begin{equation}
\begin{split}
&H^{deco. BFG}=H_0+H_1,\\
&H_0=J_z\sum\limits_{\hexagon}(S^z_{\small\hexagon})^2-\sum\limits_{\bindingterm}\lambda S_i^z(s_{IJ}\sigma^z_I\sigma^z_J),\\
&H_1=J_{\perp}\sum\limits_{(i,j)}S_i^+S_j^-+\sum\limits_{I}h\sigma_I^x,
\end{split}
\end{equation}
where $(i,j)$ runs over first, second, third neighbors within a hexagon of Kagome plaquette.

Let's take the limit $J_z,\lambda\gg J_{\perp},h$ and only focus on the low energy Hamiltonian in the ground state manifold of $H_0$. Then we can treat $H_1$ as a small perturbation and use the conventional Brillouin-Wigner perturbation to derive the effective Hamiltonian. The effective Hamiltonian is then given by (take $E_0$ as the ground state energy of $H_0$),
\begin{equation}
H_{eff}=E_0+P_g(H_1+H_1G_0'H_1+H_1G_0'H_1G_0'H_1+\cdots )P_g,
\end{equation}
where $G_0'=P_e(E_0-H_0)^{-1}P_e$ and $P_g/P_e$ are the projector onto the the ground/excited states of $H_0$. 

We will work under the limit $J_z\gg \lambda$ and calculate the effective Hamiltonian order by order to find the leading non-constant terms in $J_{\perp}$ and $h$ since we have not specified the relation between them yet. Let's denote $N$ as the number of Kagome unit-cell. The zeroth order energy is $E_0=-\frac{3}{2}N\lambda$. Higher order terms are as follows:
\begin{enumerate}
\item $H^{(1)}_{eff}=P_gH_1P_g=0.$
\item $H^{(2)}_{eff}=P_gH_1G_0'H_1P_g=-9N\frac{J_{\perp}^2}{2J_z+2\lambda}-2N\cdot \frac{h^2}{3\lambda}$. The first term comes from the process of switching a pair of spin-up and spin-down and then switching back within a hexagon. The second term comes from the process of flipping a Ising d.o.f. twice. To this order, we only have constant terms.
\item 
$
H^{(3)}_{eff}=P_gH_1G_0'H_1G_0'H_1P_g=\frac{J_{\perp}^3}{(2J_z+2\lambda)^2}\sum\limits_{(i,j,k)}(S_i^+S_k^-S_k^+S_j^-S_j^+S_i^-+S_j^+S_k^-S_i^+S_j^-S_k^+S_i^-),$ where the summation runs over all ordered triplets $(i,j,k)$ with $(i,j),(j,k),(k,i)$ appearing in $H_1$. The term $S_i^+S_k^-S_k^+S_j^-S_j^+S_i^-=(1/2+S_i^z)(1/2-S_j^z)(1/2-S_k^z)$ measures the energy of the configuration with $S_i^z=1/2,S_j^z=-1/2,S_k^z=-1/2$. And the term $S_j^+S_k^-S_i^+S_j^-S_k^+S_i^-=(1/2+S_i^z)(1/2+S_j^z)(1/2-S_k^z)$ measures the energy of the configuration with $S_i^z=1/2,S_j^z=1/2,S_k^z=-1/2$. To this order, the term does depend on the spin configuration and is the leading non-constant term in $J_{\perp}$.
\item $H^{(4)}_{eff}=P_gH_1G_0'H_1G_0'H_1G_0'H_1P_g=-\frac{10J_{\perp}^2h^2}{9J_z\lambda^2}\sum\limits_{\bowtie}(\big|\raisebox{-4mm}\bowtieB\big\rangle \big\langle\raisebox{-4mm}\bowtieA\big|+h.c.)+\mathcal{O}(\frac{h^4}{\lambda^3})+\mathcal{O}(\frac{J_{\perp}^2h^2}{J_z\lambda^2})+\mathcal{O}(\frac{J_{\perp}^4}{J_z^3})$, where we have used the limit $J_z\gg \lambda$. The first term is a kinetic term which is the leading non-constant contribution in $h$. The latter three terms are not written out explicitly due to the following reason. Terms proportional to $\frac{J_{\perp}^4}{J_z^3}$ are less significant compared to that from the 3-rd order perturbation. Terms proportional to $\frac{h^4}{\lambda^3}$ (process of flipping two different Ising d.o.f. twice) is a constant. And the potential term proportional to $\frac{J_{\perp}^2h^2}{J_z\lambda^2}$ (process of separately flipping Ising d.o.f twice and exchanging spin-up and down twice) is also a constant in the limit $J_z\gg \lambda$.
\end{enumerate}
\end{widetext}

The leading non-constant terms are terms of order $\frac{J_{\perp}^3}{J_z^2}$ and terms of order $\frac{J_{\perp}^2h^2}{J_z\lambda^2}$, where the latter is what we want. So we further require $\frac{h^2}{\lambda^2}\gg \frac{J_{\perp}}{J_z}$ such that the term obtained from the 3rd-order perturbation can be neglected.

Then we achieve the decorated BFG model
\begin{equation}
H_{eff}=-\frac{10J_{\perp}^2 h^2}{9J_z\lambda^2}\sum\limits_{\bowtie}(\big|\raisebox{-4mm}\bowtieB\big\rangle \big\langle\raisebox{-4mm}\bowtieA\big|+h.c.)
\end{equation}
in the parameter regime where $J_z\gg \lambda \gg J_{\perp},h$ and $\frac{h^2}{\lambda^2}\gg\frac{J_{\perp}}{J_z}$.

\section{Theorem-I as a special case of Theorem-II}\label{app:theorem-i}
\subsection{Necessary condition for the existence of SRE state: constraints on the on-site projective representation}
First we prove that only when the on-site projective representation $\alpha$ satisfies the following 2 conditions is a SRE ground state possible.
\begin{enumerate}
\item $\alpha^N\simeq \textbf{1}\in H^2(G,U(1)). $
\item $\gamma^{\alpha}_g(a)\simeq \textbf{1}\in H^1(G,U(1)).$
\end{enumerate}

Suppose the unit-cell is enlarged along $x$-direction to include $N$ original unit-cell, then we have $T_x^NT_yT_x^{-N}T_y^{-1}=g^N=\textbf{1}$,\textit{i.e.}, we have usual translation $T_x^N,T_y$ in the enlarged unit-cell. From Hastings' theorem we know that for a SRE ground state to exist, the enlarged unit-cell must carry usual representation. Hence we know $\alpha^N$ is a trivial 2-cocycle.

Next, we know from Theorem-II that for such a SRE state to exist, there must exist a 3-cocycle $\omega\in H^3(G,U(1))$, such that $\delta^{\omega}_g(a,b)=\alpha(a,b)^{-1}$ up to a 2-coboundary. By tuning the 2-coboundary of $\alpha(a,b)$, we are tuning the 1-coboundary of $\gamma^{\alpha}_g(a)$. Therefore we have $\gamma^{\alpha}_g(a)\simeq \gamma^{\omega}_g(a)$, where $\gamma^{\omega}_g(a)\equiv \frac{\delta^{\omega}_g(a,g)}{\delta^{\omega}_g(g,a)}$.

\begin{equation}
\begin{split}
&\delta^{\omega}_g(a,g)=\frac{\omega(a,g,g)\omega(g,a,g)}{\omega(a,g,g)}=\omega(g,a,g),\\
&\delta^{\omega}_g(g,a)=\frac{\omega(g,a,g)\omega(g,g,a)}{\omega(g,g,a)}=\omega(g,a,g).
\end{split}
\end{equation}
Therefore we always have $\gamma^{\omega}_g(a)=1$, which means $\gamma^{\alpha}_g(a)\simeq \textbf{1}\in H^1(G,U(1))$.

\subsection{Sufficient condition for the existence of SRE state: explicit construction of the 3-cocycle}
We shall show that the necessary condition given in the last section is also sufficient. To be more specific, we will construct a 3-cocycle $\omega\in H^3(G,U(1))$ out of $\alpha$ given $\alpha^N\simeq \textbf{1}\in H^2(G,U(1))$ and $\gamma^{\alpha}_g(a)\simeq \textbf{1}\in H^1(G,U(1))$, such that $\delta^{\omega}_g(a,b)=\alpha(a,b)^{-1}$. From Theorem-II, we know that such an SRE state described by 3-cocycle $\omega\in H^3(G,U(1))$ always exist, which completes our proof of Theorem-I.

\subsubsection{Canonical gauge choice for $\alpha(a,b)$}
Let's first fix a canonical gauge of $\alpha(a,b)$. Due to the direct product structure $G=G_1\times Z_N$, we denote a general group element $a\in G$ as 
\begin{equation}
a=g^{n_a}h_a, n_a=0,1\cdots N-1, h_a\in G_1.
\end{equation}

We are given the condition that $\alpha^N$ is a trivial 2-cocycle in $H^2(G,U(1))$. Let's first tune the 2-coboundary of $\alpha(a,b)$ such that $\alpha^N=1$. Then $\alpha(a,b)\in Z_N$.

We also know that 
\begin{equation}
\gamma^{\alpha}_g(a)\in B^1(G,U(1)).
\end{equation}

If $G_1$ is a unitary group, then $\gamma^{\alpha}_g(a)\equiv 1$. If $G_1$ has anti-unitary operations, we should generally represent $\gamma^{\alpha}_g(a)$ as the 1-coboundary $\frac{\gamma}{^a\gamma}$.

Therefore we know that 
\begin{equation}
\frac{\alpha(g,a)}{\alpha(a,g)}=\frac{\gamma}{^a\gamma}\in Z_N,
\rightarrow \gamma\in Z_{2N}.
\end{equation}
We choose the 2-coboundary $\delta(a)=\delta(g)^{n_a}$ where $\delta(g)=\gamma^{N-1}$.

Then under the 2-coboundary $\delta(a)$ we have
\begin{equation}
\alpha(a,b)\rightarrow\frac{\delta(g)^{n_a}\cdot ^a(\delta(g)^{n_b})}{\delta(g)^{\langle n_a+n_b\rangle_N}}\alpha(a,b),
\end{equation}
where $\langle n\rangle_N=n$ for $n<N$ and $\langle n\rangle_N=n-N$ for $n\geq N$.

Here the change of $\alpha(a,b)$ is always a $Z_N$ element since
\begin{equation}\label{eq:2coboundary}
\begin{split}
&\frac{\delta(g)^{n_a}\cdot ^a(\delta(g)^{n_b})}{\delta(g)^{\langle n_a+n_b\rangle_N}}\\
&=\begin{cases}
(\frac{^a\gamma}{\gamma})^{(N-1)n_b}\in Z_N, \text{if } n_a+n_b<N.\\
(\frac{^a\gamma}{\gamma})^{(N-1)n_b}\cdot \gamma^{N(N-1)} \in Z_N,\text{if } n_a+n_b\geq N.
\end{cases}
\end{split}
\end{equation}
Then after the change of 2-coboundary, we still have $\alpha^N=1$.

But $\gamma^{\alpha}_g(a)$ is changed as follows
\begin{equation}
\gamma^{\alpha}_g(a)=\frac{\alpha(g,a)}{\alpha(a,g)}\rightarrow \frac{\delta(g)}{^a\delta(g)}\cdot \frac{\gamma}{^a\gamma}=\frac{\gamma^N}{^a\gamma^N}=1,
\end{equation}
where we have used Eq.~\eqref{eq:2coboundary} and the fact that $\gamma^{2N}=1$. Then after the change of 2-coboundary we always have $\gamma^{\alpha}_g(a)=1$.

In summary, we have fixed $\alpha^N=1$ and $\alpha(g,a)=\alpha(a,g)$ as the canonical gauge choice.
\subsubsection{Explicit construction of 3-cocycle}
With the condition $\alpha^N=1$ and $\alpha(g,a)=\alpha(a,g)$, we can explicitly construct the 3-cocycle as follows:
\begin{equation}
\omega(a,b,c)=[\alpha(b,c)^{-1}]^{n_as_a}, s_a=1/-1\text{ for $a$ unitary/anti-unitary}.
\end{equation}

First let's prove $\omega(a,b,c)$ is indeed a 3-cocycle. We have
\begin{equation}\label{eq:3cocycle}
\begin{split}
&\omega(a,b,c)\omega(a,bc,d)\omega(b,c,d)^{s_a}\\
&=[\alpha(b,c)^{-1}]^{n_as_a}[\alpha(bc,d)^{-1}]^{n_as_a}[\alpha(c,d)^{-1}]^{n_bs_as_b}\\
&=[\alpha(c,d)^{-1}]^{n_as_as_b}[\alpha(b,cd)^{-1}]^{n_as_a}[\alpha(c,d)^{-1}]^{n_bs_as_b},
\end{split}
\end{equation}
where in the last equality we have used the 2-cocycle condition of $\alpha$, \textit{i.e.}, 
\begin{equation}
\alpha(b,c)\alpha(bc,d)=\alpha(c,d)^{s_b}\alpha(b,cd).
\end{equation}
And we also have
\begin{equation}
\begin{split}
&\omega(ab,c,d)\omega(a,b,cd)\\
&=[\alpha(c,d)^{-1}]^{(\langle n_a+n_b\rangle_N)s_as_b}\cdot[\alpha(b,cd)^{-1}]^{n_as_a},
\end{split}
\end{equation}
which equals Eq.~\eqref{eq:3cocycle} since $\alpha^N=1$. Therefore $\omega$ satisfies the 3-cocycle condition 
\begin{equation}
\omega(a,b,c)\omega(a,bc,d)\omega(b,c,d)^{s_a}=\omega(ab,c,d)\omega(a,b,cd).
\end{equation}

Next we show that the slant product of $\omega$ with respect to $g$ gives us $\alpha^{-1}$,
\begin{equation}
\begin{split}
&\delta^{\omega}_g(a,b)=\frac{\omega(a,b,g)\omega(g,a,b)}{\omega(a,g,b)}=\frac{[\alpha(b,g)^{-1}]^{n_as_a}[\alpha(a,b)^{-1}]}{[\alpha(g,b)^{-1}]^{n_as_a}}\\
&=\alpha(a,b)^{-1},\\
\end{split}
\end{equation}
where we have used $\alpha(b,g)=\alpha(g,b)$.

\section{A brief introduction to symmetric tensor network representation of SPT phases}
In this appendix we want to briefly summarize the symmetric tensor network representation of SPT phases and fix the notations for future convenience. More details of the general formalism can be found in Ref.~\onlinecite{jiang2015symmetric,jiang2017anyon}.
\subsection{Basic set-up}
Let's consider a PEPS state formed by infinite numbers of site tensors and discuss the symmetry implementation on such state\cite{perez2010characterizing,Zhao:2010p174411,Singh:2010p50301,Singh:2011p115125, Singh:2012p195114,Bauer:2011p125106,Weichselbaum:2012p2972,jiang2015symmetric}. We assume that for a symmetric PEPS the symmetry transformed tensors and the original tensors are related by a gauge transformation:
\begin{equation}
W_gg\circ \mathbb{T}=\mathbb{T},
\end{equation}
where $\mathbb{T}$ is the tensor states with all internal legs uncontracted and $W_g$ is the product of the gauge transformation acting on all internal legs of the tensor network.

The invariant gauge group (IGG) is the group of all the gauge transformations leaving the uncontracted tensor $\mathbb{T}$ invariant. These are denoted as global IGG in contrast to the plaquette IGG introduced later. The global IGG naturally arises from the following tensor equations:
\begin{equation}
\mathbb{T}=W_{a}aW_bb\circ \mathbb{T}=W_{ab}ab\circ \mathbb{T},
\end{equation}
from which we know that 
\begin{equation}
W_a\cdot ^aW_b=\eta(a,b)W_{ab},
\end{equation}
where $\eta(a,b)$ should leave the tensor invariant and hence is an IGG element.

And we have the associativity condition for $\eta(a,b)$:
\begin{equation}\label{associativity}
\eta(a,b)\eta(ab,c)=^{W_aa}\eta(b,c)\eta(a,bc).
\end{equation}

\begin{figure}[h]
\includegraphics[scale=0.4]{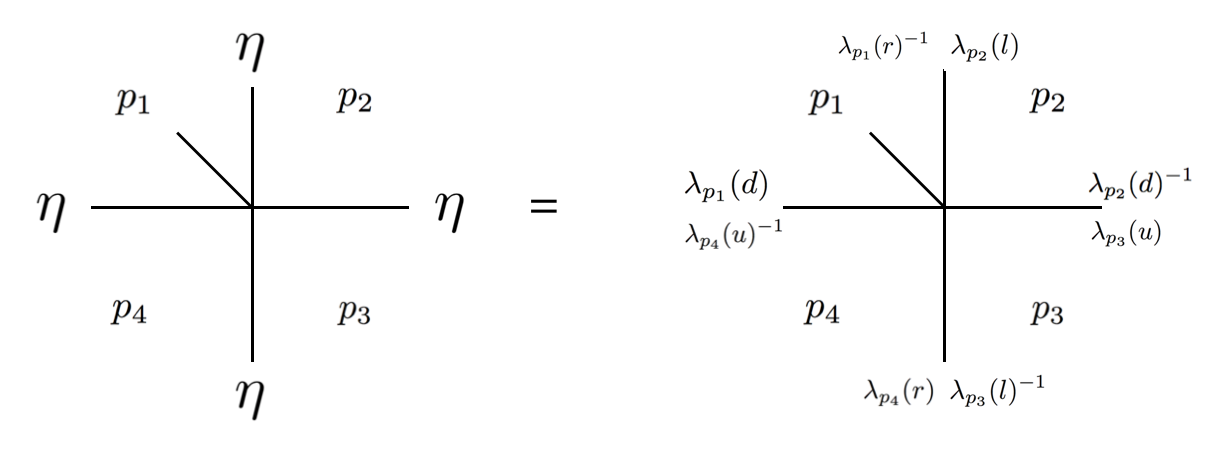}\\
\caption{The decomposition of global IGG into plaquette IGG. $\lambda$'s from different plaquettes commutes with each other, and the action of any two $\lambda$'s in the same plaquette leave the tensor invariant.}
\label{original_IGG}
\end{figure}

The global IGG elements are a characteristic of symmetry breaking or topological order. In order to obtain an SPT state, we require all the global IGG elements can be decomposed into the product of plaquette IGG elements as shown in Fig~\ref{original_IGG}, \textit{i.e.}, $\eta(a,b)=\prod_{p}\lambda_p(a,b)$. There is a global phase ambiguity in such decomposition, namely we have $\prod_p\lambda_p=\prod_p\chi_p\lambda_p$ with $\chi_p$ a global phase since $\prod_{p}\chi_p=I$. The decomposible global IGG tells us that topological order is killed and the resulting phases should be an SPT phase described by the 3-cocycle $\omega$, which arises as the phase ambiguity when lift Eq.~\eqref{associativity} to plaquette IGG,
\begin{equation}\label{threecocycle}
\lambda_p(a,b)\lambda_p(ab,c)=\omega_p(a,b,c) ^{W_aa}\lambda_p(b,c)\lambda_p(a,bc).
\end{equation}

The $\omega$ shown above is a well-defined 3-cocycle since the phase ambiguities in $\lambda$ will only modify it by a 3-coboundary.

\subsection{Representation of $\delta_g^{\omega}(a,b)$ using plaquette IGG}
In this subsection we want to give a representation of the slant product $\delta^{\omega}_g(a,b)$ in terms of plaquette IGG for an SPT state characterized by 3-cocycle $\omega$, where $g$ lies in the center of the whole symmetry group $G$. First, from definition we have
\begin{equation}
    \delta^{\omega}_g(a,b)=\frac{\omega(a,b,g)\omega(g,a,b)}{\omega(a,g,b)}
\end{equation}

The 3-cocycle arises from the decomposition of global IGG into the plaquette IGG, see Eq.~\eqref{threecocycle}. Therefore in order to compute $\delta^{\omega}_g(a,b)$, we need the following equations:
\begin{equation}\label{eq3}
\begin{split}
    & \lambda_p(a,b)\cdot\lambda_p(ab,g)=\omega(a,b,g)\cdot^a\lambda_p(b,g)\cdot\lambda_p(a,bg),\\
    & \lambda_p(g,a)\cdot\lambda_p(ga,b)=\omega(g,a,b)\cdot^g\lambda_p(a,b)\cdot\lambda_p(g,ab),\\
    & \lambda_p(a,g)\cdot\lambda_p(ag,b)=\omega(a,g,b)\cdot^a\lambda_p(g,b)\cdot\lambda_p(a,gb).
\end{split}
\end{equation}

Writing Eq.~\eqref{eq3} in a more convenient way (we ignore the subscript $p$ henceforth):
\begin{equation}
\begin{split}
    &\omega(a,b,g)=\lambda^{-1}(a,bg)\cdot ^{W_aa}\lambda^{-1}(b,g)\cdot\lambda (a,b)\cdot\lambda(ab,g),\\
    &\omega(g,a,b)=\lambda^{-1}(g,ab)\cdot ^{W_gg}\lambda^{-1}(a,b)\cdot\lambda(g,a)\cdot\lambda(ga,b),\\
    &\omega^{-1}(a,g,b)=\lambda^{-1}(ag,b)\lambda^{-1}(a,g)\cdot ^{W_aa}\lambda(g,b)\cdot\lambda (a,gb).
\end{split}
\end{equation}

We have 
\begin{equation}\label{slantprod}
\begin{split}
&\delta_g^{\omega}(a,b)=[\omega(g,a,b)]\cdot[\omega^{-1}(a,g,b)]\cdot[\omega(a,b,g)]\\
&=\lambda^{-1}(g,ab)\cdot ^{W_gg}\lambda^{-1}(a,b)\cdot [\lambda(g,a)\cdot\lambda^{-1}(a,g)]\\
&\cdot ^{W_aa}[\lambda(g,b)\cdot\lambda^{-1}(b,g)]\cdot \lambda(a,b)\cdot\lambda(ab,g)
\end{split}
\end{equation}

We can simplify Eq.~\eqref{slantprod} by defining $^{W_gg}W_aa=\xi_a(g)W_aa, a\in G$, where $\xi_a(g)=\prod \lambda_a(g)$. Another way of computing $\xi_a(g)$ is 
\begin{equation}
\begin{split}
&\xi_a(g)=W_ggW_aa(W_gg)^{-1}(W_aa)^{-1}=\eta(g,a)\eta^{-1}(a,g)\\
&\rightarrow \lambda_a(g)=\lambda(g,a)\lambda^{-1}(a,g).
\end{split}
\end{equation}

Then Eq.~\eqref{slantprod} becomes
\begin{equation}\label{eq:slantprod_plaq_IGG}
\begin{split} 
\lambda_a(g)\cdot ^{W_aa}\lambda_b(g)=\delta_g^{\omega}(a,b)\cdot ^{W_gg}\lambda(a,b)\cdot\lambda_{ab}(g)\cdot\lambda^{-1}(a,b).
\end{split}
\end{equation}

\section{The projective representation carried by a $g$-symmetry-defect}\label{app:defect_proj_rep}
In this section we want to give a tensor proof of the following fact\cite{zaletel2014detecting}: for an SPT state characterized by the 3-cocycle $\omega(a,b,c)\in H^3(G,U(1))$, the projective representation carried by the symmetry $g$-defect is represented by the inverse of the slant product $[\delta_g^{\omega}(a,b)]^{-1}$. 

\begin{figure}[h]
\includegraphics[scale=0.4]{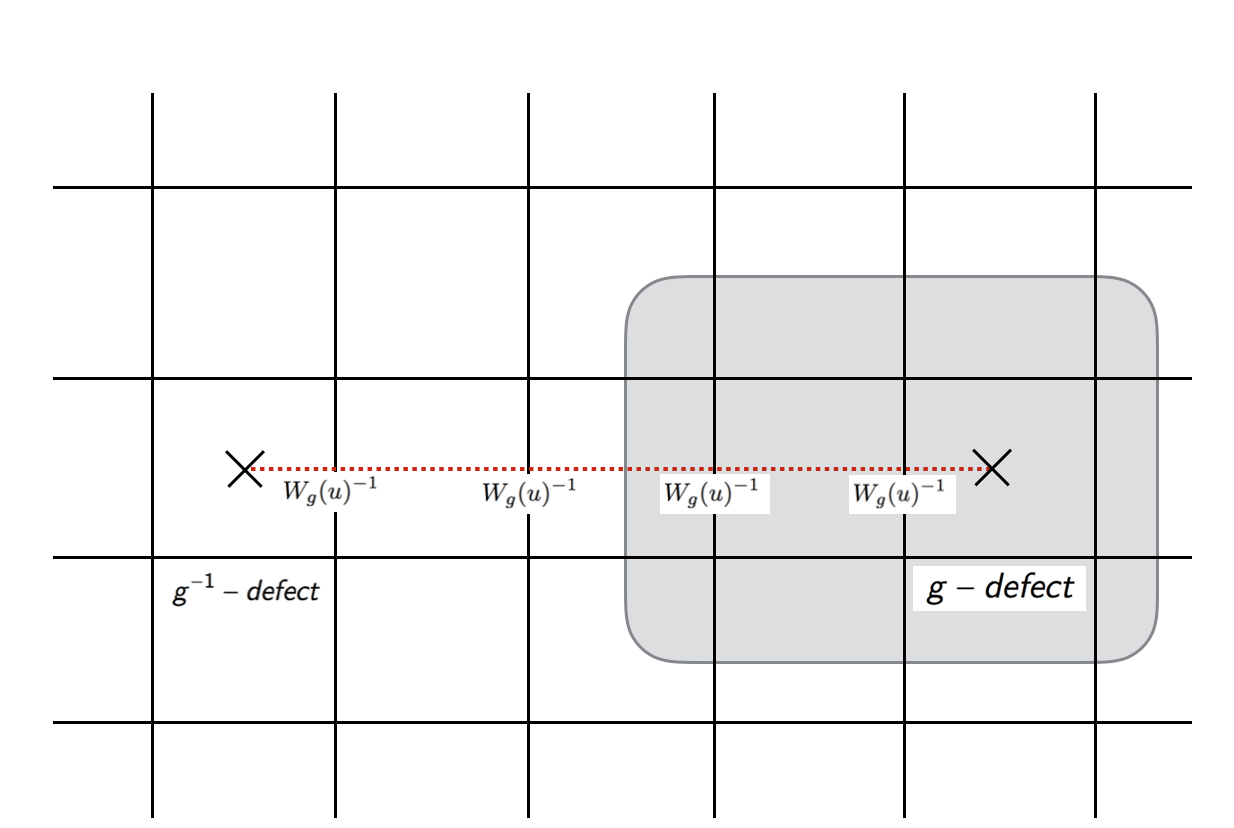}\\
\caption{An example of $g$-defect line. The $g$-defect line is obtained by inserting $W_g$ on only one side of the virtual legs crossed by the red dashed line. The tensors close to the defect core should be revised in order to make the tensor wave-function symmetric and non-vanishing. Following the usual convention, we say that the defect line always points from $g^{-1}$-defect to $g$-defect, and we always insert $W_g$ to the left when one goes forward along the line. Therefore in the figure we can identify the right end as the $g$-defect (remember $W_g(d)=W_g(u)^{-1}$). The grey area encloses a $g$-defect and we can to measure its projective representation through the action of $\eta'(a,b)$ on the boundary virtual legs, see the discussion in the main text.
}
\label{g_defect}
\end{figure}

To this end, we first create an open $g$-defect string with a pair of $g$-defects on the two ends in the given ground-state SPT wave-function $\ket{\Psi}$. The wave-function is denoted as $\ket{\Psi_{defect}}$. This is done in the tensor language by inserting $W_g$ strings and modifying the tensors close to the defect core as shown in Fig.~\ref{g_defect}. 

Let's take a patch enclosing one of the two $g$-defects and measure the projective representation carried by the $g$-defect. Before the insertion of the $g$-defect, the local symmetry action $U(a)$ on the patch is defined as acting $W_a$ on the virtual legs on the edge and $D(a)$ on the physical legs inside the patch, \textit{i.e.},
\begin{equation}
U(a)=\prod_{\text{boundary}}W_a\prod_{\text{bulk}}D(a).
\end{equation}

The symmetry operation should leave $\ket{\Psi}$ invariant up to a phase. Then the projective representation inside the patch is measured by acting 
\begin{equation}\label{eq:projective_rep}
D(a)\cdot D(b)\cdot [D(a\circ b)]^{-1}
\end{equation}
on the physical legs inside the patch. Alternatively, we can do this by monitoring the inverse of the phase generated by acting $\eta(a,b)\equiv W_a\cdot ^aW_b\cdot(W_{ab})^{-1}$ on the boundary virtual legs since, by our assumption, the action of $U(a)U(b)U(ab)^{-1}$ leaves the patch fully invariant.

In general , acting $\eta(a,b)$ on the virtual legs of a tensor leaves the tensor invariant only up to a phase. Therefore $\eta$ itself is not a global IGG. Instead, we have
\begin{equation}
\eta(a,b)=W_x(a,b)\eta'(a,b),
\end{equation}
where $W_x(a,b)$ is a pure-phase gauge transformation which yields the extra phase for each site and $\eta'(a,b)$ leaves every tensor invariant. Now $\eta'(a,b)$ is decomposable and we denote it as $\eta'(a,b)=\prod \lambda_p(a,b)$.

As for our present case, suppose we have the action of $\prod\limits_{\text{boundary}}\eta(a,b)$ on the ground state wave-function
\begin{equation}\label{eq:projective_rep_measure}
\prod_{\text{boundary}}\eta(a,b)\ket{\Psi}=e^{i\phi}\ket{\Psi},
\end{equation}
from which we know that the projective representation inside the patch is just $\prod_{\text{bulk}}(D(a)\cdot D(b)\cdot [D(a\circ b)]^{-1})=e^{-i\phi}$.

From the previous discussion we have 
\begin{equation}\label{eq:phi}
e^{-i\phi}\prod_{\text{boundary}}\eta(a,b)=\prod_{\text{boundary}}\eta'(a,b).
\end{equation}

\begin{widetext}

\begin{figure}[h]
\includegraphics[scale=0.4]{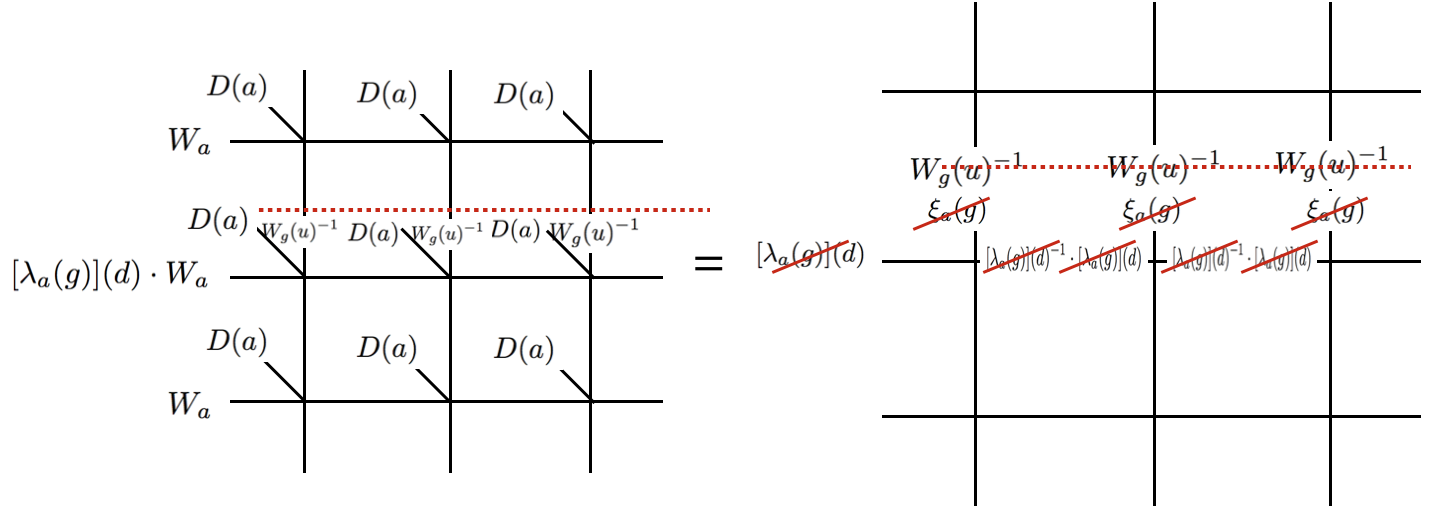}\\
\caption{Invariance of the wave-function under $U^g(a)$. In the figure we can see that $\tilde{W}_a=[\lambda_a(g)](d)\cdot W_a$ where the defect line crosses the boundary and $\tilde{W}_a=W_a$ elsewhere. Such a definition ensures that no boundary excitations are created by acting $U^g(a)$ (for the moment we do not care about what happens at the defect core). In deriving the second figure, we have used the invariance of the tensor under $W_aa$, the identity $W_aW_g^{-1}=W_g^{-1}\xi_a(g)W_a$ and invariance of the tensor under plaquette IGG $\lambda_a(g)$.}
\label{Wa_bdr}
\end{figure}

\begin{figure}[h]
\includegraphics[scale=0.6]{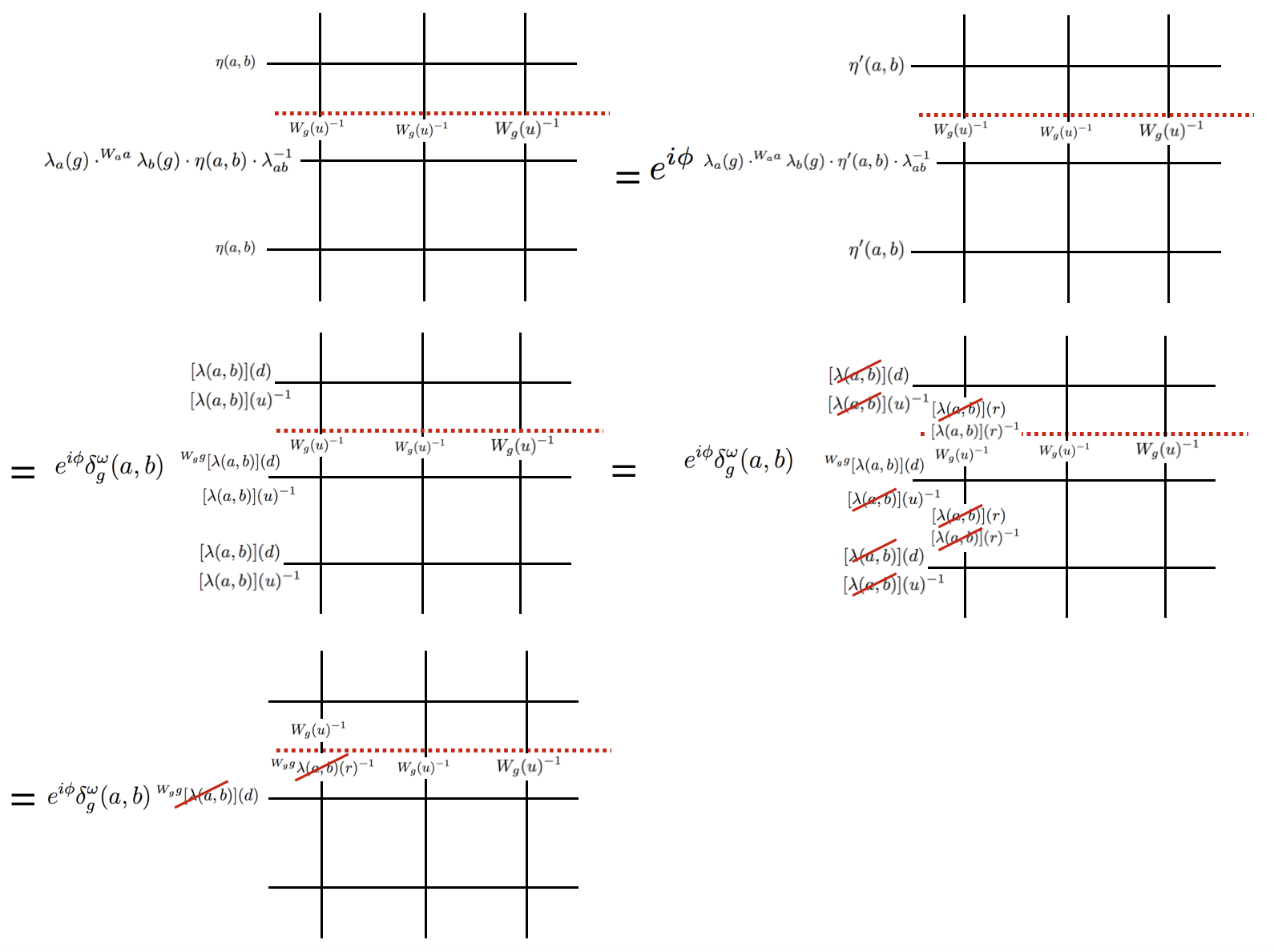}\\
\caption{Measurement of projective representation carried by $g$-defect. From Eq.~\eqref{eq:tildeWa}, we know that $\tilde{\eta}(a,b)=\lambda_a(g)\cdot ^{W_aa}\lambda_b(g)\cdot \eta(a,b)\cdot \lambda_{ab}^{-1}$ where the boundary is crossed by the defect line and $\tilde{\eta}(a,b)=\eta(a,b)$ elsewhere. In the first equality we have used Eq.~\eqref{eq:phi}. In the second equality we have used the decomposition of $\eta'(a,b)$ and Eq.~\eqref{eq:slantprod_plaq_IGG}. In the third equality we have used the tensor invariance under plaquette IGG. In the last equality we have used the identity $\lambda(r)^{-1}\cdot W_g^{-1}=W_g^{-1}\cdot ^{W_g}\lambda(r)^{-1}$ and the tensor invariance under plaquette IGG.}
\label{measurement}
\end{figure}

\end{widetext}

After the insertion of $g$-defect, the new local symmetry operation $U^{g}(a)$ should be defined as acting $\tilde{W}_a$ on the virtual legs on the edge and $D(a)$ on the physical legs inside the patch. Here we have used the $\tilde{W}_a$ as shown in Fig.~\ref{Wa_bdr} where $\tilde{W}_a$ on the bond crossing the defect line is changed to be $[\lambda_a(g)](d)\cdot W_a$ and remains $W_a$ elsewhere. This newly-defined $\tilde{W}_a$ ensures that no boundary excitations are created. The physical symmetry operation in the bulk should still be the same.

We will then use $\prod_{\text{boundary}}\tilde{\eta}(a,b)$ to measure the projective representation inside the patch, where $\tilde{\eta}(a,b)\equiv \tilde{W}_a\cdot ^a\tilde{W}_b\cdot(\tilde{W}_{ab})^{-1}$. Far away from the defect core, the tensor wave-function is basically the same as before. And $\prod_{\text{boundary}}\tilde{\eta}(a,b)$ will be the same as $\prod_{\text{boundary}}\eta(a,b)$ except at the bond crossing the defect line, see Fig.~\ref{measurement}. The $\tilde{\eta}(a,b)$ at the bond crossing the defect line should be $\tilde{W}_a\cdot ^a\tilde{W}_b\cdot (W_{ab})^{-1}$:
\begin{equation}\label{eq:tildeWa}
\begin{split}
&[\lambda_a(g)](d)\cdot W_a\cdot ^{a}[\lambda_b(g)](d)\cdot^{a}W_b\cdot ([\lambda_{ab}(g)](d)\cdot W_{ab})^{-1}\\
&=[\lambda_a(g)](d)\cdot^{W_aa}[\lambda_b(g)](d)\cdot W_a\cdot ^aW_b\cdot(W_{ab})^{-1}[\lambda^{-1}_{ab}(g)](d)\\
&=[\lambda_a(g)](d)\cdot^{W_aa}[\lambda_b(g)](d)\cdot \eta(a,b)\cdot[\lambda^{-1}_{ab}(g)](d).
\end{split}
\end{equation}

We can work with the decomposable $\eta'(a,b)$ at the boundary if we keep track of the $e^{i\phi}$ phase. After going through the calculation as shown in Fig.~\ref{measurement}, we have
\begin{equation}\label{eq:defect_projective_rep}
\prod_{bdr}\tilde{\eta}(a,b)\ket{\Psi_{defect}}=e^{i\phi}\delta^{\omega}_g(a,b)\ket{\Psi_{defect}}.
\end{equation}

Comparing Eq.~\eqref{eq:projective_rep_measure} with Eq.~\eqref{eq:defect_projective_rep} and note that projective representation is defined through the inverse of Eq.~\eqref{eq:projective_rep_measure}, we know that the $g$-defect carries $[\delta^{\omega}_g(a,b)]^{-1}$ projective representation.

\section{Consequence of the magnetic translation symmetry in tensor-network formulation}\label{app:mag_transl_proj_rep}
The magnetic translation symmetry and on-site projective representation constrain the possible symmetric short-range entangled states in a system. Specifically, we have the following fact: for a system with on-site projective representation of the on-site symmetry group $G$ characterized by $\alpha(a,b)$ and magnetic translation symmetry $T_xT_yT_x^{-1}T_y^{-1}=g$, an SPT ground state described by the 3-cocycle $\omega\in H^3(G,U(1))$ can be realized as its ground state only when $\delta^{\omega}_g(a,b)=\alpha(a,b)^{-1}$.

\subsection{Basic set-up}
We have $T_xT_yT_x^{-1}T_y^{-1}=g$, which leads to 
\begin{equation}\label{eq:commutation}
W_{T_x}T_xW_{T_y}T_y (W_{T_x}T_x)^{-1}(W_{T_y}T_y)^{-1}=W_gg.
\end{equation}

\begin{figure}[h]
\includegraphics[scale=0.6]{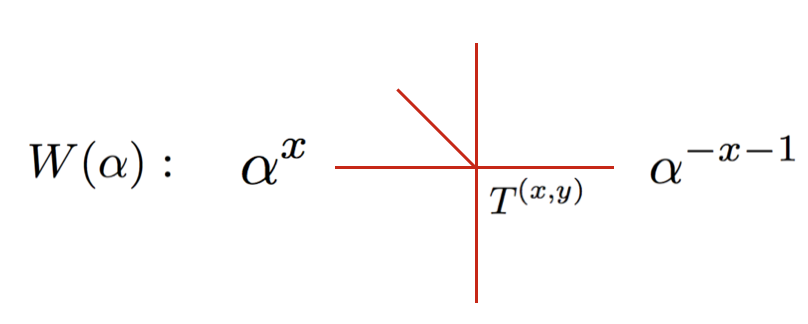}\\
\caption{The definition of phase-gauge transformation $W(\alpha(a,b))$.
}
\label{figure1}
\end{figure}

We define $\eta(a,b)$ as
\begin{equation}\label{eq:twococycle}
W_a\cdot ^aW_b=\eta(a,b)W_{ab}.
\end{equation}

By acting $\eta(a,b)$ on $\mathbb{T}$, we will get an extra phase $\alpha(a,b)^{-1}$ per unit-cell, therefore it is not a global $IGG$. However, we can define a pure-phase gauge transformation $W(\alpha(a,b))$ which yields exactly the phase $\alpha(a,b)^{-1}$ for every site tensor, see Fig.~ for an illustration.

Then we can write $\eta(a,b)$ as
\begin{equation}
\eta(a,b)=W(\alpha(a,b))\eta'(a,b),
\end{equation}
where $\eta'(a,b)=\prod \lambda(a,b)$ is an IGG and is decomposable.

Similarly since we have $D(g)\circ D(a)=\gamma_a(g)D(a)\circ D(g)$ on physical leg, where $\gamma_a(g)=\alpha(g,a)/\alpha(a,g)\in H^1(G,U(1))$. We write the action of $W_gg$ on $W_aa$ as
\begin{equation}\label{eq:g_a_commutator}
^{W_gg}W_aa=W(\gamma_a(g))\xi_a(g)\cdot W_aa,
\end{equation}
where $\xi_g(a)=\prod\lambda_g(a)$ is a decomposable $IGG$ element and the definition of $W(\gamma_a(g))$ is the same as $W(\alpha(a,b))$.

And as for $T_x, T_y$, we have
\begin{equation}\label{eq:T_a_commutator}
^{W_{T_x}T_x}W_aa=W(\gamma_a(T_x))\xi_a(T_x)W_aa,\textit{etc.}.
\end{equation}

We act Eq.~\eqref{eq:commutation} on $W_aa$ and obtain an IGG equation
\begin{equation}
^{W_gg}\xi_a^{-1}(T_y)\cdot^{W_ggW_{T_y}T_y}\xi_a^{-1}(T_x)\cdot ^{W_{T_x}T_x}\xi_a(T_y)\cdot\xi_a(T_x)=\xi_a(g),
\end{equation}
which, when lift to plaquette IGG, should give us (we have absorbed the phase ambiguity into the definition of $\lambda_a(g)$)
\begin{equation}\label{eq:lambdaidentity}
^{W_gg}\lambda^{-1}_a(T_y)\cdot^{W_ggW_{T_y}T_y}\lambda_a^{-1}(T_x)\cdot^{W_{T_x}T_x}\lambda_a(T_y)\cdot\lambda_a(T_x)=\lambda_a(g).
\end{equation}

\subsection{Acting $W_gg$}
We first act $W_gg$ on Eq.~\eqref{eq:twococycle}, then we have
\begin{equation}
\begin{split}
&^{W_gg}[W_aaW_bb]=^{W_gg}[\eta(a,b)W_{ab}ab]\\
&\Rightarrow W(\gamma_a(g))\cdot ^{a}W(\gamma_b(g))\xi_a(g)\cdot ^{W_aa}\xi_b(g)\cdot W_aaW_bb\\
&=^{W_gg}\eta(a,b)\cdot W(\gamma_{ab}(g))\xi_{ab}(g)W_{ab}ab,
\end{split}
\end{equation}
which then leads to
\begin{equation}\label{eq:globalIGGeqn}
\xi_a(g)\cdot^{W_aa}\xi_b(g)=^{W_gg}\eta'(a,b)\cdot\xi_{ab}(g)\cdot\eta'^{-1}(a,b),
\end{equation}
where extra phase factors $W(\alpha(a,b))$, $W(\gamma_a(g))$ all cancel. 

 When lifting Eq.~\eqref{eq:globalIGGeqn} to plaquette, we have (from Eq.~\eqref{eq:slantprod_plaq_IGG})
\begin{equation}\label{eq:acting_W_g}
\lambda_a(g)\cdot^{W_aa}\lambda_b(g)=\delta^{\omega}_g(a,b)\cdot^{W_gg} \lambda(a,b)\cdot \lambda_{ab}(g)\cdot\lambda(a,b)^{-1}.
\end{equation}

\subsection{Acting translation}
We have another way of deriving Eq.~\eqref{eq:acting_W_g}. We first act $W_{T_x}T_x$ on the two sides of Eq.~\eqref{eq:twococycle} and obtain an IGG equation,
\begin{equation}\label{eq:Txcommutator}
\xi_a(T_x)\cdot ^{W_aa}\xi_b(T_x)=^{W_{T_x}T_x}\eta(a,b)\cdot \xi_{ab}(T_x)\eta(a,b)^{-1},
\end{equation}
where the extra $W(\gamma_a(T_x)), W(\gamma_b(T_x)),W(\gamma_{ab}(T_x))$ cancel since we have $\gamma_a(T_x)\cdot ^a\gamma_b(T_x)=\gamma_{ab}(T_x)$.

\begin{figure}[h]
\includegraphics[scale=0.34]{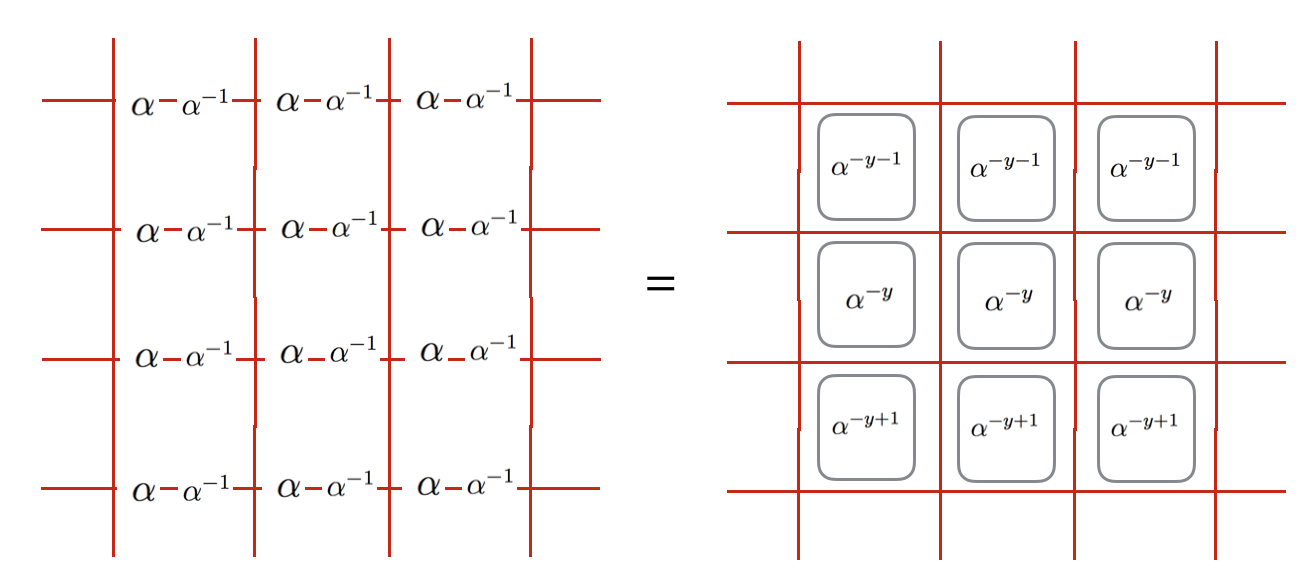}\\
\caption{The decomposition rule of $^{W_{T_x}T_x}W(\alpha(a,b))\cdot W(\alpha(a,b))^{-1}$ (LHS) as a product of plaquette IGG $\lambda(\alpha(a,b))$ (RHS).
}
\label{phase_plaq_IGG}
\end{figure}

When lift Eq.~\eqref{eq:Txcommutator} to plaquette IGG, we have
\begin{equation}\label{eq:T_x_commutator}
\lambda_a(T_x)\cdot ^{W_aa}\lambda_b(T_x)=[\alpha(a,b)]^{-y}\cdot ^{W_{T_x}T_x}\lambda(a,b)\lambda_{ab}(T_x)\lambda^{-1}(a,b),
\end{equation}
where $\prod[\alpha(a,b)]^{-y}=^{W_{T_x}T_x}W(\alpha(a,b))\cdot W(\alpha(a,b))^{-1}$ and the plaquette IGG $[\alpha(a,b)]^{-y}$ are just loops of phases as shown in Fig.~\ref{phase_plaq_IGG}.

Similarly we have
\begin{equation}\label{eq:T_y_commutator}
\lambda_a(T_y)\cdot ^{W_aa}\lambda_b(T_y)=^{W_{T_y}T_y}\lambda(a,b)\lambda_{ab}(T_y)\lambda^{-1}(a,b),
\end{equation}
where there is no extra factor coming from $^{W_{T_y}T_y}W(\alpha(a,b))\cdot W(\alpha(a,b))^{-1}$ since it is $T_y$ invariant.

\begin{widetext}
With Eq.~\eqref{eq:T_x_commutator} and Eq.~\eqref{eq:T_y_commutator} we can explicitly calculate the action of $W_{T_x}T_xW_{T_y}T_y(W_{T_x}T_x)^{-1}(W_{T_y}T_y)^{-1}$ on Eq.~\eqref{eq:twococycle} in terms of plaquette IGG. The action of $W_{T_x}T_xW_{T_y}T_y(W_{T_x}T_x)^{-1}(W_{T_y}T_y)^{-1}$ on LHS of Eq.~\eqref{eq:twococycle} is 
\begin{equation}\label{eq:W_aW_b_commutator}
\begin{split}
&^{W_gg}[\lambda_a(T_y)\cdot ^{W_aa}\lambda_b(T_y)]^{-1}\cdot ^{W_ggW_{T_y}T_y}[\lambda_a(T_x)^{W_aa}\lambda_b(T_x)]^{-1}\cdot^{W_{T_xT_x}}[\lambda_a(T_y)^{W_aa}\lambda_b(T_y)]\cdot[\lambda_a(T_x)^{W_aa}\lambda_b(T_x)]\\
&=^{W_ggW_aa}\lambda_b^{-1}(T_y)\cdot ^{W_gg}[\lambda_a^{-1}(T_y)\cdot^{W_{T_y}T_yW_aa}\lambda_b^{-1}(T_x)^{W_{T_y}T_y}\lambda^{-1}_a(T_x)]\cdot ^{W_{T_x}T_x}\lambda_a(T_y)\cdot ^{W_{T_x}T_xW_aa}\lambda_b(T_y)\cdot \lambda_a(T_x)^{W_aa}\lambda_b(T_x)\\
&=\lambda_a(g)^{W_aaW_gg}\lambda_b^{-1}(T_y)\cdot \lambda_a^{-1}(g)\cdot ^{W_gg}[^{W_aaW_{T_y}T_y}\lambda_b^{-1}(T_x)\lambda_a^{-1}(T_y)^{W_{T_y}T_y}\lambda^{-1}_a(T_x)]\cdot ^{W_{T_x}T_x}\lambda_a(T_y)\cdot \lambda_a(T_x)\\
&\cdot^{W_aaW_{T_x}T_x}\lambda_b(T_y)\cdot^{W_aa}\lambda_b(T_x)\\
&=\lambda_a(g)\cdot ^{W_aaW_gg}\lambda_b^{-1}(T_y)\cdot ^{W_aaW_ggW_{T_y}T_y}\lambda_b^{-1}(T_x)\cdot[\lambda_a^{-1}(g)^{W_gg}\lambda_a^{-1}(T_y)^{W_ggW_{T_y}T_y}\lambda_a^{-1}(T_x)\cdot ^{W_{T_x}T_x}\lambda_a(T_y)\lambda_a(T_x)]\\
&\cdot^{W_aaW_{T_x}T_x}\lambda_b(T_y)^{W_aa}\lambda_b(T_x)\\
&=\lambda_a(g)\cdot ^{W_aa}[^{W_gg}\lambda_b^{-1}(T_y)\cdot ^{W_ggW_{T_y}T_y}\lambda_b^{-1}(T_x)\cdot^{W_{T_x}T_x}\lambda_b(T_y)\lambda_b(T_x)]\\
&= \lambda_a(g)\cdot ^{W_aa}\lambda_b(g),
\end{split}
\end{equation}
where we have used Eq.~\eqref{eq:g_a_commutator} and Eq.~\eqref{eq:T_a_commutator} in the first three equalities and we have used Eq.~\eqref{eq:lambdaidentity}
in the last two equalities.

The action of $W_{T_x}T_xW_{T_y}T_y(W_{T_x}T_x)^{-1}(W_{T_y}T_y)^{-1}$ on RHS of Eq.~\eqref{eq:twococycle} is 
\begin{equation}\label{eq:W_ab_commutator}
\begin{split}
&^{W_gg}[^{W_{T_y}T_y}\lambda(a,b)\lambda_{ab}(T_y)\lambda^{-1}(a,b)]^{-1}\cdot^{W_ggW_{T_y}T_y}[\alpha(a,b)^{-y}\cdot ^{W_{T_x}T_x}\lambda(a,b)\lambda_{ab}(T_x)\lambda^{-1}(a,b)]^{-1}\\
&\cdot^{W_{T_x}T_x}[^{W_{T_y}T_y}\lambda(a,b)\lambda_{ab}(T_y)\lambda^{-1}(a,b)]\cdot \alpha(a,b)^{-y}\cdot ^{W_{T_x}T_x}\lambda(a,b)\lambda_{ab}(T_x)\lambda^{-1}(a,b)\\
&=^{W_{T_y}T_y}[\alpha(a,b)^{-y}]^{-1}\cdot\alpha(a,b)^{-y}\cdot^{W_gg}\lambda(a,b)^{W_gg}\lambda_{ab}^{-1}(T_y)\cdot ^{W_ggW_{T_y}T_y}\lambda_{ab}^{-1}(T_x)\cdot ^{W_{T_x}T_x}\lambda_{ab}(T_y)\cdot \lambda_{ab}(T_x)\cdot\lambda^{-1}(a,b)\\
&=\alpha(a,b)^{-1}\cdot^{W_gg}\lambda(a,b)\cdot\lambda_{ab}(g)\cdot\lambda^{-1}(a,b),
\end{split}
\end{equation}
where we have used Eq.~\eqref{eq:lambdaidentity} and $\alpha^{-1}(a,b)$ is just a plaquette IGG with loop of phases $\alpha^{-1}(a,b)$.

\end{widetext}

Combining Eq.~\eqref{eq:W_aW_b_commutator} with Eq.~\eqref{eq:W_ab_commutator}, we have
\begin{equation}\label{eq:acting_W_T}
\lambda_a(g)\cdot ^{W_aa}\lambda_b(g)=\alpha^{-1}(a,b)^{W_gg}\lambda(a,b)\cdot\lambda_{ab}(g)\cdot \lambda^{-1}(a,b).
\end{equation}

Comparing Eq.~\eqref{eq:acting_W_T} with Eq.~\eqref{eq:acting_W_g}, we have
\begin{equation}\label{eq:constraint_3_cocycle}
\alpha^{-1}(a,b)=\delta^{\omega}_g(a,b).
\end{equation}

It is easy to see that the global phase ambiguities in Eq.~\eqref{eq:lambdaidentity},\eqref{eq:T_x_commutator},\eqref{eq:T_y_commutator} will at most modify the LHS of Eq.~\eqref{eq:constraint_3_cocycle} up to a 2-cobounday, therefore it should be understood as the 2-cycle equivalence $\delta^{\omega}_g\simeq\alpha^{-1}\in H^2(G,U(1))$.

\section{Generic constructions of symmetry-enforced SPT tensor-network wavefunctions}\label{app:generic_construction}
\begin{figure}
\centering
   \includegraphics[scale=0.45]{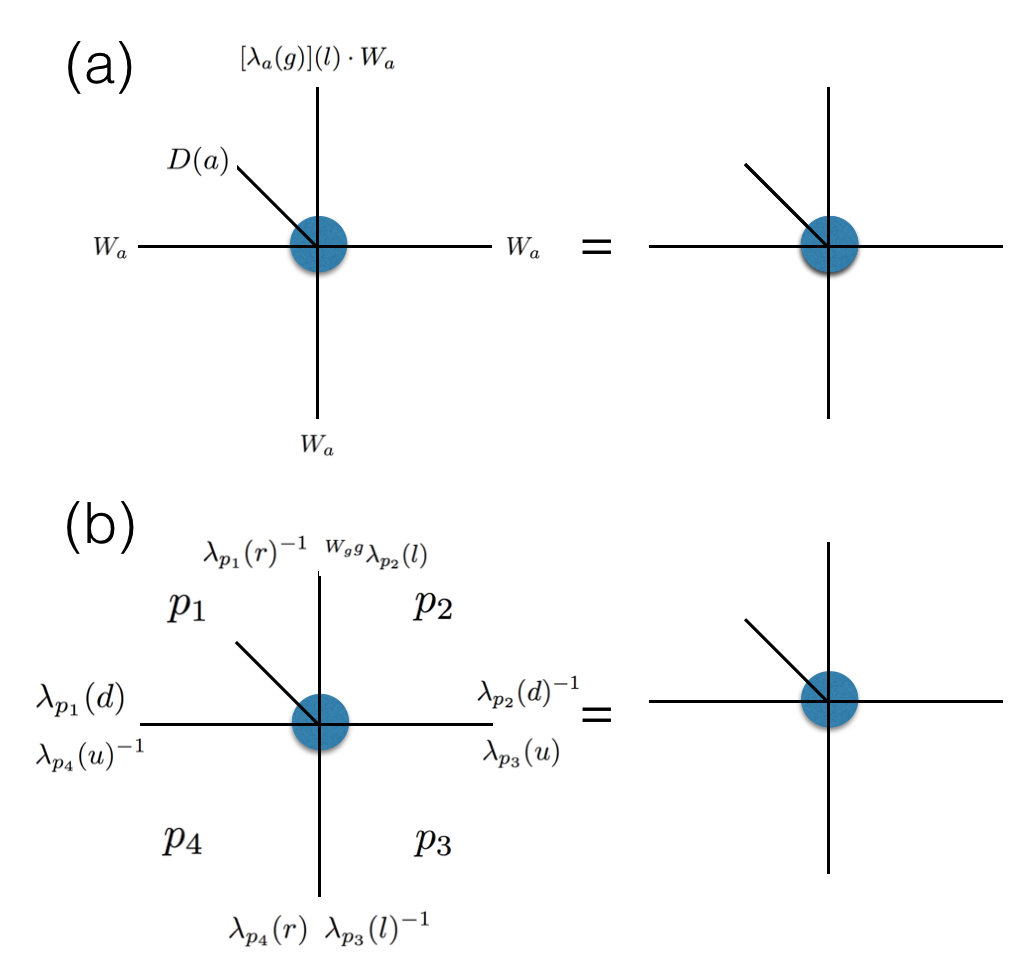} 
  \caption{The original tensor before insertion of $g$-defect is required to be invariant under the revised symmetry operation and the revised plaquette IGG.}
\label{tensor_invariance}
\end{figure}

In this section we want to construct an SPT state with on-site symmetry group $G$, magnetic translation symmetry satisfying $T_xT_yT_x^{-1}T_y^{-1}=g$ where the SPT is characterized by the 3-cocycle $\omega\in H^3(G,U(1))$ and the on-site symmetry group is represented projectively with the 2-cocycle $\alpha(a,b)$ equal to the inverse of the slant product of $\omega$ with respect to $g$,\textit{i.e.},
\begin{equation}
\alpha(a,b)\simeq[\delta_g^{\omega}(a,b)]^{-1}\in H^2(G,U(1)).
\end{equation}

To achieve this goal, we will use the tensor network formalism. Let's start from a SPT tensor wavefunction with the symmetry group $\mathbb{Z}^2\times G$ described by a 3-cocycle $\omega\in H^3(G,U(1))$, where $\mathbb{Z}^2$ represents the usual translation $T^{orig.}_x,T^{orig.}_y$. Then we know that every tensor is invariant under the action $D(a)$ on physical leg together with $\prod W_a$ on all the virtual legs, from which we have a set of tensor equations. Here we require $D(a)$ to be a direct sum of usual representation $D_1(a)$ and projective representation $D_2(a)$ with 2-cocycle $[\delta_g^{\omega}(a,b)]^{-1}$. We choose the gauge transformation $W_{T^{orig.}_x},W_{T^{orig.}_y}$ to be identity for simplicity. The global IGG $\eta(a,b)$ comes from the following tensor equation:
\begin{equation}
W_a\cdot ^aW_b=\eta(a,b)W_{ab},
\end{equation}
where $\eta(a,b)$ is decomposable, \textit{i.e.}, $\eta(a,b)=\prod \lambda_p(a,b)$. We require tensors to be fully invariant under $\eta(a,b)$ without even generating phases. This condition ensures $D(a)$ on the physical legs to be projected onto $D_1(a)$ sector.

\begin{figure}
\centering
   \includegraphics[scale=0.45]{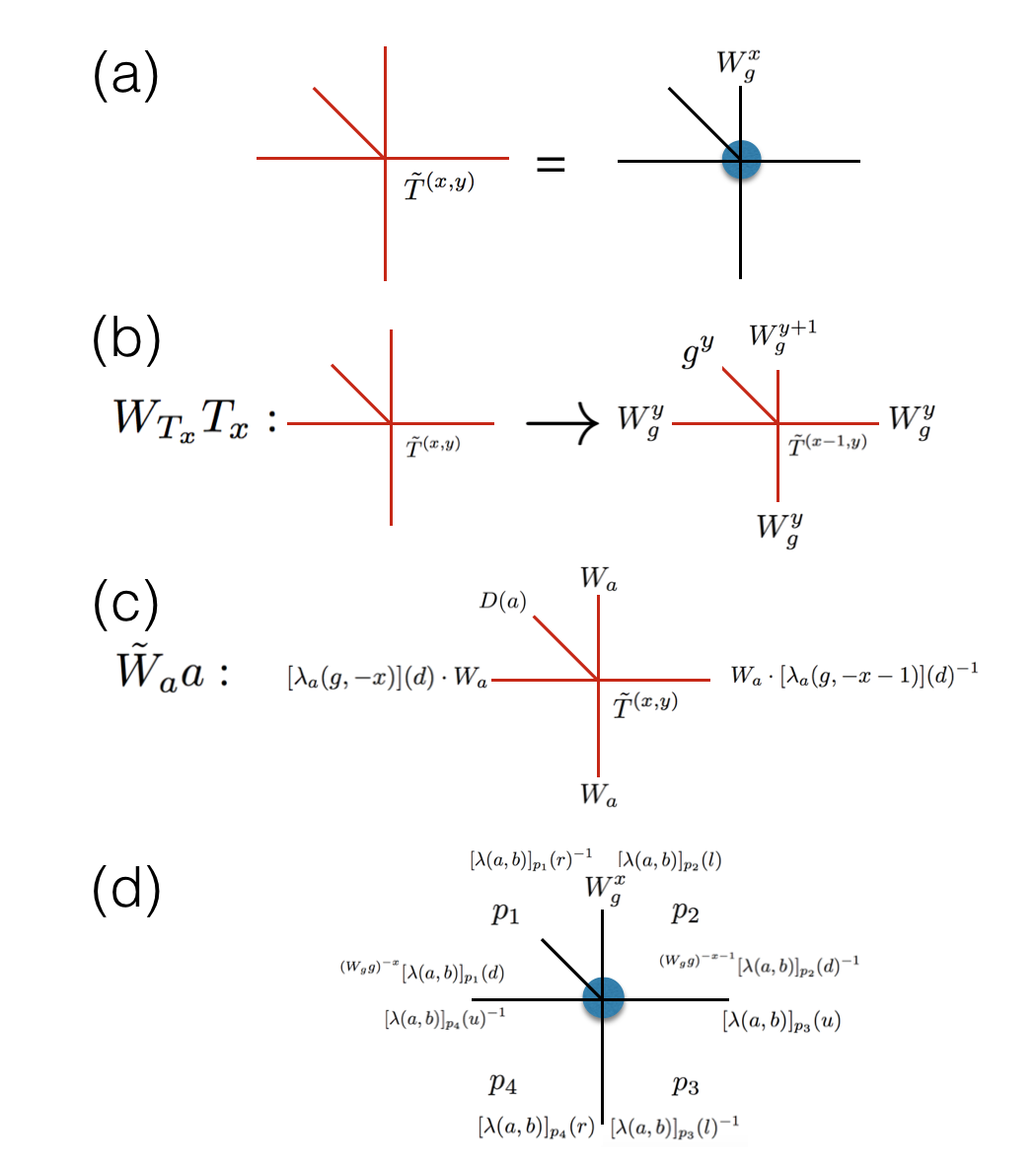} 
  \caption{(a) The definition of the new tensor $\tilde{T}^{(x,y)}$ after the insertion of $[W_g(u)]^x$ to the upper leg of every original tensor $T^{(x,y)}$.
  (b) The new translation operation $W_{T_x}T_x$. It can be readily checked that $\tilde{T}^{(x,y)}$ is invariant under such translation. Note that we have $T_x=g^yT^{orig.}_x$, $T_y=T^{orig.}_y$ and $W_{T_y}=\textbf{1}$. (c) The new on-site symmetry operation $\tilde{W}_aa$. It is shown in Fig.~\ref{symm_invariance} that $\tilde{T}^{(x,y)}$ is invariant under such symmetry operation. (d) The new plaquette IGG for the new tensor $\tilde{T}^{(x,y)}$. As before, $\lambda$'s from different plaquettes commute with each other, and the action of any two $\lambda$'s in the same plaquette leave the tensor invariant. The tensor $\tilde{T}^{(x,y)}$ invariance under plaquette IGGs follows trivially from Fig.~\ref{tensor_invariance}.}
\label{fig:new_tensor}
\end{figure}

We define $^{W_gg}W_aa=\xi_a(g)W_aa$, where $\xi_a(g)=\prod\lambda_a(g)$ is a decomposable global IGG. More generally we define $^{(W_gg)^x}W_a=\xi_a(g,x)W_a$, where $\xi_a(g,x)=\prod \lambda_a(g,x)$. Then we have the following relation:
\begin{equation}
\lambda_a(g,x+1)=^{W_gg}\lambda_a(g,x)\cdot \lambda_a(g).
\end{equation}

Then we change our tensor wave-function in the following way:
\begin{enumerate}
\item We revise the original tensor such that it is invariant under the symmetry operation and the new plaquette IGG defined in Fig.~\ref{tensor_invariance}. Note that this step is necessary for us to obtain a symmetric and non-vanishing tensor wave-function after the insertion of $W_g$.
\item We insert $[W_g(u)]^x$ on the upper leg of every tensor as shown in Fig.~\ref{fig:new_tensor}. Physically it means inserting one $g$-defect per unit-cell.
\item We define the new on-site symmetry operation $\tilde{W}_{a}$ and translation symmetry $W_{T_x}$ as shown in Fig.~\ref{fig:new_tensor}. We will show that the new tensor is invariant under such symmetry transformations.
\end{enumerate}

From the last section we have shown that every $g$-defect carries a projective representation represented by $\delta_g^{\omega}(a,b)^{-1}$, therefore one would expect after insertion of $g$-defects, we now have one $\delta_g^{\omega}(a,b)^{-1}$ projective representation per unit-cell. Let's show it more clearly through explicit calculations.

\begin{widetext}

\begin{figure}[h]
\includegraphics[scale=0.6]{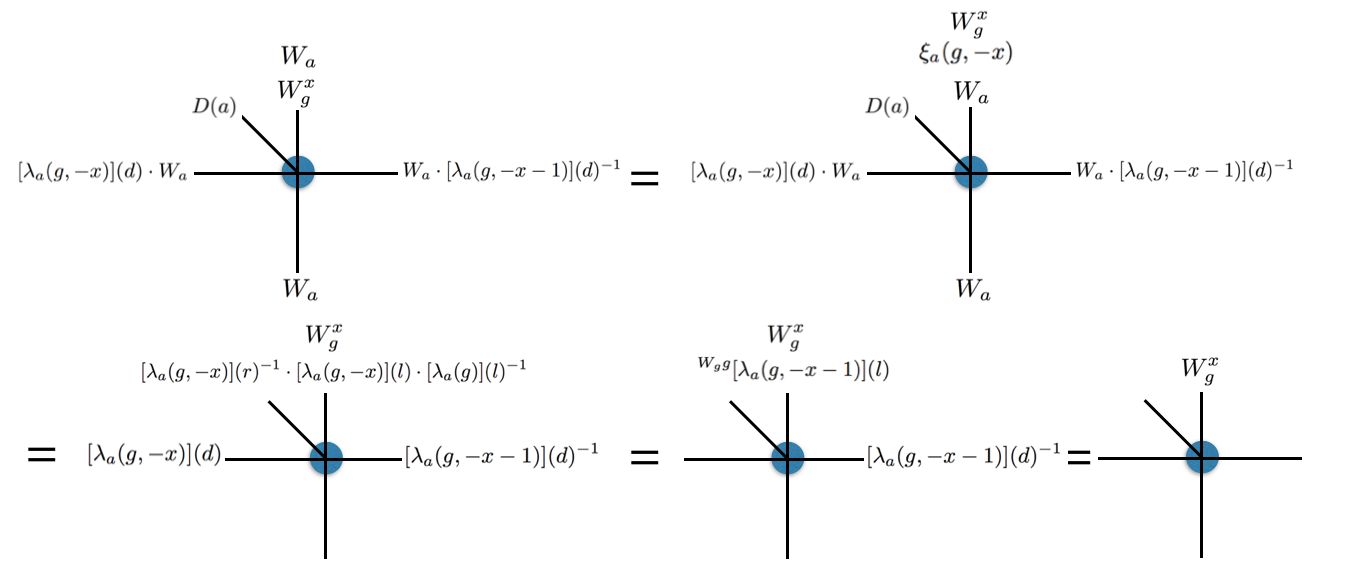}\\
\caption{The revised tensor $\tilde{T}(x,y)$ is invariant under the newly-defined symmetry operation. The first equality comes from the commutation relation $W_aW_g^x=W_g^x\xi_a(g,-x)W_a$. In the second equality we have used the invariance of tensor under $W_a$ as shown in Fig.~\ref{tensor_invariance} and the decomposition of $\xi_a(g,-x)$. In the third equality we have used the identity $\lambda_a(g,-x)(l)=^{W_gg}[\lambda_a(g,-x-1)](l)\cdot \lambda_a(g)(l)$. And we have also used the invariance of tensor under plaquette IGG as in Fig.~\ref{fig:new_tensor} in the third and fourth equalities.}
\label{symm_invariance}
\end{figure}
\end{widetext}

First, let's show that the revised tensor wave-function satisfies all the required symmetries. It's apparent that the new tensor has magnetic translation symmetry $W_{T_x}T_x,W_{T_y}T_y$ defined in Fig.~\ref{fig:new_tensor}, \textit{i.e.}
\begin{equation}
W_{T_x}T_xW_{T_y}T_y(W_{T_x}T_x)^{-1}(W_{T_y}T_y)^{-1}=W_gg.
\end{equation}

It can be proven that the new tensor is invariant under the new symmetry transformation $\tilde{W}_aa$ as shown in Fig.~\ref{symm_invariance}. The invariance of the new tensor $\tilde{T}^{(x,y)}$ under the new plaquette IGG is also apparent as shown in Fig.~\ref{fig:new_tensor}. Then we know that Eq.~\eqref{threecocycle} still holds for this state, which means the new state obtained is still an SPT state described by the same 3-cocycle $\omega$.

Finally we want to show that the new tensor now carries a projective representation $[\delta_g^{\omega}(a,b)]^{-1}$ per unit-cell. It can be proven that by acting 
$
\tilde{W}_a\cdot^a\tilde{W}_b\cdot\tilde{W}_{ab}^{-1}
$
 on all the virtual legs of a tensor we will get a phase $\delta_g^{\omega}(a,b)$ for every tensor as shown in Fig.~\ref{projective_rep}.
 \begin{widetext}
In doing so we need the following identity
\begin{equation}\label{identity}
\lambda_a(g,x)^{W_aa}\lambda_b(g,x)\lambda(a,b)\lambda_{ab}(g,x)^{-1}\cdot ^{(W_gg)^x}[\lambda(a,b)^{-1}]=[\delta_g^{\omega}(a,b)]^x.
\end{equation}

\begin{figure}[h]
\includegraphics[scale=0.6]{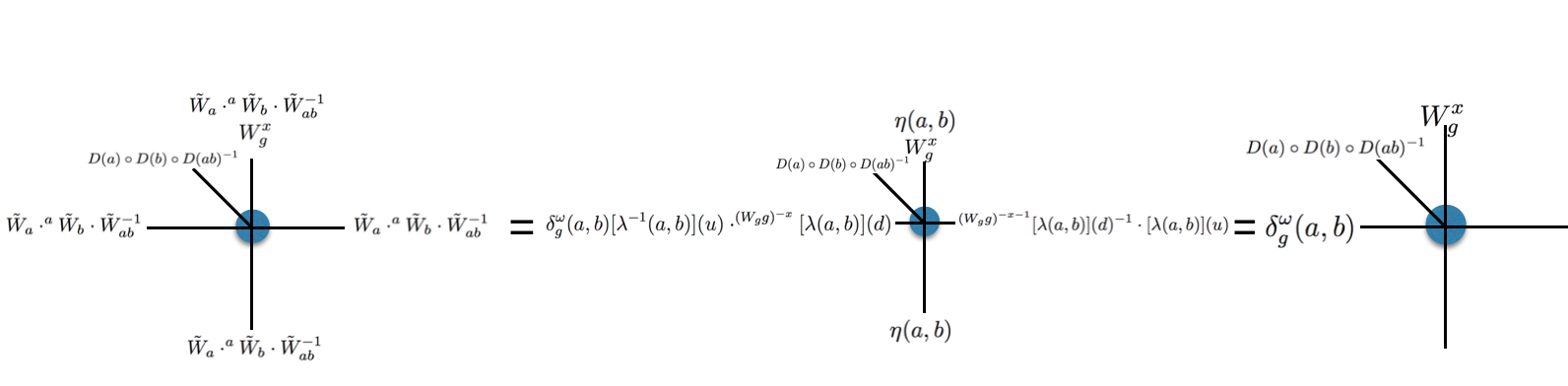}\\
\caption{Every site-tensor carries a projective representation characterized by $[\delta_g^{\omega}(a,b)]^{-1}$. We show this by acting $\tilde{W}_aa\tilde{W}_bb(\tilde{W}_{ab}ab)^{-1}$ on both the physical legs and the virtual legs of tensor $\tilde{T}^{(x,y)}$, which should leave the tensor invariant without generating any phase. But from the calculation we find that the action on vitrual legs will contribute a factor $\delta_g^{\omega}(a,b)$, therefore the representation on the physical legs are $D(a)\cdot D(b)=[\delta_g^{\omega}(a,b)]^{-1}D(ab)$, \textit{i.e.}, they are projected onto the $D_2(a)$ sector with projective representation. In the calculation above, we have used Eq.~\eqref{identity} in the first equality. And we have used the invariance of the tensor under the plaquette IGG defined in Fig.~\ref{fig:new_tensor} in the second equality.}
\label{projective_rep}
\end{figure}

Let's denote the LHS of Eq.~\eqref{identity} as $f(x)$. From Eq.~\eqref{eq:slantprod_plaq_IGG} we have $f(1)=\delta^{\omega}_g(a,b)$, then we need to find out the relation between $f(x)$ and $f(x+1)$. Using $\lambda(g,x)=^{W_gg}\lambda(g,x-1)\cdot \lambda(g)$, we can rewrite Eq.~\eqref{identity} as
\begin{equation}
\begin{split}
&[^{W_gg}\lambda_a(g,x-1)\lambda_a(g)]^{W_aa}[^{W_gg}\lambda_b(g,x-1)\cdot \lambda_b(g)]\lambda(a,b)[^{W_gg}\lambda_{ab}(g,x-1)\cdot\lambda_{ab}(g)]^{-1}\cdot ^{(W_gg)^x}[\lambda(a,b)^{-1}]\\
&=[^{W_gg}\lambda_a(g,x-1)\lambda_a(g)]^{W_aaW_gg}\lambda_b(g,x-1)\cdot ^{W_aa}\lambda_b(g)\lambda(a,b)\lambda_{ab}(g)^{-1}\cdot[^{W_gg}\lambda_{ab}(g,x-1)]^{-1}\cdot ^{(W_gg)^{x}}[\lambda(a,b)^{-1}]\\
&=^{W_gg}\lambda_a(g,x-1)[\lambda_a(g)\xi^{-1}_a(g)]^{W_ggW_aa}\lambda_b(g,x-1)\xi_a(g)^{W_aa}\lambda_b(g)\lambda(a,b)\lambda_{ab}(g)^{-1}\cdot ^{W_gg}[\lambda_{ab}(g,x-1)]^{-1}\cdot ^{(W_gg)^x}[\lambda(a,b)^{-1}]\\
&=\delta_g^{\omega}(a,b)^{W_gg}\lambda_a(g,x-1)[\lambda_a(g)\xi^{-1}_a(g)]^{W_ggW_aa}\lambda_b(g,x-1)[\xi_a(g)\lambda_a(g)^{-1}]^{W_gg}\lambda(a,b)\cdot ^{W_gg}[\lambda_{ab}(g,x-1)]^{-1}\cdot ^{(W_gg)^x}[\lambda(a,b)^{-1}]\\
&=\delta_g^{\omega}(a,b)^{W_gg}[\lambda_a(g,x-1)^{W_aa}\lambda_b(g,x-1)\lambda(a,b)\cdot \lambda_{ab}(g,x-1)^{-1}\cdot ^{(W_gg)^{x-1}}\lambda(a,b)^{-1}].
\end{split}
\end{equation}

The above derivation tells us that $f(x)=\delta_g^{\omega}(a,b)^{W_gg}f(x-1)$, therefore by induction we have $f(x)=[\delta_g^{\omega}(a,b)]^x$.

With the help of Eq.~\eqref{identity}, we can readily calculate the new IGG $\tilde{\eta}(a,b)\equiv \tilde{W}_a\cdot^a\tilde{W}_b\cdot\tilde{W}_{ab}^{-1}$. The $\tilde{\eta}(a,b)$ on the up and down virtual legs are just $\eta(a,b)$ defined before. The $\tilde{\eta}(a,b)$ on the left leg is computed as follows:
\begin{equation}
\begin{split}
&[\lambda_a(g,-x)](d)\cdot W_a\cdot ^{a}[\lambda_b(g,-x)](d)\cdot^{a}W_b\cdot ([\lambda_{ab}(g,-x)](d)\cdot W_{ab})^{-1}\\
&=[\lambda_a(g,-x)](d)\cdot^{W_aa}[\lambda_b(g,-x)](d)\cdot W_a\cdot ^aW_b\cdot(W_{ab})^{-1}[\lambda_{ab}(g,-x)](d)^{-1}\\
&=[\lambda_a(g,-x)](d)\cdot^{W_aa}[\lambda_b(g,-x)](d)\cdot \eta(a,b)\cdot[\lambda_{ab}(g,-x)](d)^{-1}\\
&=[\delta_g^{\omega}(a,b)]^{-x}[\lambda(a,b)](u)^{-1}\cdot ^{(W_gg)^{-x}}[\lambda(a,b)](d).
\end{split}
\end{equation}
\end{widetext}

Similarly, we have on the right leg 
\begin{equation}
[\delta_g^{\omega}(a,b)]^{x+1}\cdot ^{(W_gg)^{-x-1}}[\lambda(a,b)](d)^{-1}\cdot[\lambda(a,b)](u).
\end{equation}

 Therefore, as shown in Fig.~\ref{projective_rep}, we know that $\tilde{\eta}(a,b)$ is just $\delta_g^{\omega}(a,b)$ times the product of plaquette IGG shown in Fig.~\ref{fig:new_tensor} which leaves the tensor invariant up to a phase $\delta_g^{\omega}(a,b)$. Then on the physical leg we are forced to have $D(a)D(b)=[\delta_g^{\omega}(a,b)]^{-1}D(ab)$, \textit{i.e.}, $D(a)$ is projected onto $D_2(a)$ sector. So the desired on-site projective representation is achieved.

\bibliography{filling_enforced_spt}

\end{document}